\documentclass[iop,numberedappendix,appendixfloats]{emulateapj}

\usepackage{amssymb}
\usepackage{amsmath}
\usepackage{float}
\usepackage[font=footnotesize,caption=false]{subfig}
\usepackage{booktabs}
\usepackage{ulem}
\usepackage[usenames,dvipsnames]{color}
\usepackage{appendix}

\usepackage{xspace}
\usepackage{graphicx}
\usepackage[colorlinks,breaklinks, citecolor=blue,linkcolor=blue,urlcolor=blue]{hyperref}
\usepackage{apjfonts}
\usepackage{etoolbox}

\newcommand{\spark}{\small{SPARK}\normalsize{}}
\newcommand{\sersic}{S\'{e}rsic }

\def\Msun{\hbox{M$_{\odot}$}}
\def\sfrunits{\Msun\ yr$^{-1}$}
\def\halpha{H$\alpha$  }

\def\halphans{H$\alpha$}

\def\NII{[N\hspace{.03cm}II]}
\def\SII{[S\hspace{.03cm}II]}

\def\OIII{[O\hspace{.03cm}III]}

\def\kms{km s$^{-1}$}

\def\kmostd{KMOS$^\mathrm{3D}$~}
\def\kmostdns{KMOS$^\mathrm{3D}$}

\makeatletter
\patchcmd{\NAT@citex}
    {\@citea\NAT@hyper@{\NAT@nmfmt{\NAT@nm}\NAT@date}}
    {\@citea\NAT@nmfmt{\NAT@nm}\NAT@hyper@{\NAT@date}}
    {}
    {}
\patchcmd{\NAT@citex}
    {\@citea\NAT@hyper@{%
         \NAT@nmfmt{\NAT@nm}%
         \hyper@natlinkbreak{\NAT@aysep\NAT@spacechar}{\@citeb\@extra@b@citeb}%
         \NAT@date}}
    {\@citea\NAT@nmfmt{\NAT@nm}%
     \NAT@aysep\NAT@spacechar%
     \NAT@hyper@{\NAT@date}}
    {}
    {}
\patchcmd{\NAT@citex}
    {\@citea\NAT@hyper@{%
         \NAT@nmfmt{\NAT@nm}%
         \hyper@natlinkbreak
         {\NAT@spacechar\NAT@@open\if*#1*\else#1\NAT@spacechar\fi}%
         {\@citeb\@extra@b@citeb}%
         \NAT@date}}
    {\@citea\NAT@nmfmt{\NAT@nm}%
        \NAT@spacechar\NAT@@open\if*#1*\else#1\NAT@spacechar\fi%
        \NAT@hyper@{\NAT@date}}
    {}
    {}
\makeatother

\shorttitle{The KMOS$^\mathrm{~3D}$ survey}
\shortauthors{Wisnioski et al.}

\begin{document}
\title{The KMOS$^\mathrm{3D}$ Survey: design, first results, and the evolution of galaxy kinematics \\
from $0.7 \leq \lowercase{z} \leq 2.7$ \altaffilmark{$\dagger$}}
\author{E. Wisnioski\altaffilmark{1}, N.M. F\"orster Schreiber\altaffilmark{1},  
S. Wuyts\altaffilmark{1}, 
E. Wuyts\altaffilmark{1}, 
K. Bandara\altaffilmark{1},
D. Wilman\altaffilmark{1,2}, 
R. Genzel\altaffilmark{1,3,4}, \\
R. Bender\altaffilmark{1,2},
R. Davies\altaffilmark{1},  
M. Fossati\altaffilmark{2,1},
P. Lang\altaffilmark{1}, 
J. T. Mendel\altaffilmark{1}, 
A. Beifiori\altaffilmark{1,2}, 
G. Brammer\altaffilmark{5}, 
J. Chan\altaffilmark{1}, 
M. Fabricius\altaffilmark{1},  \\
Y. Fudamoto\altaffilmark{1},
S. Kulkarni\altaffilmark{1},  
J. Kurk\altaffilmark{1}, 
D. Lutz\altaffilmark{1}, 
E. J. Nelson\altaffilmark{6}, 
I. Momcheva\altaffilmark{6},
D. Rosario\altaffilmark{1},
R. Saglia\altaffilmark{1}, 
S. Seitz\altaffilmark{1,2}, \\ 
L.J. Tacconi\altaffilmark{1},
P. G. van Dokkum\altaffilmark{6} } 
\affil{\altaffilmark{1}{Max-Planck-Institut f\"{u}r extraterrestrische Physik (MPE), Giessenbachstr. 1, D-85748 Garching, Germany; \href{mailto:emily@mpe.mpg.de}{emily@mpe.mpg.de}}\\
\altaffilmark{2}{Universit\"ats-Sternwarte, Ludwig-Maximilians-Universit\"at, Scheinerstrasse 1, D-81679 M\"unchen, Germany}\\
\altaffilmark{3}{Department of Physics, Le Conte Hall, University of California, 94720 Berkeley, USA}\\
\altaffilmark{4}{Department of Astronomy, Hearst Field Annex, University of California, Berkeley, USA}\\
\altaffilmark{5}{Space Telescope Science Institute, 3700 San Martin Drive, Baltimore, MD 21218, USA}\\
\altaffilmark{6}{Department of Astronomy, Yale University, New Haven, CT 06511, USA}}

\altaffiltext{$\dagger$}{Based on observations obtained at the Very Large Telescope (VLT) of the European Southern Observatory (ESO), Paranal, Chile (ESO program IDS 092A-0091, 093.A-0079)}
\begin{abstract}
We present the \kmostd survey, a new integral field survey of over $600$ galaxies at $0.7<z<2.7$ using KMOS at the Very Large Telescope (VLT). The \kmostd survey utilises synergies with multi-wavelength ground and space-based surveys to trace the evolution of spatially-resolved kinematics and star formation from a homogeneous sample over 5 Gyrs of cosmic history. Targets, drawn from a mass-selected parent sample from the 3D-HST survey, cover the star formation$-$stellar mass ($M_*$) and rest-frame $(U-V)-M_*$ planes uniformly. We describe the selection of targets, the observations, and the data reduction. In the first year of data we detect \halpha emission in 191 $M_*=3\times10^{9}-7\times10^{11}$ \Msun~galaxies at $z=0.7-1.1$ and $z=1.9-2.7$.   In the current sample $83$\% of the resolved galaxies are rotation-dominated, determined from a continuous velocity gradient and $v_\mathrm{rot}/\sigma_0>1$, implying that the star-forming `main sequence' (MS) is primarily composed of rotating galaxies at both redshift regimes.  When considering additional stricter criteria, the \halpha kinematic maps indicate at least $\sim70$\% of the resolved galaxies are disk-like systems.
Our high-quality KMOS data confirm the elevated velocity dispersions reported in previous IFS studies at $z\gtrsim0.7$. For rotation-dominated disks, the average intrinsic velocity dispersion decreases by a factor of two from $50$ \kms~at $z\sim2.3$ to $25$ \kms~at $z\sim0.9$. Combined with existing results spanning $z\sim 0-3$, we show that disk velocity dispersions follow an evolution that is consistent with the dependence of velocity dispersion on gas fractions predicted by marginally-stable disk theory.\\
\end{abstract}



\section{Introduction}
\setcounter{footnote}{0}

The baryonic growth of galaxies near the peak of the galaxy formation epoch was dominated by in-situ star formation maintained through an equilibrium of gas accretion from the cosmic web and (mainly) minor mergers, with star formation and gas recycling through the circum-galactic medium driven by stellar and AGN feedback  (e.g., \citealt{2006MNRAS.370..645B,2009ApJ...703..785D,2010ApJ...718.1001B,2010MNRAS.405.1690D,2010MNRAS.404.1111G,2012MNRAS.421...98D,2013ApJ...772..119L,2013MNRAS.436.3031V,2014arXiv1402.2283D,2014arXiv1407.7040S,2014arXiv1409.0009S}).
The balance between these mechanisms maintains galaxies on a tight ``main sequence'' (MS) in  star formation rate (SFR) vs. stellar mass ($M_*$), with $\sim0.3$ dex scatter. The decreasing zero point of the MS reflects the evolution of the cosmic SFR density from the ``peak activity'' at $z\sim1-2.5$ to the ``winding down'' epochs at $z<1$ (e.g. \citealt{2007ApJ...660L..43N,2007ApJ...670..156D,2007A&A...468...33E,2011ApJ...739L..40R,2014arXiv1407.1843W}). This scenario is further empirically motivated by the dominance of disk-like systems among MS star-forming galaxies (SFGs) out to $z\sim2.5$ (e.g., \citealt{2006ApJ...645.1062F,2009ApJ...706.1364F,2006Natur.442..786G,2008ApJ...687...59G,2008ApJ...682..231S,2009A&A...504..789E,2012A&A...539A..92E,2011ApJ...742...96W}) and the evolution of molecular gas mass fractions (e.g. \citealt{2010Natur.463..781T,2013ApJ...768...74T,2010ApJ...713..686D,2013ApJ...778....2S,2014arXiv1409.1171G}).

In this picture, galaxies grow in stellar mass mostly while on the MS (e.g. \citealt{2011ApJ...739L..40R}). Rare occasional bursts of star-formation can lead to temporary offsets above the MS (e.g. through mergers). Above $M_*\sim 10^{11}$ \Msun, most galaxies appear to be rapidly quenched dropping below the MS at $z\lesssim2.5$ \citep{2010ApJ...721..193P,2012ApJ...745..179W}. Spatially resolved information of the kinematics, star-formation and nebular conditions provide key insights into the processes that drive galaxy growth, bursts, and quenching by probing their dynamical state and key signatures of secular- or merger-driven growth.

Resolved kinematic surveys that utilise integral field spectroscopy (IFS) $-$ ranging from a handful to $\sim100$ SFGs $-$ have been critical in establishing this picture. By revealing the kinematic nature and prevalence of rotating disks among luminous SFGs at $z\sim1-4$, these surveys have contributed key evidence in support of ``smoother'' mass accretion and for the importance of internal dynamical processes in the early evolution of massive galaxies (e.g. \citealt{2006Natur.442..786G, 2006ApJ...645.1062F,2008ApJ...687...59G,2009ApJ...699..421W,2009ApJ...697.2057L,2009ApJ...706.1364F,2009A&A...504..789E,2012A&A...539A..92E,2011MNRAS.417.2601W,2012A&A...539A..91C,2012ApJ...760..130S,2013ApJ...779..139S,2014MNRAS.443.2695S}). 
However due to practical limitations of sample selection and telescope time, large statistical and cohesive data sets at $z\gtrsim1$ have been unattainable especially at the depth required to measure subtle kinematic and emission line features (see \citealt{2013arXiv1305.2469G} for a review of IFS results). Currently, a new multiplexed generation has begun for high-redshift galaxy studies of dynamics and chemical evolution with the advent of deployable integral field unit (IFU) systems and multi-object spectrographs operating in the near-infrared where key emission lines are redshifted at $0.5\lesssim z\lesssim3$. 

In this paper we introduce the \kmostd Survey $-$ hereafter simply \kmostdns $-$ a survey that leverages new rest-frame-optical redshift catalogs with the multiplex of a deployable IFU system for a 20 fold increase in efficiency over previous single-IFU surveys. \kmostd is a guaranteed time program using the \textit{K-band Multi-Object Spectrograph} (KMOS; \citealt{2004SPIE.5492.1179S,2013Msngr.151...21S}) on the VLT to map kinematics, star formation, metallicity and the physical conditions of the ISM of a mass-selected sample. Primary emission lines of interest are \halphans, \NII, and \SII, observed through the $YJ$, $H$, and $K$ atmospheric windows probing galaxies at $0.7<z<2.7$. 

Our strategy is designed to achieve a wide and uniform coverage of the SFG population at $M_*\gtrsim10^{9.5}$ \Msun~while still obtaining high signal to noise (S/N) data of individual galaxies to determine good quality line ratios and line profiles in individual galaxies. The baseline sample is 600 galaxies $-$ balancing observational depth with the statistics required to explore trends of kinematic properties as a function of e.g. $M_*$, SFR, $(U-V)_\mathrm{rest}$ color, and redshift with 10-20 galaxies in individual $M_*$, SFR, $(U-V)_\mathrm{rest}$ bins. \kmostd targets are drawn from the 3D-HST space-based near-infrared (IR) grism survey \citep{brammer:2012:04,2014arXiv1403.3689S}. The 3D-HST sample, with grism- and spectroscopic- based redshifts, forms a more representative sample of the full SFG population (including dusty and low sSFR galaxies) than rest-UV spectroscopic samples, from which many past IFS targets were drawn.

Capitalizing on crucial synergies with multi-wavelength ground and space-based surveys, \kmostd will draw connections between ionized gas properties (spatially resolved kinematics, star formation, outflows, excitation and metallicity) and stellar properties (stellar structure, stellar populations, and environment) to provide constraints on the physical mechanisms driving mass growth, feedback and star-formation shutdown.  These connections have already been proved possible with the first year data from \kmostdns, which reveal nuclear outflows in a high fraction of $\log(M_*$[\Msun]$)>10.9$ galaxies \citep{2014ApJ...796....7G} and  confirm an evolution to lower \NII/\halpha at earlier times but find no correlation between \NII/\halpha and SFR \citep{2014ApJ...789L..40W}.

This paper focuses on the kinematic properties derived from \halphans, near-IR continuum, velocity, and velocity dispersion maps. We combine galaxy dynamics, structural parameters, and multi-band imaging to make robust kinematic determinations for $>100$ galaxies between $0.7<z<2.7$, expanding the redshift range of previous individual surveys with homogeneous selection and ancillary data. The wide redshift range and large dataset of the full sample will consistently track the evolution of galaxies from the peak in cosmic SFR density into the epochs of its decline.

Previous IFS and long-slit observations of high-redshift SFGs have revealed large ionized gas velocity dispersions $-$ $5-10\times$ local galaxies $-$ after correcting for instrumental resolution and rotational broadening (e.g. \citealt{2006ApJ...653.1027W,2006ApJ...645.1062F,2009ApJ...706.1364F,2006Natur.442..786G,2007ApJ...669..929L,2009ApJ...697.2057L,2009ApJ...699..421W,2011A&A...528A..88G,2011MNRAS.417.2601W,2012A&A...539A..92E,2012ApJ...760..130S,2012ApJ...758..106K}). Various origins have been proposed for the high velocity dispersions such as feedback \citep{Dib:2006fk,Green:2010fk}, gas accretion \citep{2006ApJ...645.1062F,2006Natur.442..786G,2010ApJ...712..294E,2010ApJ...719..229G,2012ApJ...754...48F}, conversion of gravitational potential energy into random motions either at the outer disk/cosmic web boundary \citep{2006ApJ...645.1062F,2006Natur.442..786G,2008ApJ...687...59G,2012MNRAS.421..818C}, or in the inner disk by torques in clump-clump interactions \citep{2004A&A...413..547I,2010MNRAS.409.1088B,2010ApJ...719.1230A}. We exploit here our first year results to set tighter constraints on the evolution and origin of disk velocity dispersions. 

This paper is organized as follows. Section~\ref{sec.obs} describes the \kmostd sample selection, first year observations, and data reduction. Section~\ref{sec.results} describes the \halpha detected galaxies, presents their kinematic maps and the methods used to derive their kinematic properties, which are used in Section~\ref{sec.galclass} to determine galaxy classifications. In Section~\ref{sec.analysis} we utilise the new \kmostd results at $z\sim1$ and $z\sim2$ to investigate the evolution of disk velocity dispersion of ionized gas and possible dependencies on other derived properties. This paper is summarized in Section~\ref{sec.conclusions}. 
We assume a $\Lambda-$CDM cosmology with $H_0 = 70$ km s$^{-1}$ Mpc$^{-1}$, $\Omega_m = 0.3$, and $\Omega_\Lambda = 0.7$. For this cosmology, $1''$ corresponds to $\sim7.8$ kpc at $z = 0.9$ and $\sim8.2$ kpc at $z = 2.3$. Magnitudes are given in the AB photometric system. A \cite{2003PASP..115..763C} initial mass function (IMF) is adopted throughout.

\section{Observations and data}
\label{sec.obs}

\subsection{Survey design \& sample selection}
\label{subsec.selection}
We select \kmostd targets from the 3D-HST Treasury Survey \citep{brammer:2012:04,2014arXiv1403.3689S} in the fields accessible from the VLT (GOODS-S, COSMOS, UDS). The 3D-HST Treasury Survey is a \textit{Hubble Space Telescope} WFC3/G141 grism survey, which provides spectra with resolution of $R\sim130$ over $\lambda = 1.1 - 1.7~\mu$m in five `deep fields' (COSMOS, GOODS-S, GOODS-N, UDS, and AEGIS). These spectra provide redshifts from continuum breaks and/or emission lines with expected precision of $\sim700-1000$ km s$^{-1}$ \citep{brammer:2012:04,2013ApJ...770L..39W}.  Where only continuum is measured in the grism the continuum is used to constrain photometric redshifts, based on broadband photometry, by contributing a highly sampled portion of the SED. The grism redshifts rely on rest-frame optical continuum and spectral features and do not require a priori emission line detections.
Thus target selection using 3D-HST grism redshifts effectively reduces the inherent bias towards blue, star-forming, dust-free galaxies of previous rest-frame-UV spectroscopically-selected samples at $z>1.5$. 

The 3D-HST survey overlaps with the imaging fields of the CANDELS survey, which contributes high-resolution WFC3 near-IR imaging along with ACS imaging for all the \kmostd targets \citep{2011ApJS..197...35G,2011ApJS..197...36K}. The fields further benefit from multi-wavelength coverage from the X-ray to far-IR and radio (e.g. \citealt{2008ApJS..179..124U,lutz:2011aa,2011ApJS..195...10X,2012ApJS..201...30C,2013A&A...553A.132M}). The consistency of deep infrared photometry for all targets yields a homogeneous set of spectral energy distribution (SED) parameters including stellar mass, UV+IR SFRs, and correction for global dust extinction following \cite{2011ApJ...738..106W}. We assume solar metallicity, the \cite{2000ApJ...533..682C} reddening law, and either constant or exponentially declining SFRs. Star-formation rates are determined from the same SED fits or, for objects observed and detected in at least one of the mid- to far-IR (24$\mu$m to 160$\mu$m) bands with the Spitzer/MIPS and Herschel/PACS instruments, from rest-UV+IR luminosities through the Herschel-calibrated ladder of SFR indicators of \cite{2011ApJ...738..106W}. High resolution (FWHM$\sim0.15-0.20$'') four-band imaging ($VIJH$) in UDS and COSMOS and seven-band imaging ($BVizYJH$) in GOODS-S provides resolved information of stellar populations, dust extinction and stellar mass maps that will complement the kinematics, star formation and nebular emission data derived from KMOS for a combined view of resolved gas and stellar profiles of individual galaxies \citep{2012ApJ...753..114W,2013ApJ...779..135W,2013ApJ...763L..16N,2014arXiv1402.0866L}.

\kmostd galaxies are selected within three redshift bands that cover $0.7 \lesssim z \lesssim 2.7$, where \halpha emission is located in the $YJ$, $H$, and $K$ band filters of KMOS, with a magnitude cut of $Ks<23$. This corresponds to targets being drawn from a 95 per cent mass complete sample at stellar masses, $\log(M_*[\Msun])>9.65, 10.22,10.53$ for the redshift ranges $0.7<z<1.1$, $1.2<z<1.8$, $1.9<z<2.7$ respectively. We note that the mass completeness is dependent on SFR and thus the values quoted are conservative. At higher SFRs we reach down to lower masses than quoted above. 

We use redshift probability functions to avoid contamination of the \halphans$-$\NII~complex with OH night skylines and atmospheric absorption windows. For any individual object the effectiveness of the skyline-avoidance criterion is dependent on redshift precision. With the $Ks$ magnitude cut all galaxies in the resulting sample have either a spectroscopic or grism redshift.  For the objects with grism-based redshifts the redshift probability function is conservatively convolved with a $\pm1000$ km s$^{-1}$ Gaussian. The location and density of OH skylines is compared with the redshift probability functions for an estimate of the likelihood that \halpha will be contaminated. Galaxies with the highest probability of contamination are excluded from the sample\footnote{The exact value of this criterion was determined based on previous observations by team members and was found to be appropriate after the first set of KMOS observations.}. The OH and atmosphere avoidance criteria removes $\sim$70\% of possible targets in the full redshift range.The availability of existing spectroscopic redshifts in the planned \kmostd sample is highly dependent on the field. In GOODS-S where a wealth of spectroscopic data are available, 77\% of \kmostd target galaxies have a spectroscopic redshift. In contrast, the percentage of target galaxies in COSMOS and UDS that have spectroscopic redshifts is 10\% and 13\% respectively.

Further cuts are made based upon grism quality flags as described by \cite{brammer:2012:04}. These additional cuts remove $\sim$15\% of possible targets after the OH avoidance cut. Finally, based upon visual inspection of the grism spectra, $\lesssim10$\% objects are removed due to low S/N detections of the continuum ($z_\mathrm{grism}\sim z_\mathrm{phot}$) or overlapping targets in the grism (contaminated continuum flux). The fraction of objects removed for grism contamination is higher in the fields that are dominated by grism redshifts.

\subsubsection{Survey strategy}
The survey is designed to reach a balance between total number of galaxies and data quality resulting from deep observations necessary to extract high-quality science. All $z\sim1, 1.5, 2$ galaxies will be observed for a minimum of four, six, and eight hours respectively. The large dynamic range in expected \halpha luminosity of our target sample (factor of $\sim1000$) requires that longer exposure times for some individual objects to ensure an unbiased sampling of the underlying population.  We therefore adopt an observing strategy such that there is significant overlap between adjacent pointings, allowing us to re-observe faint targets and guaranteeing high angular completeness.

In this paper we present the first year of \kmostd observations, which have largely focused on massive galaxies ($M_*>10^{10}$\Msun) with SFRs $>0.1$ \sfrunits~and $>0.2$ \sfrunits~in redshift ranges $0.7<z<1.1$ and $1.9<z<2.7$. The full redshift ranges of our first year data are $0.67<z<1.04$ and $2.00<z<2.68$ with medians of $z=0.90$ and $z=2.30$ respectively. These datasets hereafter will be referred to as the $z\sim1$ and $z\sim2$ samples.

\subsection{Observations}
\label{subsec.kmos}
 KMOS is a multiplexed near-infrared IFS system with 24 deployable $2.8''\times2.8''$ image slicers over an $7.2'$ diameter patrol field. The IFS units connect to three cryogenic grating spectrometers with 2k$\times$2k Hawaii-2RG HgCdTe detectors. The typical spectral resolution, $R$, in the $YJ$, $H$, and $K$-band filters used for \kmostd is 3400, 4000, and 4200 respectively. KMOS is a seeing-limited instrument with square $0.2''$ spatial pixels comprising $14\times14$ pixel IFS units.

Observations were prepared with the KMOS Arm Allocator (KARMA; \citealt{2008SPIE.7019E..27W}). Hereafter an individual KARMA setup, or 24 arm allocation, will be referred to as a ``pointing''. Each pointing was observed for either 300s or 600s using a standard object-sky-object dither pattern, where sky exposures were offset to a clear sky position. Additional subpixel/pixel shifts were included in the object-sky dithering to average over bad pixels.  One IFU in each detector is allocated to a star during science observations. The stars are used to monitor the variations in the PSF and photometric conditions between the observed frames and in each of the three detectors.

\begin{figure*}
\includegraphics[scale=0.85,  trim=0.2cm 0cm 0.2cm 10cm, clip ]{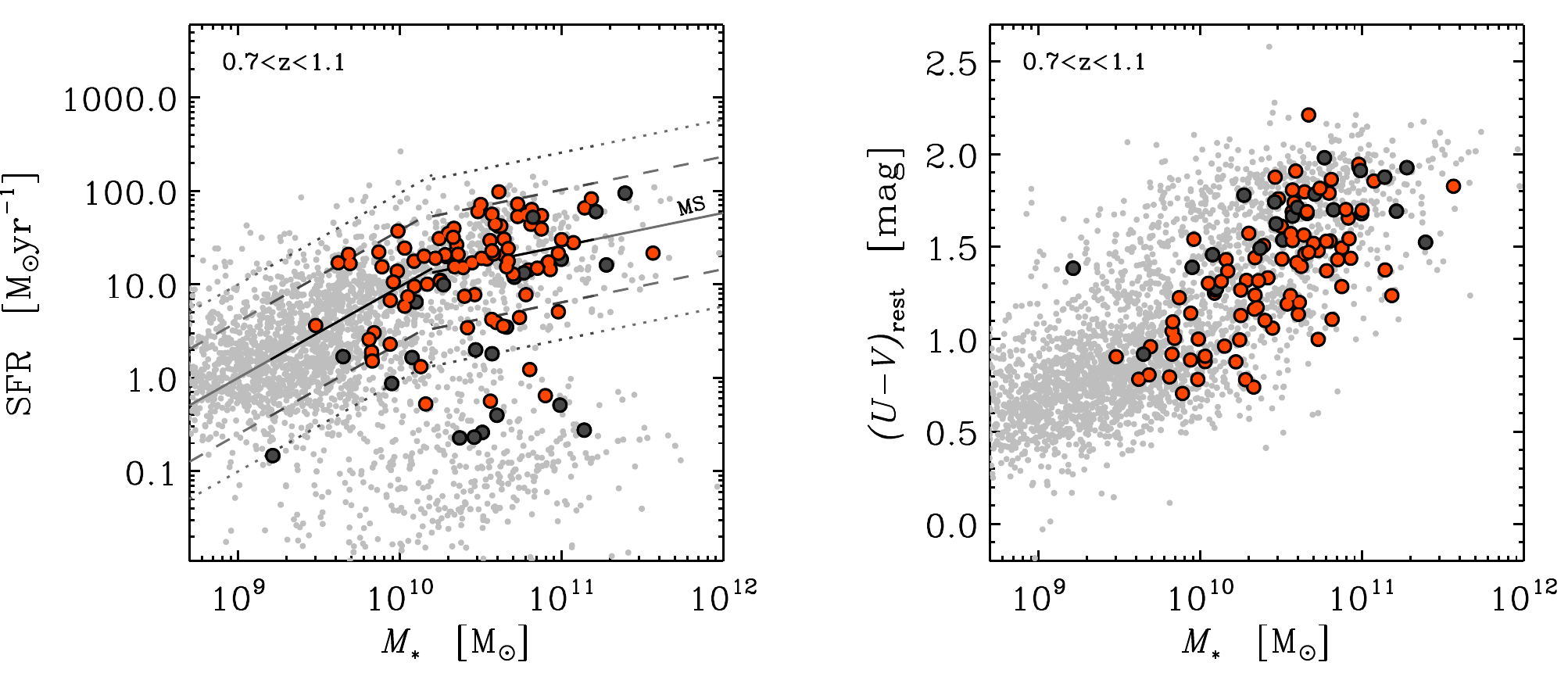}
\includegraphics[scale=0.85,  trim=0.2cm 0cm 0.2cm 10cm, clip ]{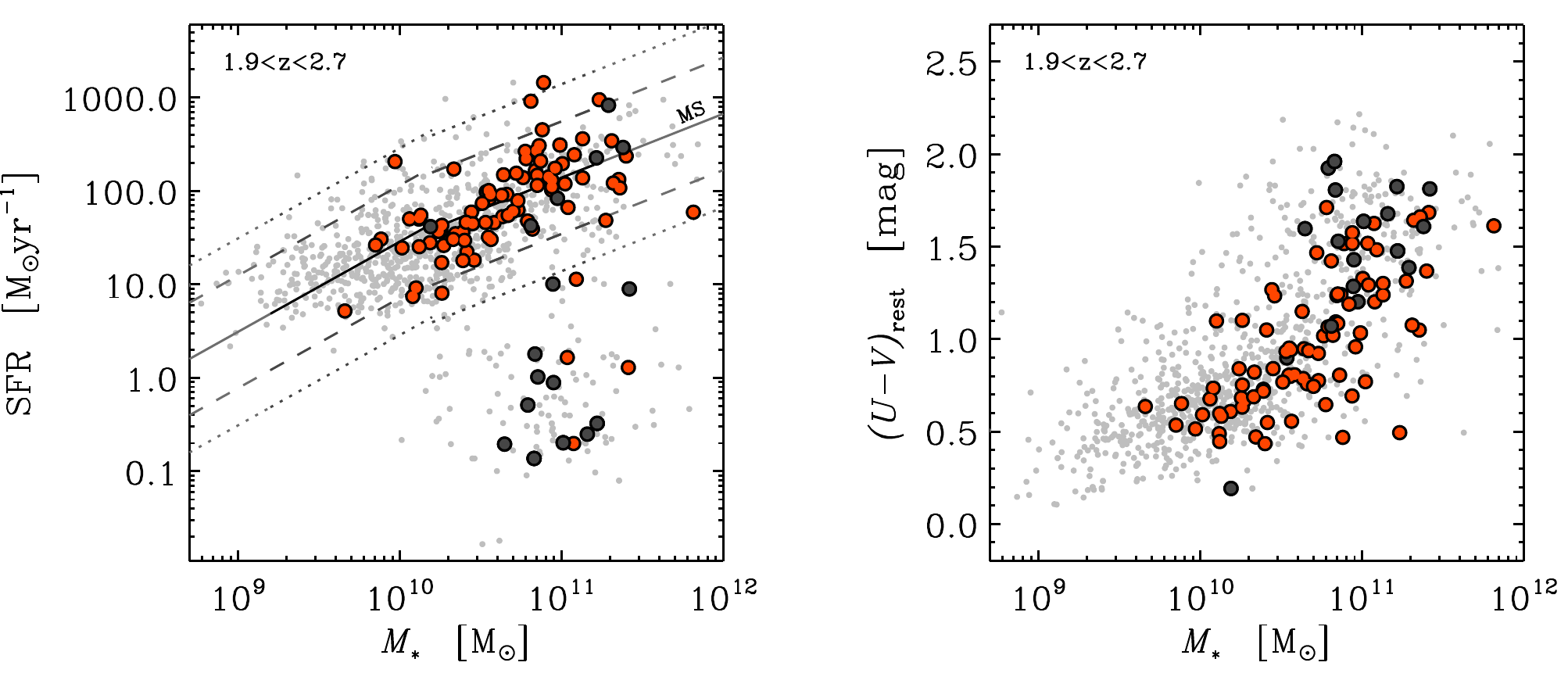}
\caption[...]{Properties of the observed \kmostd sample at $0.7<z<1.1$ \textit{(top)} and $1.9<z<2.7$ (\textit{bottom}) in the SFR$-M_*$ plane \textit{(left)} and ($U-V$)$_\mathrm{rest}-M_*$ plane \textit{(right)}. Small grey points show the 3D-HST sample without the magnitude and OH contamination selection criterion imposed (Section~\ref{subsec.selection}). Large symbols are the \kmostd galaxies that have been observed thus far. Large red symbols are galaxies with \halpha detections from KMOS and large gray symbols are not detected with the current KMOS data. SFRs are derived from a Herschel calibrated ladder of SFR indicators \citep{2011ApJ...738..106W} and $M_*$ are derived from SED fits. The broken power-law parameterization, valid between $\log{M_*}$[\Msun]$=9.2-11.2$, is shown by the solid lines, as defined using 3D-HST data in all CANDELS fields from $0.5<z<2.5$ using UV+IR SFRs \citep{2014arXiv1407.1843W}. Power law coefficients for redshifts between the bins given in \cite{2014arXiv1407.1843W} are obtained through interpolation of the coefficients as a function of stellar mass. Dashed lines and dotted lines show $4\times$ and $10\times$ above and below the canonical MS respectively. 
}
\label{fig.ms}
\end{figure*}

Observations were taken during Commissioning on 2013 January 24-25, 29-30, March 29-31,  during P92 on 2014 October 30 - November 15, December 7-8, and January 6-10, 2014, and during P93 on 2014 April 19-23 and May 9-10. The observations were taken in good conditions with typical seeing of $0.6''$ in $YJ$ and $K$-band. Fourteen pointings have been observed, 7 in $YJ$ and 7 in $K$. By utilizing overlapping pointings, to allow longer exposure times on certain objects, 37 galaxies have exposure times between 11 and 20 hours. In total 223 galaxies were directly targeted, 106 galaxies at $z\sim1$ and 117 galaxies at $z\sim2$. In some cases nearby galaxies were observed within the IFU of the primary target. As a result, 11 additional targets at $z\sim1$ and 12 at $z\sim2$ were observed for a total of 246 galaxies observed. Of the 23 serendipitous galaxies 6 fit the \kmostd selection criteria outlined in Section~\ref{subsec.selection}, the remaining 17 fall out of the selection due to likely OH contamination,  $Ks>23$, or a bad grism flag.
Figure~\ref{fig.ms} shows the observed galaxies in the SFR$-M_*$ and $(U-V)_\mathrm{rest}-M_*$ planes.

\subsection{Data reduction}
\label{subsec.dr}
All data were reduced with the Software Package for Astronomical Reduction with KMOS ({\sc spark}; \citealt{2013arXiv1308.6679D}) using recipes outlined in the {\sc spark} Instructional Guide\footnote{ftp://ftp.eso.org/pub/dfs/pipelines/kmos/kmos-pipeline-cookbook-0.9.pdf}. The reduction steps include flat fielding, illumination correction, wavelength calibration, and the sky subtraction technique developed by \cite{2011ApJ...741...69D}. Additional processing was done on the raw and reduced data to address known detector issues including removal of the Odd-Even Effect and correction for level offsets in the readout channels.

Individual frames were median combined into final cubes using spatial shifts measured from the average center of the stars within the same pointings. Variations in flux and seeing among the combined frames were typically $\leq10$\% and  $\leq0.1$'', respectively.

Estimates of the PSF size for each pointing are made from the average FWHM of the stars included in the pointings. The FWHM of the stars are measured from the combined data cubes using a 2D Gaussian. The star and science observations are simultaneous giving a measurement of the PSF that is consistent with the total time and conditions corresponding to the galaxy data. The mean and median PSF size for the sample presented here is 0.62$''$ and 0.58$''$ respectively.  A more complete analysis of the PSF for \kmostd observations will be presented by Wilman et al. (\textit{in preparation}). 

\subsection{Kinematic mapping}
\label{subsec.kinmaps}
The IDL  emission line fitting code {\sc linefit} is used to derive the kinematic maps from the reduced data cubes \citep{2009ApJ...706.1364F,2011ApJ...741...69D}. {\sc linefit}, originally developed for SINFONI, has been adapted to be used with KMOS data. In short, {\sc linefit} fits a 1D Gaussian model that is convolved with the instrument's line spread function (see below). The fits are performed on all individual continuum subtracted spaxels in the final combined KMOS cubes. The continuum level is determined from two line-free spectral windows around the \halphans-\NII~complex. The estimate of the continuum is the mean of the values between the 40th and 60th percentiles, in each window. {\sc linefit} takes into account the three-dimensional noise properties of the input data via weighting in the fits.  The code determines robust and realistic uncertainties on the derived flux and kinematic properties using Monte Carlo techniques, for more details see Appendix B of \cite{2011ApJ...741...69D}.

To correct for instrumental broadening a Gaussian line profile is created in {\sc linefit} at the instrumental resolution and then convolved with the model emission line to match the observed profile. Due to differences of resolution in each IFU and along the wavelength axis \citep{2013arXiv1308.6679D}, a unique value for the resolution is derived for each object based on which IFU and at which wavelength \halpha emission is observed. The resolution is derived from the arc lines and OH sky emission at the same wavelength for the same IFU. For objects observed in multiple IFUs the resolution is determined from the skylines in the un-sky-subtracted combined cube for that particular object.

All observations are median filtered spatially with a $2\times2$ pixel box to slightly increase the S/N per pixel without significant loss of spatial resolution. No spectral smoothing or filtering is applied to the data. 

In the resulting \halpha kinematic maps, pixels where the S/N of \halpha drops below $\sim5$ are masked out. The kinematic maps on average reach $2\times$ the effective H-band radius. Further pixels are masked for which the line center and width were clearly unreliable based on inspection of the velocity and velocity dispersion maps. The latter masking generally corresponds to the following criteria: velocity and dispersion of a given pixel exceeded by a factor of at least two the typical maximum value over the maps, a velocity uncertainty of $>100$ km s$^{-1}$, and/or a relative velocity dispersion uncertainty of $>50$\%. Continuum maps are constructed for all galaxies.

\section{Results}
\label{sec.results}

From the initial observations presented in this paper, 179 targeted galaxies have \halpha emission line detections, 85 detections at $z\sim1$ and 94 detections at $z\sim2$, translating into a 80\% success rate for targeted galaxies at both redshifts.  An additional 12 serendipitous galaxies were detected within the IFUs of targeted galaxies bringing the total number of detections to 191 galaxies (90 at $z\sim1$ and 101 at $z\sim2$).  In 77\% of the galaxies with \halpha emission detected \NII$\lambda6584$~emission is also detected. 

Figure~\ref{fig.ms} shows the location of the \kmostd galaxies with the detected galaxies identified in red. The observed galaxies are shown relative to the underlying 3D-HST parent catalog in SFR, stellar mass, and $(U-V)_\mathrm{rest}$ color for the $z\sim1$ and $z\sim2$ samples, demonstrating the selection techniques outlined in Section~\ref{subsec.selection}. At the massive end, where galaxies have been targeted in our initial observations, they closely follow the underlying galaxy population in the SFR$-M_*$ and $(U-V)_\mathrm{rest}-M_*$ planes.

The normalization of the MS is shown at each redshift with a solid line as defined by the broken power-law MS parametrization from \cite{2014arXiv1407.1843W}. The detection fraction is higher on the MS in the SFR$-M_*$ plane, particularly within a factor of $4\times$ in SFR of the MS, and for bluer galaxies, $(U-V)_\mathrm{rest}<1.3$. However,  with current integration times we have also detected a number of galaxies in \halpha $10\times$ below the MS and with $(U-V)_\mathrm{rest}=1.5-2.0$ at $z\sim1$ and $z\sim2$.

\subsection{Redshift accuracy}
The agreement between $z_\mathrm{kmos}$ and $z_\mathrm{3D-HST}$, $| z_\mathrm{kmos} - z_\mathrm{3D-HST} |/(1+z_\mathrm{kmos})$, of the detected \kmostd galaxies reflects the redshift accuracy expected from the 3D-HST survey \citep{brammer:2012:04}. For galaxies selected on a prior spectroscopic redshift and detected with KMOS the median redshift difference, $\Delta z/(1+z)$, is 96 \kms. For galaxies selected on a grism redshift and detected with KMOS the median $\Delta z/(1+z)$ is 703 \kms.  Even galaxies with no significant line detection in the grism provide useful constraints for redshifts \citep{2013ApJ...770L..39W}. Within the \kmostd sample these galaxies have median $\Delta z/(1+z)$ of 1650 \kms. We conclude that grism redshifts have an accuracy well suited for \kmostdns, and provide a significant improvement over photometric redshifts, and thus a reliable target sample for our survey.

The detection fraction is marginally higher for galaxies with spectroscopic redshifts in the parent catalog, 85\%, than for galaxies with grism redshifts, 78\%.   While galaxies with grism redshifts are fainter on average and are more often from continuum based redshifts, we do not find a strong trend of the frequency of non-detections or $\Delta z/(1+z)$ with target brightness as probed by the F140W magnitudes. The non-detections may be a result of either incomplete observing time with future observations planned (50\% of non-detections have incomplete observations) or of large redshift errors shifting \halpha outside of the observed band. Within the detected sample 19 galaxies have redshifts from \kmostd deviant from the expected redshift from 3D-HST by $>10,000$ \kms~(2 galaxies at $z\sim1$ and 17 galaxies at $z\sim2$).

\subsection{Kinematic measurements}
\label{subsec.vel}
\label{subsec.disp}
\begin{figure*}
\begin{center}
\includegraphics[scale=0.42,  trim=0.5cm 0cm 0cm 0cm, clip ]{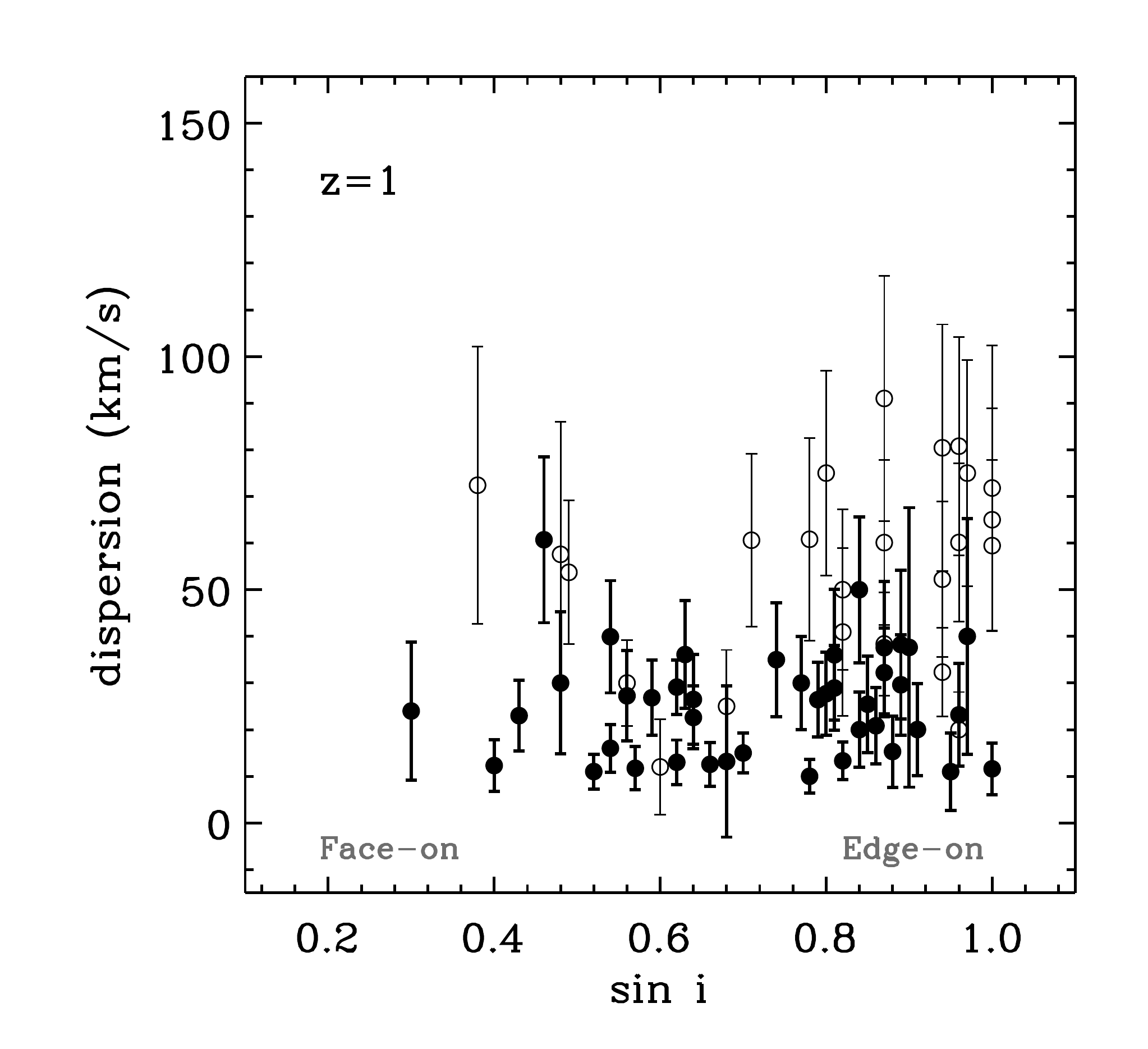}
\includegraphics[scale=0.42,  trim=0.5cm 0cm 0cm 0cm, clip ]{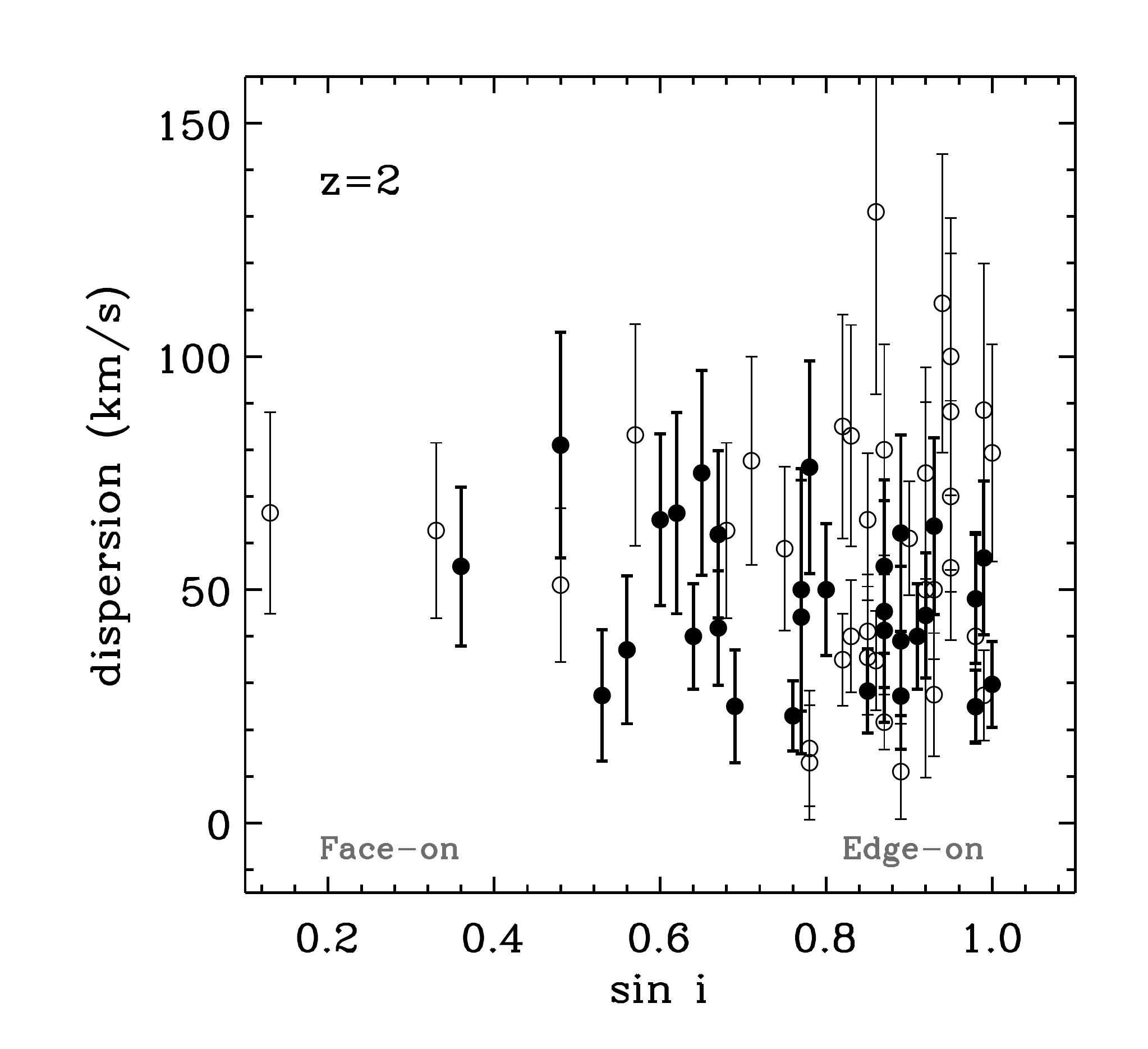}
\caption[...]{Velocity dispersion, $\sigma_0$, as a function of inclination for \kmostd galaxies at $z\sim1$ (left) and $z\sim2$ (right). Disk galaxies, as described in Section~\ref{sec.galclass}, with sufficiently high S/N per resolution element to constrain the velocity dispersion in the outer regions of the galaxies are shown by solid circles, the remaining \kmostd galaxies are shown by open circles. }
\label{fig.dispsini}
\end{center}
\end{figure*}

In the following sections, we examine the kinematic properties of our present sample. For all galaxies we derive global properties from the integrated spectrum.  For galaxies with \halpha emission extending beyond one resolution element (70\% of detected galaxies) \halpha emission, velocity and velocity dispersion maps are produced following the procedure in Section~\ref{subsec.kinmaps}. For resolved galaxies we derive additional parameters from  one-dimensional axis profiles extracted along the kinematic major axis, used for kinematic classification in Section~\ref{sec.galclass}. The kinematic axis is determined from the 2D velocity field as the direction of the largest observed velocity difference, with the kinematic center defined as the spatial location of the velocity midpoint between the velocity extrema. We measure velocity and velocity dispersion profiles by extracting spectra in apertures equivalent to the size of the average PSF along the kinematic major axis and fit for \halpha emission using the {\sc linefit} code described in Section~\ref{subsec.kinmaps}. For all galaxies with velocity gradients we measure an observed velocity difference, $v_\mathrm{obs}$, from the maximal and minimal velocities of the velocity axis profile, such that
\begin{equation}
v_\mathrm{obs}= \frac{1}{2} (v_\mathrm{max}-v_\mathrm{min}).
\label{eq.Vshear}
\end{equation} 
In some cases, typically galaxies that do not show ordered motions, the maximal velocity difference is not along a unique axis passing through the galaxy center. In these cases the maximal and minimal 5\% of pixels in the velocity map are used to determine $v_\mathrm{max}$ and $v_\mathrm{min}$. Kinematic maps with axis profiles are shown for a subset of the \kmostd first year sample in Figure~\ref{afig.disks} of Appendix~A.


We define two measures of velocity dispersion, the total velocity dispersion and the intrinsic velocity dispersion. The integrated or total velocity dispersion,  $\sigma_\mathrm{tot}$, sometimes referred to as $\sigma_\mathrm{net}$ (e.g. \citealt{2009ApJ...697.2057L}), is measured from a single Gaussian fit to the integrated spectrum, or the sum of all the unmasked spaxels and corrected for instrumental broadening. This measurement includes any velocity motions within the galaxy $-$ both resolved and un-resolved. The presence of a possible broad (FWHM $\geq$ 500 \kms) emission components from large scale winds, common in high-redshift galaxies, may inflate $\sigma_\mathrm{tot}$ in such single-Gaussian fits but extensive simulations show that it would be of order 30\% or less (F\"orster Schreiber et al. 2014b).

To obtain a intrinsic measure of velocity dispersion within the galaxies $-$ free of resolved motions and where possible un-resolved motions $-$ we measure $\sigma_0$ from the outer regions of galaxies, typically along the major-kinematic axis where beam-smearing is negligible (e.g. \citealt{2009ApJ...706.1364F}, F\"orster Schreiber et al. 2014b). Beam smearing is most significant in compact galaxies ($r_\mathrm{eff}\lesssim$ PSF) and in the central regions of disk galaxies \citep{2013ApJ...767..104N}.  Low S/N in outer regions of rotationally-supported galaxies can prohibit the measurement of the intrinsic velocity dispersion where it would be least affected by beam-smearing of the steep inner velocity curve $-$ motivating further the prioritization of deep observations for the \kmostd design. Examples of $\sigma$ axis profiles can be seen in Fig.~\ref{afig.disks} of Appendix~A, which confirm for many well-resolved \kmostd galaxies that the velocity dispersion axis profiles flatten at the same radius that the corresponding velocity axis profile flattens. The axis profiles allow measurements slightly beyond the extent of individual spaxels as they are measured from summed spectra within an aperture comparable to the PSF.

We obtain low velocity dispersion measurements for some galaxies in the \kmostd sample ($\sigma_0\sim10$ \kms) that are at or below the resolution limit of the observations ($\sigma_\mathrm{instr}=27-46$ \kms) forcing the question:  how far below the spectral resolution limit can the dispersion be reliably measured with KMOS? To investigate this issue a suite of model emission spectra were created with a uniform distribution of input intrinsic velocity dispersions from $\sigma_\mathrm{input} =1-100$ \kms. The model spectra include noise and skyline residuals matched to the same features in the data. The wavelength position of the emission line is varied in the many iterations. The model spectra are produced with a range of S/N of the emission line, from S/N of 3 to 100, and are convolved with the typical instrumental resolution of the \kmostd observations. To test the ability to recover the input velocity dispersions, the models are fit with Gaussians using {\sc linefit} following the procedure outlined in Section~2.4.

The recovered dispersion values are in general agreement with the input values of high S/N models. However, when moving below the resolution element the absolute difference between recovered and input dispersion increases, obtaining on average an overestimate of $\sigma_\mathrm{input}$. When $\sigma_\mathrm{input}> \sigma_\mathrm{instr}$ the recovered dispersions match the input dispersions with a 20\% error. When $\sigma_\mathrm{input}\approx\sigma_\mathrm{instr}$ the recovered dispersions match the input dispersions with a 30\% error. However, when moving from $\sigma_\mathrm{input}\approx \sigma_\mathrm{instr}$ to $\sigma_\mathrm{input}\approx 0.3\sigma_\mathrm{instr}$ the typical error rises from 30\% to 60\%. These errors are added in quadrature to the statistical errors of $\sigma_0$ and $\sigma_\mathrm{tot}$.

\begin{figure*}
\begin{center}
\includegraphics[scale=0.42,  trim=0.5cm 0cm 0cm 0cm, clip ]{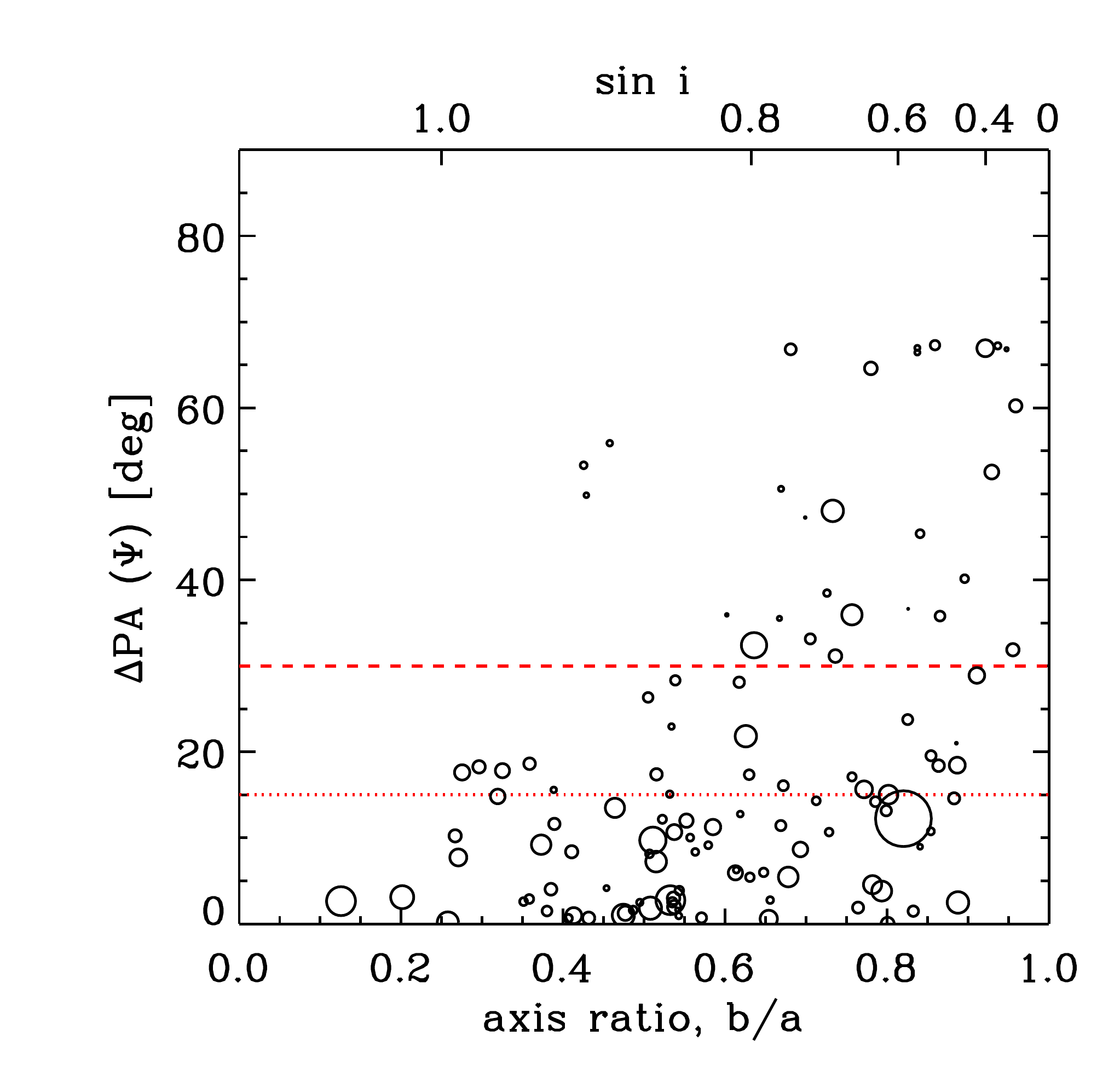}
\includegraphics[scale=0.345,  trim=0.5cm -2.9cm 0cm 0cm, clip ]{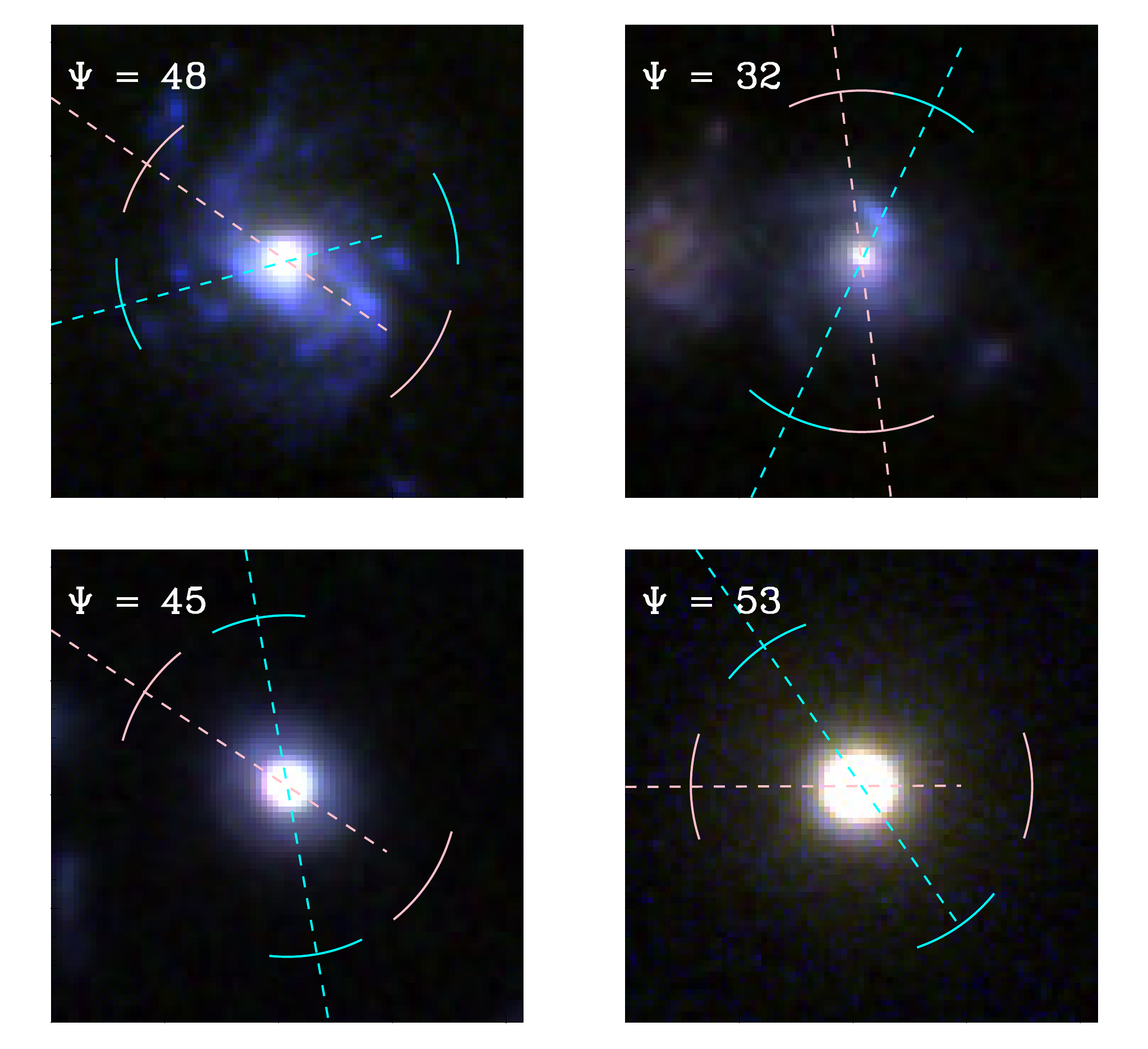}
\caption[...]{Misalignment between photometric and kinematic position angles, $\Psi$, as a function of axis ratio, $b/a$, measured from F160W HST imaging and KMOS kinematic maps respectively. Symbol size shows the relative size of galaxies in the F160W band.  Kinematic and photometric position angles agree within $15^\circ$ (dotted line) for $\sim60$\% of measurements, while misalignments greater than $30^\circ$ (dashed lines) are determined for $\sim20$\% of measurements and typically have axis ratios $>0.6$. The right panel shows four examples where $\Psi>30^\circ$, the pink dashed line shows the expected PA$_\mathrm{phot}$ measured from rest-frame imaging and the cyan dash line shows the kinematic PA$_\mathrm{kin}$ measured from KMOS data. The arcs are $36^\circ$ wide, showing $2\times$ the average difference between kinematic and photometric angles}.
\label{fig.pa}
\end{center}
\end{figure*}
We investigate the possible dependency of measured velocity dispersion with inclination at $z\sim1$ and $z\sim2$ in Figure~\ref{fig.dispsini}. At any given point in disk galaxies we observe the line-of-sight velocity dispersion ($\sigma_\mathrm{LOS}$), a mixing of the radial component ($\sigma_\mathrm{R}$) and vertical component ($\sigma_\mathrm{z}$) such that for face-on galaxies $\sigma_\mathrm{LOS}\sim\sigma_\mathrm{z}$. As such, $\sigma_0$ in the outer regions of more edge-on systems may yield larger measurements along the line-of-sight for disks with $\sigma_\mathrm{R}>\sigma_\mathrm{z}$ or unresolved rotation, possibly inflating the $\sigma_\mathrm{LOS}$ (e.g. \citealt{2010ApJ...719.1230A}).  This effect may be enhanced in seeing-limited data due to the larger beam size or compact galaxies where the beam size is a significant fraction of the observed galaxy.

In the low-redshift THINGS survey \citep{2008AJ....136.2782L} of HI gas in disk galaxies, a clear increase in dispersion values by $2-3\times$ the main locus ($10-20$ \kms~) is seen at high inclinations ($\sin{i}>0.87$).
Previous analyses of these properties in $z\gtrsim1$ kinematic data sets have been either inconclusive \citep{2004ApJ...612..122E} or do not show a strong trend \citep{2011ApJ...733..101G}. Figure~\ref{fig.dispsini} shows the observed relation for all \kmostd galaxies (open cirlces) and disk galaxies (as described in Section~\ref{sec.galclass}; closed circles). Inclination, $i$ is derived from the F160W images, where $\cos^{2}{i}=(b/a^2-q_0^2)/(1-q_0^2)$ and $q_0=0.25$ for a thick disk.

We find no trend between $\sigma_0$ and inclination at either redshift for the sub-sample of disk galaxies. Inspection of galaxies with dispersions above the main locus at both redshifts reveal compact below-MS galaxies and galaxies where $\sigma_0$ is an upper-limit due to low S/N at large radii. The lack of a correlation for disk galaxies may be a consequence of a higher fraction of galaxies departing from thin disks at $z>1$ (e.g. \citealt{2003A&A...399..879R,2006ApJ...650..644E,2012ApJ...745...85L,2014arXiv1407.4233V}) such that at all viewing angles $\sigma_\mathrm{LOS}$ is a mixing of $\sigma_\mathrm{R}$ and $\sigma_\mathrm{z}$. Models of disk galaxies of similar size, mass, and intrinsic dispersion to the \kmostd galaxies show that, at the typical S/N of $\sim20-40$ of our data sets, the velocity dispersion is well recovered for highly inclined systems but may be underestimated by $\sim$10\% at $\sin{i}< 0.8$ \citep{2011ApJ...741...69D}.  Accounting for this possible effect would however not change the conclusions from our measurements.

\subsection{Kinematic to photometric misalignments}
\label{sec.misalign}
We compare the derived kinematic position angle from KMOS, PA$_\mathrm{kin}$, and the position angle determined from the rest-frame optical imaging (HST F160W; \citealt{2012ApJS..203...24V}), PA$_\mathrm{phot}$ to determine the reliability of photometric predictions for the kinematic axis and to use for kinematic classification in Section~\ref{sec.galclass}. The comparison utilises the misalignment diagnostic, $\Psi$, from \cite{1991ApJ...383..112F};
\begin{equation}
\sin{\Psi} =  \mid\sin (\mathrm{PA_{phot}}-\mathrm{PA_{kin}})\mid,
\label{eq.psi}
\end{equation}
such that $\Psi$ is defined as a value between $0^\circ$ and $90^\circ$ insensitive to $180^\circ$ differences between measurement systems. The distribution of misalignments is shown in Fig.~\ref{fig.pa} as a function of photometric axis ratio, $b/a$, and $\sin{i}$, for objects where a kinematic major axis can be determined. 

The mean and median $\Psi$ of all galaxies are $18^\circ$ and $12^\circ$ respectively, within the expected errors of both measurements. For $60$\% of the galaxies the agreement is better than $15^\circ$ and for $80$\% of the galaxies the agreement is better than $30^\circ$. This comparison demonstrates that the PA$_\mathrm{phot}$ derived from rest-frame optical photometry provides a reasonable approximation to the PA$_\mathrm{kin}$ in $\sim80$\% ($\Psi<30^\circ$) of cases. Of galaxies with $\Psi>30^\circ$, 64\% are at $z\sim2$, where galaxy morphologies are often more irregular (the F160W images are probing bluer wavelengths; see Fig.~\ref{fig.msijh}) and galaxies are often more compact \citep{2014ApJ...788...28V}. The average effective F160W sizes of the resolved samples are $r_\mathrm{eff}=5.0$ kpc and $r_\mathrm{eff}=3.2$ kpc at $z\sim1$ and $z\sim2$ respectively.

Among the $\Psi>30^\circ$ cases with larger sizes the misalignment can generally be attributed to possible extinction or the influence of sub-structure in morphology such as spiral arms, a central bar, or clump-like features, particularly in face-on systems ($b/a>0.6$), as seen for two examples on the right-hand side of Fig.~\ref{fig.pa}. For the remaining $\Psi>30^\circ$ cases that have compact morphologies with little if any low surface brightness features, our data do reveal several cases with clear differential motions of the ionized gas along an axis that would not be predicted based on morphological information alone. These objects in our sample tend to lie at redder colors and below the main sequence $-$ a so far poorly explored part of the $z=1-2.5$ galaxy population in IFS surveys.

\section{Galaxy classification}
\label{sec.galclass}
\begin{figure*}
\begin{center}
\includegraphics[scale=0.60,  trim=1.5cm 4cm 0cm 14cm, clip ]{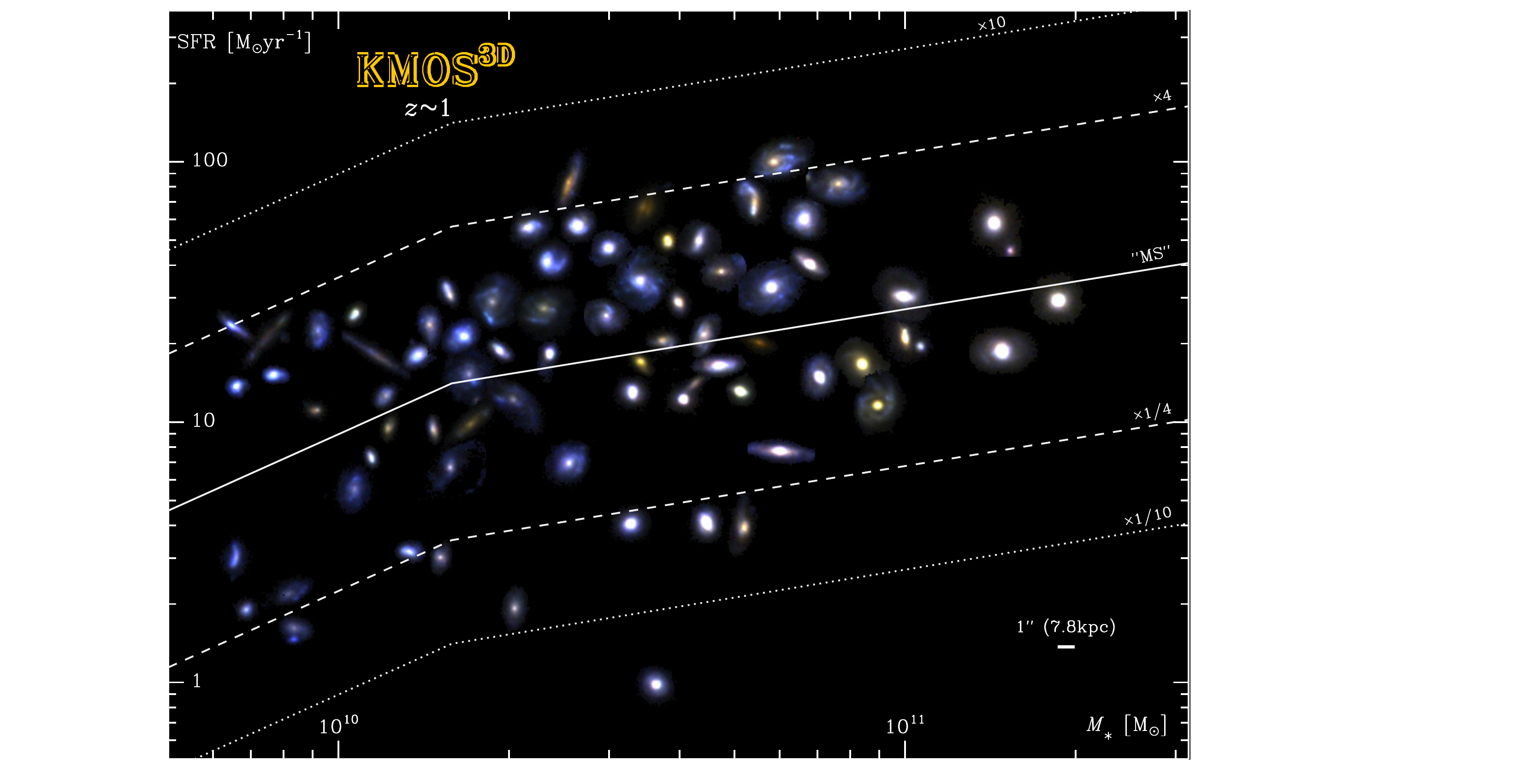}\\
\includegraphics[scale=0.60,   trim=1.5cm 4cm 0cm 14cm, clip ]{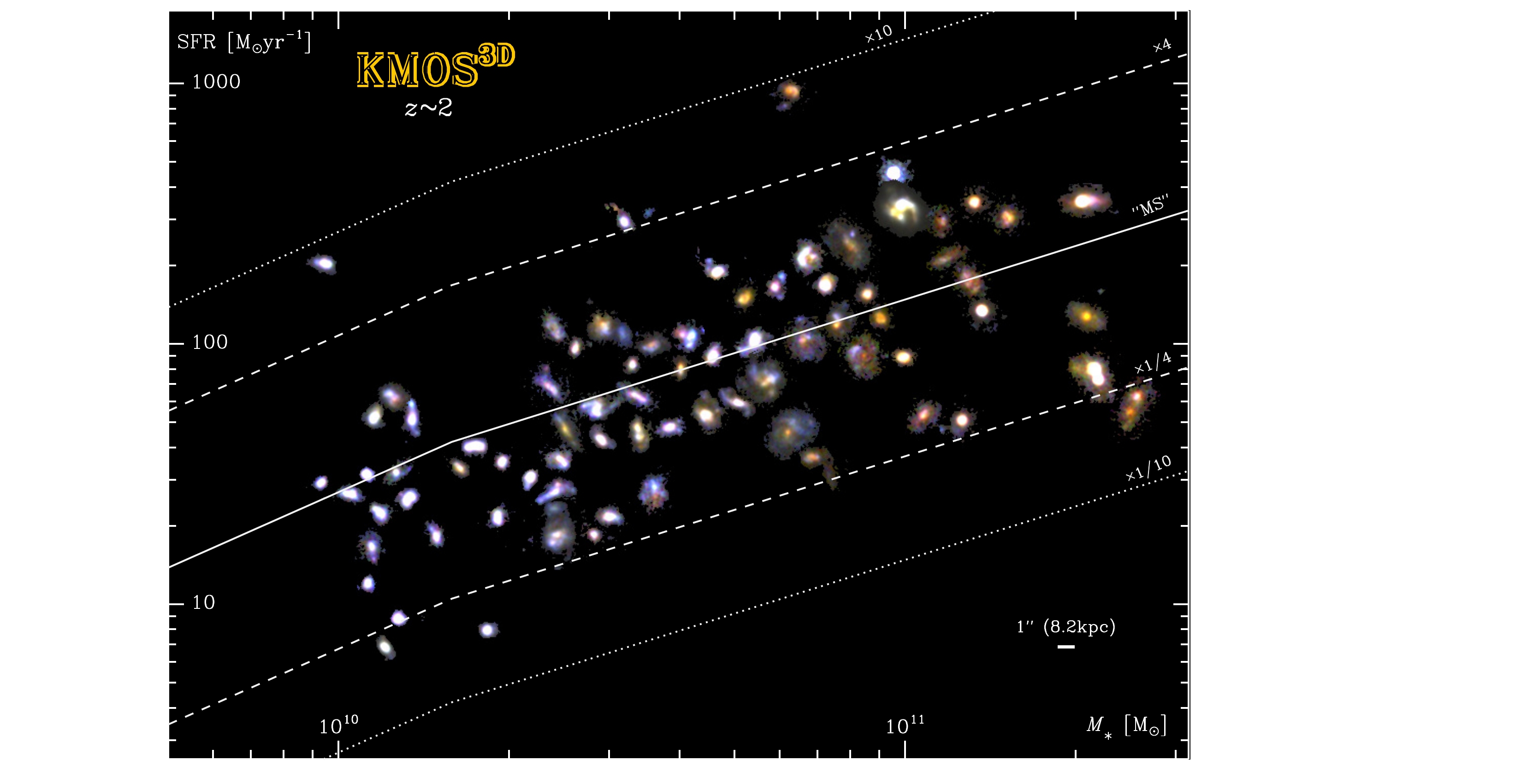}
\caption[...]{  Observed-frame $IJH$ composite images of the resolved \kmostd galaxies are shown at their approximate locations in the SFR$-M_*$ plane for the $z\sim1$ (top) and $z\sim2$ (bottom) samples. To avoid overlapping images small offsets are made in the two values to a clean area of parameter space. Offsets are always less than 0.2 dex in either $M_*$ or SFR, within the typical uncertainties in these properties. The solid line shows the canonical main sequence at $z\sim1$ and $z\sim2$ respectively from \cite{2014arXiv1407.1843W} adjusted for evolution by interpolating between redshift bins. The dashed and dotted lines show this main sequence scaled up or down by factors of $\times4$ and $\times10$ respectively. All sources are shown on the same angular scale, as denoted by the $1''$ scale bar at the bottom right of the plots; the orientation is North up, East left for all objects. }
\label{fig.msijh}
\end{center}
\end{figure*}

\begin{figure*}
\begin{center}
\includegraphics[scale=0.60, trim=1.5cm 4cm 0cm 14cm, clip ]{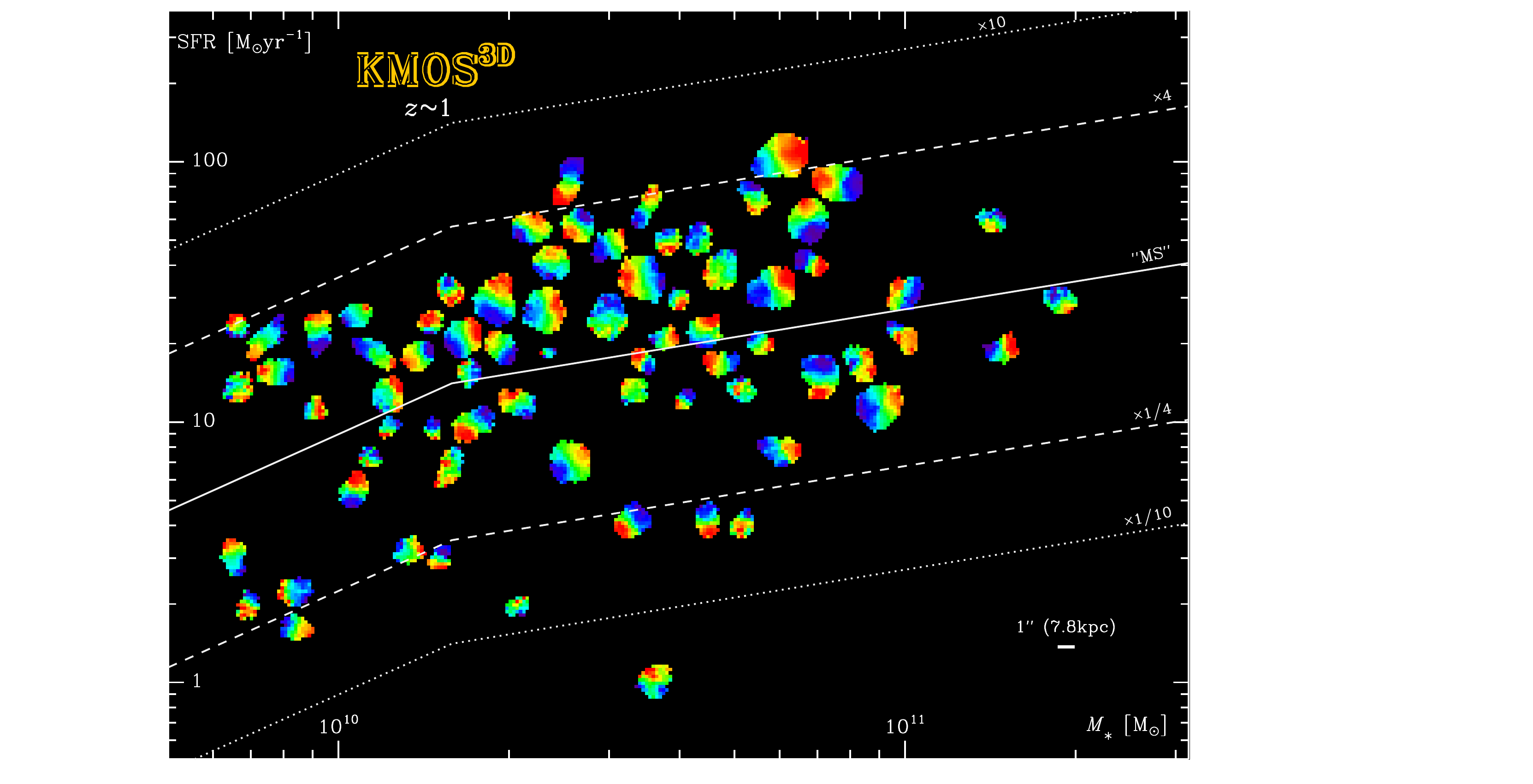}\\
\includegraphics[scale=0.60,   trim=1.5cm 4cm 0cm 14cm, clip ]{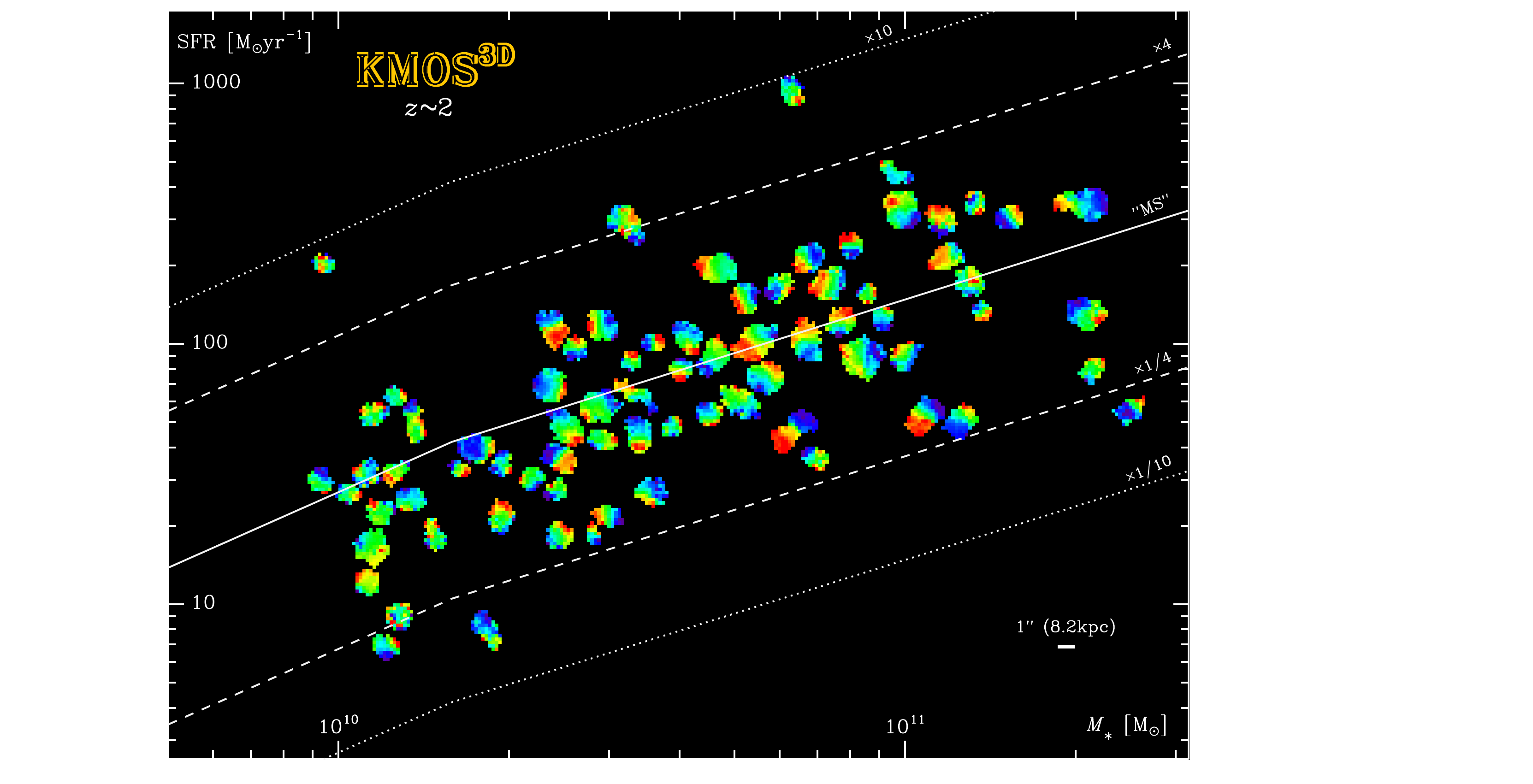}
\caption[...]{KMOS \halpha velocity fields of the resolved \kmostd galaxies at their approximate locations in the SFR$-M_*$ plane for the $z\sim1$ (top) and $z\sim2$ (bottom) samples. Positioning of the maps, lines and labels are the same as in Fig.~\ref{fig.msijh}. 
 }
\label{fig.msvelo}
\end{center}
\end{figure*}

\begin{figure*}
\begin{center}
\includegraphics[scale=0.60,   trim=1.5cm 4cm 0cm 14cm, clip ]{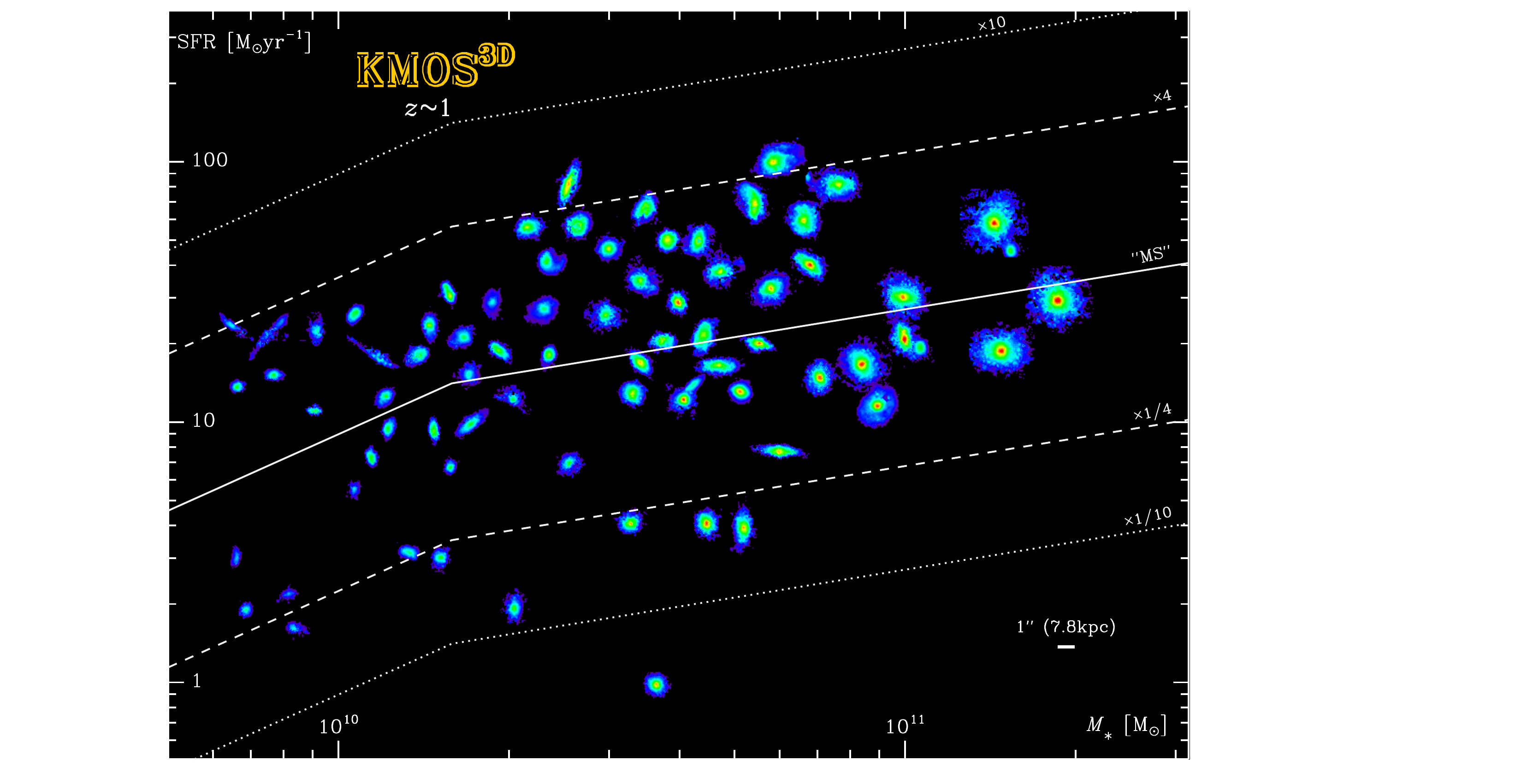}\\
\includegraphics[scale=0.60,   trim=1.5cm 4cm 0cm 14cm, clip ]{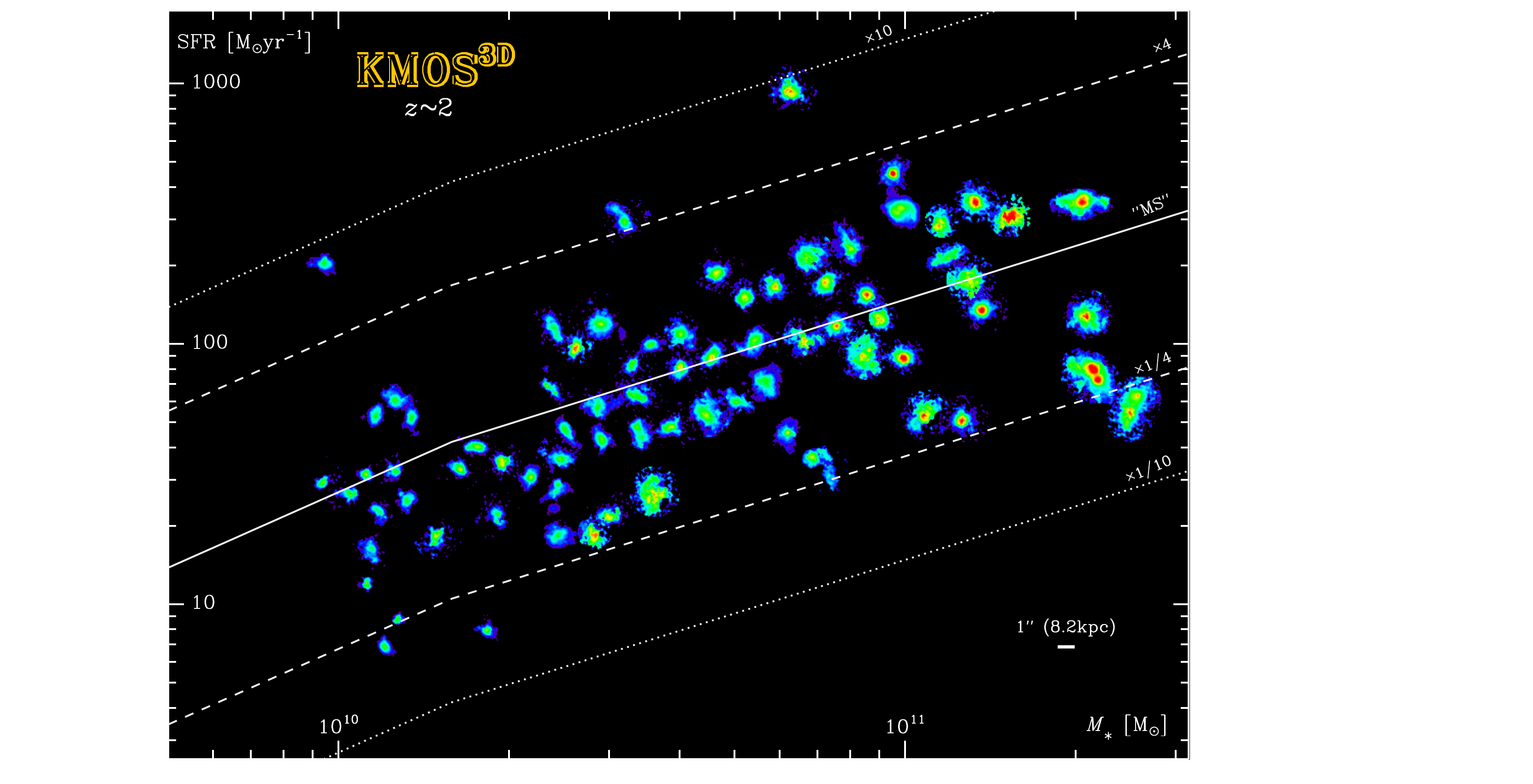}
\caption[...]{
Resolved stellar mass maps \citep{2012ApJ...753..114W} of the resolved \kmostd galaxies at their approximate locations in the SFR$-M_*$ plane for the $z\sim1$ (top) and $z\sim2$ (bottom) samples. Positioning of the maps, lines and labels are the same as in Fig.~\ref{fig.msijh}. All mass maps are represented on the same mass-color scale.  }
\label{fig.msmass}
\end{center}
\end{figure*}

The depth and sensitivity of the \kmostd observations allow for the detection of \halpha emission in the outer parts of galaxies essential to constraining kinematic classifications. Robust kinematic classifications not only allow for a better understanding of individual galaxies, but can characterise subsets of galaxies in relation to other key physical properties, e.g. sSFR and stellar structure (e.g. \citealt{2009ApJ...697.2057L,2012ApJ...759...29L}; Tacchella et al. 2014). Such analyses for large data sets ($>100$ galaxies) thus far have primarily been possible with automated morphological parameters (e.g. \sersic indices; \citealt{2011ApJ...742...96W,2012ApJ...745...85L,2014arXiv1407.4233V}) or visual classifications (e.g. \citealt{2012ApJ...757...23K,2013ApJ...778..129H}), which yield similar findings of a MS dominated by disk galaxies with the majority of mergers lying above the MS.

The full \kmostd survey will sample galaxy dynamics across the MS using direct kinematic tracers. The first year's worth of \kmostd data, nearly doubling the statistics of existing surveys, already extends kinematic observations to lower sSFR and redder colours for all kinematic types providing a more uniform coverage of the MS.  Figs.~\ref{fig.msijh}, \ref{fig.msvelo}, and  \ref{fig.msmass} show the resolved galaxies by their color-composite observed-frame $IJH$ images, velocity maps, and mass maps at their approximate locations in the SFR$-M_*$ plane  at $z\sim1$ and $z\sim2$. A wide variety of photometric morphologies are observed including galaxies that appear as edge- and face-on disks with little or no central concentration as well as systems with clear bulges. Figures~\ref{fig.msijh} and \ref{fig.msmass} show the build up of a central mass concentration, i.e. increasing bulge to total ratios when moving to higher galaxy stellar masses reflecting trends of the underlying galaxy population \citep{2014arXiv1402.0866L}. This can be connected qualitatively to the velocity maps in Fig.~\ref{fig.msvelo} as the largest ordered disk-like galaxies that are more often found at high stellar masses.

\subsection{Disk sample}
Distinguishing rotationally-supported galaxies from other kinematic classes is ideally done though a quantitative analysis, such as kinemetry (e.g. \citealt{2006MNRAS.366..787K}). The technique, relying on deviations from symmetries in kinematic maps with respect to that of an ideal rotating disk, is most reliably applied to systems with high S/N ratio over at least $\sim10$ resolution elements $-$ suitable for only a small sample of the largest \kmostd galaxies \citep{2008ApJ...682..231S}. In the absence of this technique, and in practice yielding the same results, we define disk galaxies by a series of increasingly stricter criteria applicable to the full sample, including compact galaxies (e.g. \citealt{2013ApJ...767..104N, 2014ApJ...785...75G}). The criteria are as follows: \\

1.) the velocity map exhibits a continuous velocity gradient along a single axis (in larger systems this is synonymous with the detection of a `spider' diagram; \citealt{1978ARA&A..16..103V});\\

2.) $v_\mathrm{rot} / \sigma_0 > 1$, where $v_\mathrm{rot}$ is the rotational velocity corrected for inclination, $i$, by $v_\mathrm{rot}=v_\mathrm{obs}/\sin{i}$.\\

These criteria alone are satisfied by 83\% of the resolved galaxies, 92\% at $z\sim1$ and 74\% at $z\sim2$. We apply the stricter additional criteria that;\\

3.) the position of the steepest velocity gradient, as defined by the midpoint between the velocity extrema along the kinematic axis, is coincident within the uncertainties ($\sim1.6$ pixels) with the peak of the velocity dispersion map;\\

4.) for inclined galaxies ($q<0.6$) the photometric and kinematic axes are in agreement ($<30$ degrees);\\

5.) the position of the steepest velocity gradient is coincident, within the uncertainties, with the centroid of the continuum center (a proxy for the center of the potential, i.e. in the higher mass galaxies this is usually a bulge).\\

In Table~\ref{table.disk} we present cumulative disk fractions as each additional criterion is considered. We note that the parameters underlying criteria 4 and 5, photometric PA and continuum center, are sensitive to extinction, sub-structure, and galaxy size. These features can lead to axis misalignment (see Section~\ref{sec.misalign}) and a shift of the continuum center from the true center of the galaxy potential.

\begin{table}
\caption{\% of galaxies satisfying disk criteria}
\begin{tabular*}{\linewidth}{@{\extracolsep{\fill}}lc|ccc}
\hline
{Criteria:} &  { 1,2} & {1,2,3} & { 1,2,3,4} & { 1,2,3,4,5}\\
\hline
Full Sample   & 83\% & 73\% & 71\% & 58\%  \\
$z\sim1$       & 93\% & 78\% & 78\% & 70\% \\
$z\sim2$       & 74\% & 68\% & 64\% & 47\% \\
\hline
\label{table.disk}
\end{tabular*}
\end{table}

The disk fraction is higher at $z\sim1$ than at $z\sim2$, particularly when considering criteria 4 and 5. As discussed in Section~\ref{sec.misalign}, and reflected in Table~\ref{table.disk}, axes misalignment is more common at $z\sim2$ where galaxies are more compact and irregular in the observed F160W images.  However, the main difference occurs when including criterion 5 $-$ coincidence of the kinematic and continuum center $-$ where at $z\sim2$ it is possible that extinction and sub-structure, as well as minor-mergers (e.g. \citealt{2009MNRAS.394L..51B,2010ApJ...715..202H}) more frequently influence the measured galaxy centroid. When considering only criteria 1$-$3, the significance of the different disk fractions between redshift bins, 10\%, remains unclear as the size difference between the two redshifts may hinder the detection of small disks at $z\sim2$ \citep{2013ApJ...767..104N}. 

The inclination, $i$, used to estimate the rotational velocity is measured using GALFIT from the rest-frame optical CANDELS images (F160W; \citealt{2012ApJS..203...24V}). However, we note that the inclination derived from photometry will not necessarily correspond to the kinematic inclination, in particular for cases of $\Psi>30$ degrees. At high inclinations small errors on $i$ do not have a large effect on the correction, however for systems with low inclination a small error on $i$ results in a large error on $v_\mathrm{rot}$. This uncertainty is explored in detail in a forthcoming paper (S. Wuyts et al. \textit{in prep}).

Of the resolved sample 73\% fit criteria 1$-$3, herein the `disk sample'. \kmostd disks are primarily found along the canonical `MS', extending down to $10^{9.6}$ and $10^{10}$ \Msun~at the low mass end for $z\sim1$ and $z\sim2$ respectively, with five galaxies found below $1/4\times$ the MS at $z\sim1$. At $z\sim2$ two galaxies are detected significantly above the MS, both of which are compact, consistent with expectations for star-bursting activity \citep{2011ApJ...742...96W,2011ApJ...730....4B,2011A&A...533A.119E} and are not classified as disks by the above criteria.

In comparison, without the kinematic data a disk fraction is obtained for the resolved \kmostd sample from the photometric data alone using single-component \sersic fits to the rest-optical photometry \citep{2012ApJS..203...24V}. When defining a disk galaxy as having \sersic index $n\leq2.5$ we find a 83\% disk fraction, approximately equal at both redshifts. The photometric disk fraction matches the total fraction of galaxies fulfilling the rotation criteria 1 and 2. Correlating this with our kinematic classifications, 6 galaxies classified as disks from KMOS data, using all criteria, have $n>2.5$. These galaxies are massive with average $\log{M_*}$ [\Msun]$=11.1$ and visually bulge dominated in their $IJH$ images with faint disks or spiral arms (a known effect from comparison with visual classifications; e.g. \citealt{2012ApJ...757...23K}).

\subsection{Pairs and mergers}
\label{subsec.pairs}
In the current sample 11 targeted galaxies  (5 at $z\sim1$ and 6 at $z\sim2$) have possible companions that (1) are in the 3D-HST catalog, (2) are  expected to be within the IFU of the primary target (within a projected separation of $1.5''$ or $\sim12$ kpc), and (3) have a redshift such that \halpha is expected within 500 \kms~of the primary target.  The galaxies in ``close pairs'' fall on the MS at both redshifts. The kinematic maps in which one or both members are detected in \halpha show irregularities or non-ordered motions indicating possible mergers.  The stellar mass ratios for the close pairs range from 1:1 to 1:15. 

The numbers above do not include galaxy pairs with wider spatial separations typically included in merger rate analyses (e.g. \citealt{2000MNRAS.311..565L,2014arXiv1406.2327L}) nor do they include mergers in which galaxies are too close to be resolved into multiple objects in the 3D-HST catalogs, but which nevertheless may be classified as mergers based on their morphologies (e.g. \citealt{2008ApJ...672..177L,2013MNRAS.432..285S,2013MNRAS.430.1158S}) or kinematics (e.g.  COS4\_19753 and COS3\_21583). The kinematics of mergers and galaxy pairs within the sample will be addressed in later papers as the number of observed galaxies increases.

\subsection{AGN}
No criteria are imposed to exclude galaxies hosting active galactic nuclei (AGN) from the \kmostd sample.  AGN incidence at $z>1$, particularly on the MS, is expected to be between 5-30\% at $\log (M_*$ [\Msun]) $\sim10 -12$,  with the fraction increasing with increasing mass (e.g., \citealt{2005ApJ...633..748R, 2007ApJ...670..173D,2009A&A...507.1277B,hainline:2012:06,2012ApJ...753L..30M,2012A&A...545A..45R,2012MNRAS.427.3103B}).
As reported in \cite{2014ApJ...796....7G} a similar fraction of massive galaxies in the \kmostd sample can be classified as hosting an AGN from X-ray, optical, infrared, and radio AGN indicators. These galaxies are found in a variety of kinematic morphologies, most commonly unresolved or rotating galaxies. One galaxy hosting an AGN is a member of one of the $z\sim2$ close pairs.

\section{Velocity dispersion over cosmic time}
\label{sec.analysis}
\begin{figure*}
\begin{center}
\includegraphics[scale=0.42,  trim=0.5cm 0cm 0cm 0cm, clip ]{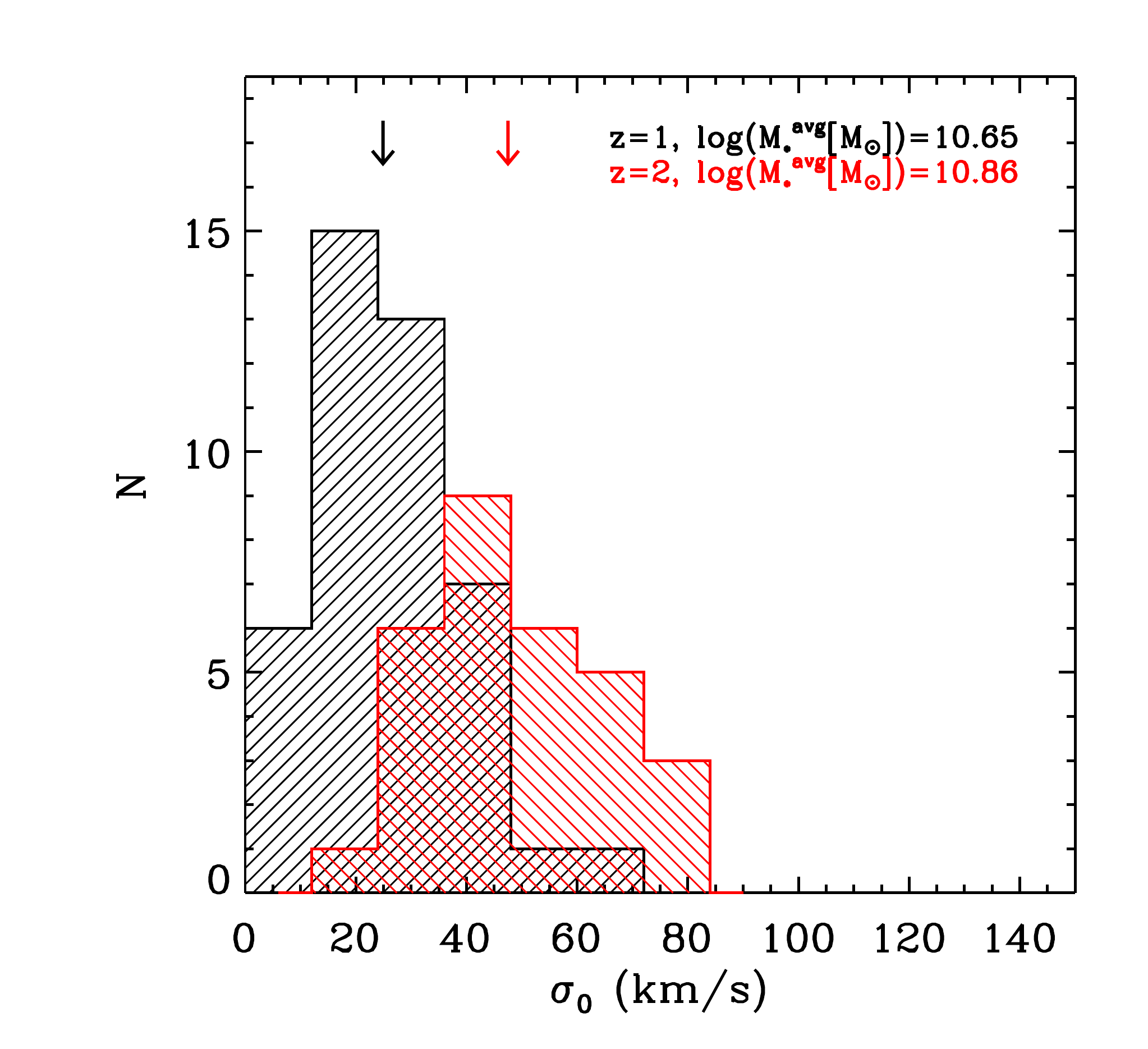}
\includegraphics[scale=0.42,  trim=0.5cm 0cm 0cm 0cm, clip ]{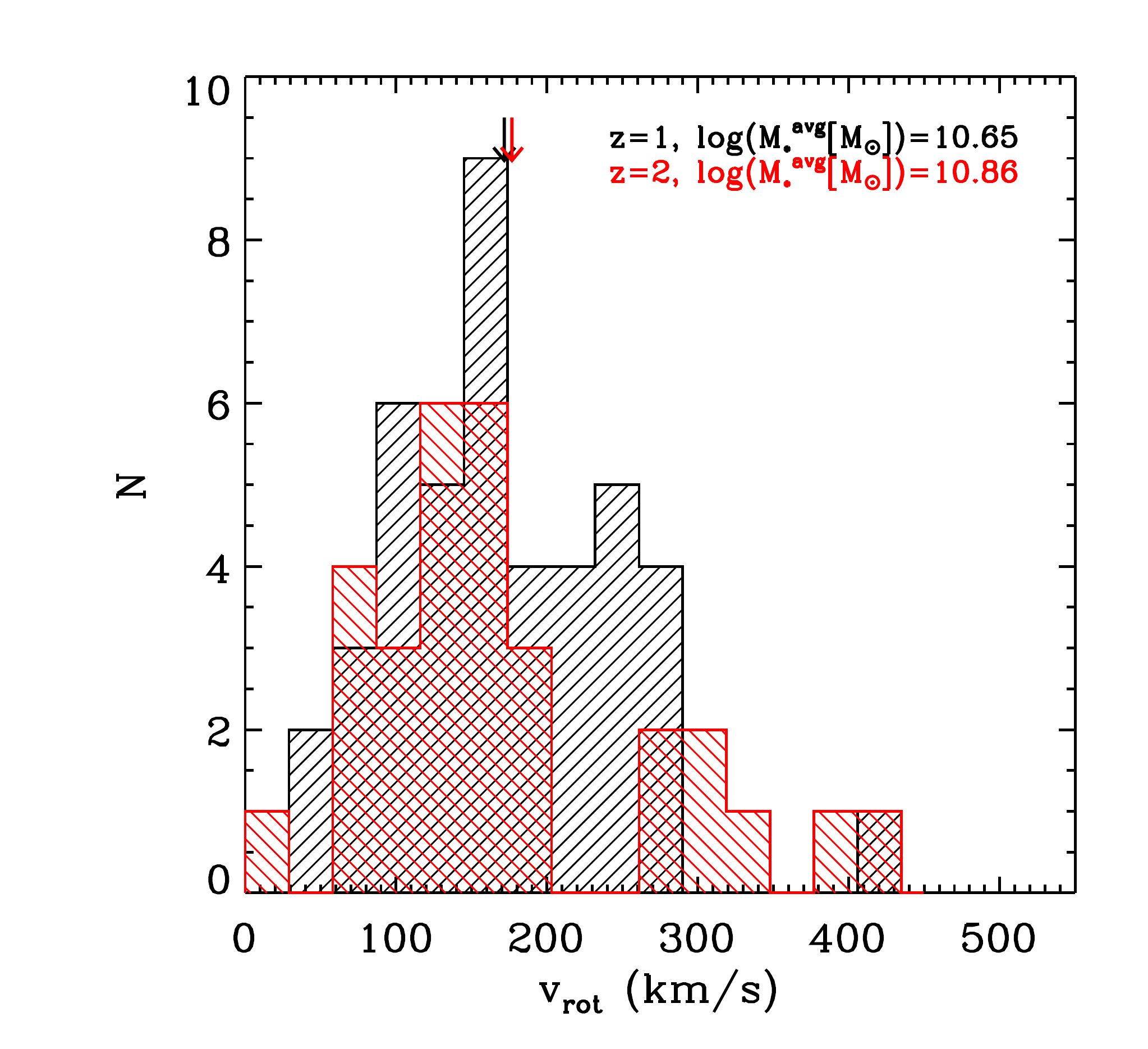}
\caption[...]{Histograms of disk galaxy velocity dispersion, $\sigma_0$ and $v_\mathrm{obs}$ measurements split into $z\sim1$ (black) and $z\sim2$ histograms (red). The velocity dispersion and velocity histograms show only galaxies that show rotation and have a reliable $\sigma_0$ measurement in the \kmostd Survey. The arrows represent the mean values of each distribution. The mean dispersion of the $z\sim2$ galaxies is $2\times$ greater than the mean dispersion of the $z\sim1$ population. While there is a 6 km s$^{-1}$ difference between the average $z\sim1$ and $z\sim2$ velocity histograms, the populations are mostly overlapping. Average stellar masses for the distributions are $5\times10^{10}$ \Msun and $7\times10^{10}$ \Msun~at $z\sim1$ and $z\sim2$. }
\label{fig.disphisto}
\end{center}
\end{figure*}

Using the two \kmostd redshift slices we quantify the evolution in ionized gas velocity dispersion from $z=2.3$ to $z=0.9$ $-$ from the peak of cosmic star formation to the rapid decline. A decrease in intrinsic velocity dispersion has been reported from $z\sim1$ to $z\sim0$ \citep{2012ApJ...758..106K}. However, between $z\sim2$ to $z\sim1$ an accurate measurement of the degree of evolution has not been possible due to a lack of consistent or sizeable datasets (e.g. \citealt{2012A&A...539A..92E,2012ApJ...758..106K}). 
For this analysis we use the disk galaxies from \kmostd with sufficiently high S/N per resolution element to constrain the velocity dispersion in the outer regions of the galaxies. Figure~\ref{fig.disphisto} shows the distribution of velocity dispersion measurements from these galaxies split into $z\sim1$ and $z\sim2$ redshift bins. The samples have comparable stellar mass distributions with average $\log(M_*[$\Msun$])$ of 10.65 and 10.86 respectively.
The means of each $\sigma_0$ distributions are 24.9 \kms~and 47.5 \kms~as shown by the downward arrows $-$ a factor of $2\times$ evolution for disk galaxies from $z\sim2$ to $z\sim1$. In contrast, the rotational velocity, $v_\mathrm{rot}$, distributions are comparable in their peaks and widths with a measured difference of the means at $z\sim1$ and $z\sim2$ of 6 km s$^{-1}$.

\begin{figure}
\includegraphics[scale=0.42,  trim=0.5cm 0cm 0cm 0cm, clip ]{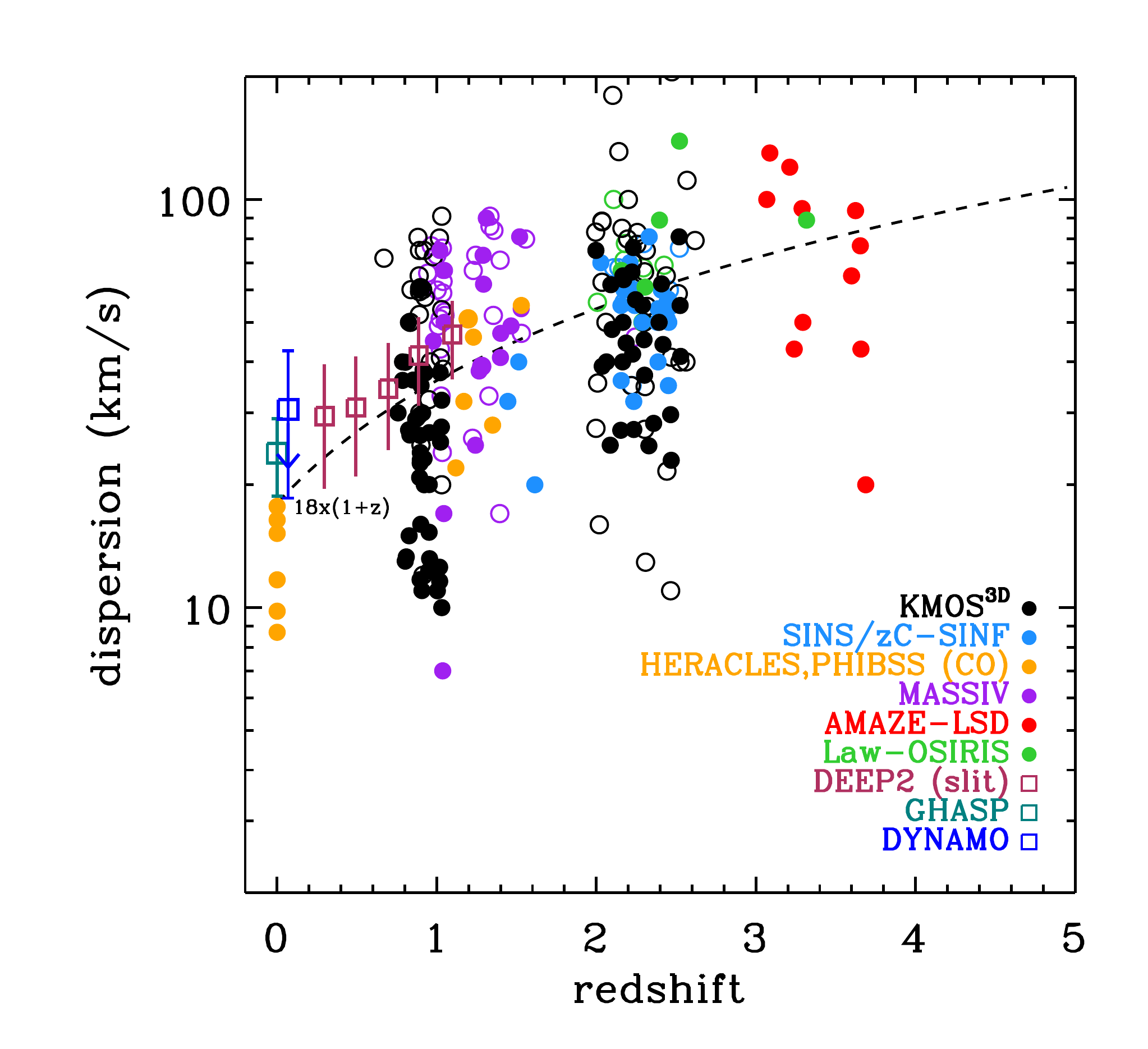}
\caption[...]{Galaxy velocity dispersion measurements from the literature at $z=0-4$ from molecular and ionized gas emission (including IFS and long-slit). \kmostd measurements at $z\sim1$ and $z\sim2$ are shown by black circles. Filled circles represent disk galaxies or `rotators', open circles represent all other kinematic categories. Open squares are averages of surveys at $z\lesssim1$. Sources for the literature data are given in Section~\ref{sec.analysis}. The dashed line shows a simple ($1+z$) evolution scaled by a factor of $18\times$ to overlap with the data.}
\label{fig.dispevo}
\end{figure}

In Fig.~\ref{fig.dispevo} we examine the evolution of velocity dispersion determined from \kmostd within the wider redshift range of IFS samples across $z=0-4$, including all galaxy types with dispersions measured from the ionized gas via \halpha or \OIII~and molecular gas via millimeter interferometric observations of CO.  While the scatter is large, there appears to be an approximate $1+z$ evolution in measured velocity dispersion across cosmic time. 

\begin{figure*}
\begin{center}
\includegraphics[scale=0.51,  trim=5cm 2.1cm 6.5cm 0cm, clip ]{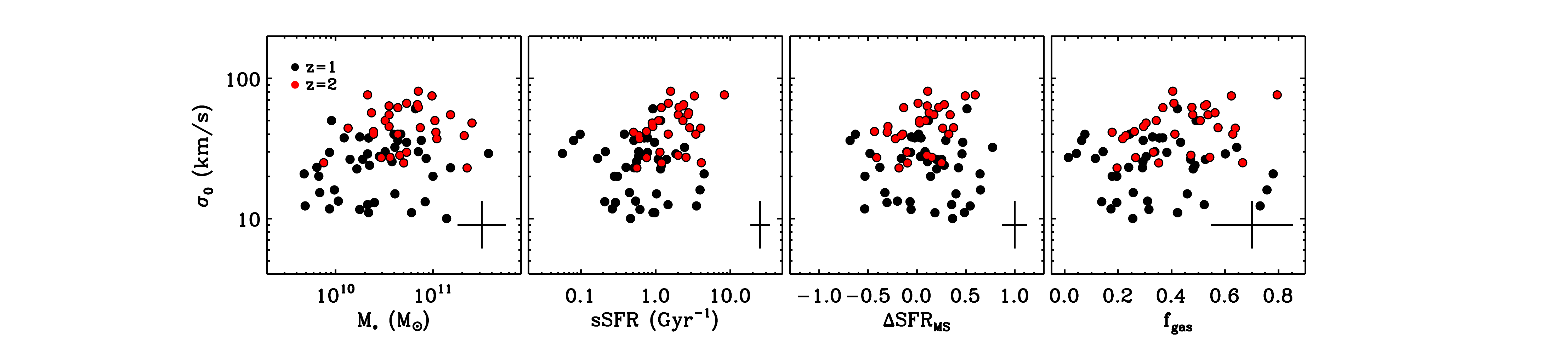}
\includegraphics[scale=0.51,  trim=5cm 0cm 6.5cm 0.9cm, clip ]{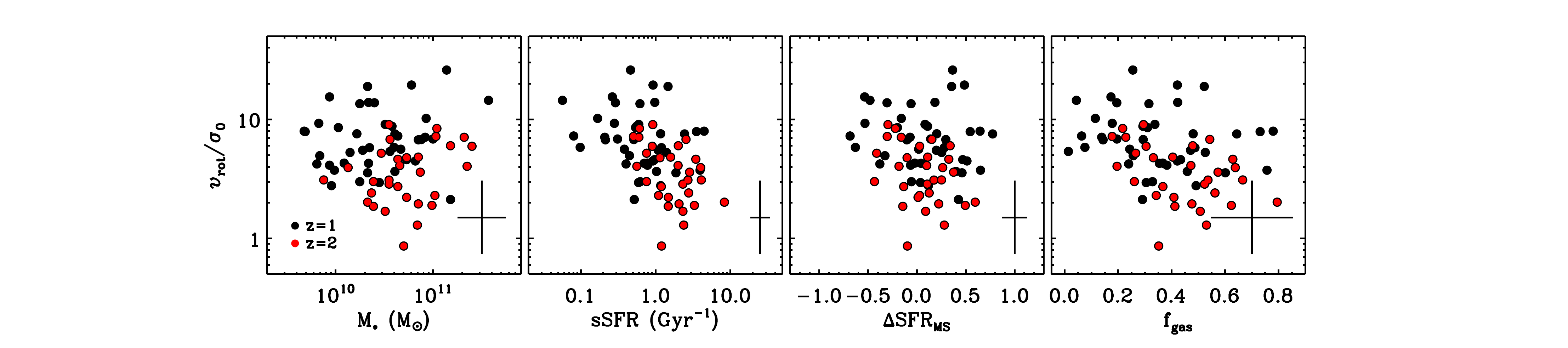}
\caption[...]{Disk velocity dispersion, $\sigma_0$, and disk stability, $v_\mathrm{rot}/\sigma_0$, for the \kmostd first year disk sample as a function of $M_*$, sSFR, $\Delta$SFR, and $f_\mathrm{gas}$ at $z\sim1$ (black points) and $z\sim2$ (red points). Representative error bars for individual points are shown at the lower right of each plot. }
\label{fig.disptrends}
\end{center}
\end{figure*}

The data included in Fig.~\ref{fig.dispevo} at $z\gtrsim1$ are from the 
MASSIV survey ($\log M^\mathrm{avg}_*[$\Msun$]=10.2$; \citealt{2012A&A...539A..92E,2012A&A...546A.118V}), 
SINS/zC-SINF survey ($\log M^\mathrm{avg}_*[$\Msun$]=10.6$; \citealt{2009ApJ...706.1364F}, F\"orster Schreiber et al. 2014b), 
OSIRIS survey from \cite{2009ApJ...697.2057L} ($\log M^\mathrm{avg}_*[$\Msun$]=10.3$), and  
AMAZE-LSD surveys ($\log M^\mathrm{avg}_*[$\Msun$]=10.1$; \citealt{2011A&A...528A..88G}). 
At $z<1$ averages are shown by open squares for the GHASP sample (\citealt{2010MNRAS.401.2113E}; $\log M^\mathrm{avg}_*[$\Msun$]=10.6$),  and the ``main sequence'' DYNAMO galaxies at $z<0.1$ (Object classes A-C and E; $\log M^\mathrm{avg}_*[$\Msun$]=10.3$; \citealt{2014MNRAS.437.1070G}). Velocity dispersions derived from molecular gas are included from the $z=1-2$ PHIBSS survey ($\log M^\mathrm{avg}_*[$\Msun$]=11.0$; \citealt{2013ApJ...768...74T}) and at $z=0$ from the HERACLES survey ($\log M^\mathrm{avg}_*[$\Msun$]=10.5$; \citealt{2008AJ....136.2782L,2009AJ....137.4670L}). 

The DEEP2 survey is the only long-slit survey included in Fig.~\ref{fig.dispevo}, due to its wide redshift range ($z=0.2-1.2$) and reported dispersion evolution for galaxies of $10^8-5\times10^{10}$ \Msun~\citep{2012ApJ...758..106K}. We note that long-slit data may report high dispersion values due to PA uncertainties and lack of spatial information but is similar in general to the ``along the slit'' method used to derive $\sigma_0$. 

A source of scatter within the samples shown in Fig~\ref{fig.dispevo} may originate from measurement uncertainties in determining the intrinsic velocity dispersion or when $\sigma_0$ is close to or below the instrumental resolution as discussed in Section~\ref{sec.results}. Furthermore, dispersions of disturbed systems or non-disk systems may be probing different physics than dispersions of disk galaxies; e.g. mergers may enhance $\sigma_0$ \citep{2011ApJ...730....4B,2013MNRAS.434.1028P,2014A&A...568A..14A}. For galaxies showing rotation but for which a flattening in velocity, as expected for disks, has not been reached the measured dispersion is an upper limit and will be higher than the intrinsic dispersion. Indeed, while considering only the disk galaxies satisfying all criteria in Section~\ref{sec.galclass} with $\sigma_0$ measurements constrained in the galaxies outer regions, the scatter of dispersion in the \kmostd samples is reduced.

Differences between samples can also arise from diverse selection criteria, spatial and spectral resolution, stellar mass ranges, and measurement methods that could bias the interpretation of evolution.  The \kmostd dispersion values are measured with the same methods as the SINS/zC-SINF, PHIBSS and HERACLES surveys, using the outside of the disk kinematics free of beam-smearing.  The MASSIV survey uses a beam-smearing corrected error-weighted mean of the dispersion map that should give values consistent with $\sigma_0$, however the difference in methods could lead to the higher mean dispersions from the MASSIV galaxies.
 The GHASP dispersions are calculated from the average of the 20\% lowest dispersion spaxels from the dispersion maps, a method consistent with $\sigma_0$ when there are a sufficient number of spaxels in the map. AMAZE/LSD subtract a dynamical disk model from the observed dispersion map to correct for galaxy rotation and give an average of the results. 
The flux weighted mean of the dispersion map was used to derive a global $\sigma$ for \cite{2009ApJ...697.2057L} and DYNAMO \citep{2014MNRAS.437.1070G} and is known to give systematically higher values \citep{2011ApJ...741...69D}, although \cite{2014MNRAS.437.1070G} do include a beam smearing correction.

Spectral and spatial resolution limits of the high-redshift surveys are comparable. The MASSIV, AMAZE/LSD and SINS/zC-SINF surveys were observed with SINFONI in both seeing limited and AO resolution modes ($R=3000-5000$, pixel scale = $0.05-0.25''$). KMOS has comparable sensitivity, spectral and spatial resolution as SINFONI in seeing-limited mode. The \cite{2009ApJ...697.2057L} sample taken with OSIRIS, a similar AO instrument on the Keck Telescope, has spatial sampling of $0.05''$ and comparable spectral resolution.  The resolution of the DEEP2 spectra, spanning from $z=0.2-1.2$, is $R\sim5000$ ($\sigma=25.5$ \kms), which may hinder some measurements of $\sigma$ in the lower-redshift regime, possibly underestimating the degree of the overall evolution. 

At $z<1$ there is more variety among the resolution of the samples. The spectral and spatial resolutions of the $z\sim0$ GHASP and HERACLES surveys are $3-13$ \kms~and $\sim0.3-0.5$ kpc respectively, providing reliable measurements of intrinsic velocity dispersions in the outer regions of disk galaxies. The $z\sim0.01$ DYNAMO sample has comparable spectral resolution of $15$ \kms, but poorer spatial resolution ($\sim3$ kpc) leading to possible beam smearing effects. 
Indeed, recent work with higher spatial resolution ($\sim1$ kpc) of two DYNAMO disk galaxies reveal 30 and 37 \kms~in the outer disks compared to the originally reported beam-smearing corrected integrated dispersions of 50 and 45 \kms~respectively (Bassett et al. 2014). The measurement technique presented in Bassett et al. is more in line with $\sigma_0$ and thus the percent reduction derived from the two galaxies is reflected in Fig.~\ref{fig.dispevo} with a downward arrow. 

\subsection{Trends with other observed properties}
\label{sec.disptrends}
As galaxies build-up their mass and size (depleting their gas reservoirs) they become more stable to small perturbations from gas accretion, galactic winds, and/or minor mergers \citep{2009ApJ...707..250M,2014ApJ...785...75G,2013ApJ...768...74T}. Assuming this scenario we investigate possible correlations between turbulence and disk stability ($v_\mathrm{rot}/\sigma_0$) with $M_*$, sSFR, $\Delta$SFR$=\log(\mathrm{SFR/SFR}_\mathrm{MS})$, and gas fractions ($f_\mathrm{gas}$) that may account for the scatter of \kmostd galaxies in Fig.~\ref{fig.dispevo}.

Spearman rank correlation coefficients and significance levels for the six panels of Fig.~\ref{fig.disptrends} reveal that $M_*$, sSFR, $\Delta$SFR, and $f_\mathrm{gas}$ are more strongly correlated with $v_\mathrm{obs}/\sigma_0$ than $\sigma_0$ when considering the full redshift range of the current disk sample. The strongest correlation with dispersion as measured by the Spearman coefficient is with $\Delta$SFR in the $z\sim2$ sample with $\rho=0.45$ with 2.5$\sigma$ significance level, hinting at a possible trend of an increasing velocity dispersion when moving above the MS at a fixed mass. This trend however is absent at $z\sim1$ putting the already low significance at $z\sim2$ in question. When both redshift regimes of the \kmostd disk sample are considered together a weak correlation ($\rho=0.31$ at $2.5\sigma$) is measured between $M_*$  and $\sigma$.  However, within individual redshift slices the dynamic range in $M_*$ of the first year data is limited. 

The strongest correlation when considering the combined $z\sim1-2$ sample is between $v_\mathrm{obs}/\sigma_0$ and sSFR ($\rho=0.48$ at $4\sigma$), which may contribute to the evolution between samples in Fig.~\ref{fig.dispevo} due to the known evolution of sSFR. The next strongest correlation is between $v_\mathrm{obs}/\sigma_0$ and $f_\mathrm{gas}$ ($\rho=0.40$ at $3\sigma$) where $f_\mathrm{gas}$ is derived from the galaxies redshift and SFR following \cite{2013ApJ...768...74T}. A correlation between $v_\mathrm{obs}/\sigma_0$ and $f_\mathrm{gas}$ was also reported in \cite{2013ApJ...767..104N} and is in line with predictions from models that suggest accretion energy would drive up both disk turbulence and gas fractions (e.g. \citealt{2010ApJ...712..294E}). Given the tight inter-relationship between $f_\mathrm{gas}$, sSFR and $\Delta$SFR (e.g. \citealt{2013ApJ...768...74T,2014arXiv1409.1171G}), it is perhaps unsurprising that kinematic properties of the gas are found to correlate with all three parameters.  The connection of disk velocity dispersion on sSFR and $f_\mathrm{gas}$ are explored more in the next section.

\subsection{Velocity dispersions in the context of galaxy evolution}

Both outflows and accretion are predicted to be prevalent at the high redshifts where large internal motions are measured. Outflows are observed to be common at high redshift ($z\sim1-3$) from star formation and AGN \citep{2008A&A...491..407N,2009ApJ...692..187W,Shapiro:2009sj,2010ApJ...717..289S,2011MNRAS.412.1559N,2012MNRAS.426.1073H,2012ApJ...752..111N,2014ApJ...787...38F,2014ApJ...796....7G}.
While accretion has yet to be directly observed, a consequence of the expected accretion is larger molecular gas content and higher sSFRs at early times, with the evolution of both quantities established observationally (molecular gas fractions: \citealt{2010Natur.463..781T,2013ApJ...768...74T,2010ApJ...713..686D,2011A&A...528A.124C,2014arXiv1409.1171G}; sSFR: \citealt{2010ApJ...718.1001B,2010MNRAS.405.1690D, 2012ApJ...754L..29W,2014arXiv1407.1843W,2013ApJ...763..129S,2014ApJ...781...34G}). However these mechanisms and their evolution are not exclusive, but rather essential elements of the equilibrium or regulator model in which star-forming galaxies are well described as being in a fairly steady equilibrium between inflows, star formation, and outflows \citep{2010ApJ...718.1001B,2012MNRAS.421...98D,2013ApJ...772..119L,2014arXiv1402.2283D}.

We use the first year of \kmostd data in tandem with data from the literature to test the scenario that the evolution of velocity dispersion is consistent with the equilibrium model and is primarily a result of the gas inflow onto galaxies, most efficient at high redshifts. 
We derive a scaling for velocity dispersion for near-critical disks as a function of redshift using the evolution of molecular gas fractions, depletion time ($t_\mathrm{dep}$), and specific SFR using recent observational results in the literature that are independent of the IFS data discussed thus far. 

We derive the expected gas fraction evolution of star-forming galaxies at a given stellar mass, where the gas fraction is defined as
\begin{equation}
f_\mathrm{gas}=  \frac{1}{1+(t_\mathrm{dep}\mathrm{sSFR})^{-1}}  
\label{eq.fgas1}
\end{equation}
\citep{2013ApJ...768...74T}. 

To rewrite equation~\ref{eq.fgas1} as a function of redshift, we use the evolution of depletion time described by
\begin{equation}
t_\mathrm{depl}[\mathrm{Gyr}]=1.5 \times (1+z)^{\alpha},
\label{eq.depltime}
\end{equation}
with $\alpha$ measured to be $-0.7$ to $-1.0$ by \cite{2013ApJ...768...74T}, and predicted to be $-1.5$ in the analytic model of \cite{2012MNRAS.421...98D}.  For simplicity and consistency with \cite{2013ApJ...768...74T} $\alpha=-1$ is adopted. The leading factor of 1.5 Gyr is a normalization to the typical depletion time observed in local galaxies \citep{2008AJ....136.2782L,2011ApJ...730L..13B,2012ApJ...758...73S}. Using the cosmic decline of specific SFR defined at $0.5<z<2.5$ from \cite{2014arXiv1407.1843W},

\begin{equation}
\mathrm{sSFR}(M_*,z) = 10^{a(M_*)}(1+z)^{b(M_*),}
\label{eq.ssfr}
\end{equation}
where we fit to the constants in Table~5 of \cite{2014arXiv1407.1843W} to describe $a$ and $b$ as a function of stellar mass, such that

\begin{align*}
&a(M_*) = -10.73 + \frac{1.26}{1 + e^{(10.49-\log{M_*})/(-0.25)}},\\
&b(M_*) =  ~~1.85 + \frac{1.57}{1 + e^{(10.35-\log{M_*})/0.19}}.\addtocounter{equation}{1}\tag{\theequation}
\label{eq.constants}
\end{align*}
Equation~\ref{eq.constants} is valid in the $M_*$ range of $\log{M_*}$[\Msun]$=9.2-11.2$ constrained by the data.

The evolution of the $t_\mathrm{depl}$ and evolution of sSFR are strongly linked (\citealt{2011MNRAS.415...61S, 2013ApJ...768...74T,2014arXiv1409.1171G}).  For simplicity we use the above relations to derive the evolution of gas fractions. We note that our results are consistent with adopting $t_\mathrm{dep}$ from \cite{2014arXiv1409.1171G}. Furthermore, at $z\gtrsim3$ the evolution of sSFR is debated \citep{2013ApJ...763..129S,2014ApJ...781...34G} with different behaviours expected from different models or extrapolating fits to data beyond where they are constrained. Equation~\ref{eq.ssfr} is unconstrained at $z>2.5$, but the resulting extrapolation is roughly consistent with published measurements out to $z\sim4$ (e.g. \citealt{2012ApJ...754...83B,2013ApJ...763..129S}).

By considering only disk galaxies a prediction for the evolution of ionized gas velocity dispersions can be derived by rewriting the Toomre stability criterion \citep{1964ApJ...139.1217T} as
\begin{equation}
\frac{v_\mathrm{rot}}{\sigma_0} = \frac{a}{f_\mathrm{gas}(z) Q_\mathrm{crit} }, 
\label{eq.vsigma_fgas}
\end{equation}
where $a=\sqrt{2}$ for a disk with constant rotational velocity and $Q_\mathrm{crit}=1.0$ for a quasi-stable thin gas disk \citep{2006ApJ...645.1062F,2011ApJ...733..101G}. As a result disk velocity dispersion is expected to evolve directly with the gas fraction;
\begin{equation}
\sigma_0 (z) = \frac{1}{\sqrt{2}}v_\mathrm{rot}f_\mathrm{gas}(z).
\label{eq.disp_evo}
\end{equation}

\begin{figure}  
\begin{center}
\includegraphics[scale=0.42,  trim=0.5cm 0cm 0cm 0cm, clip ]{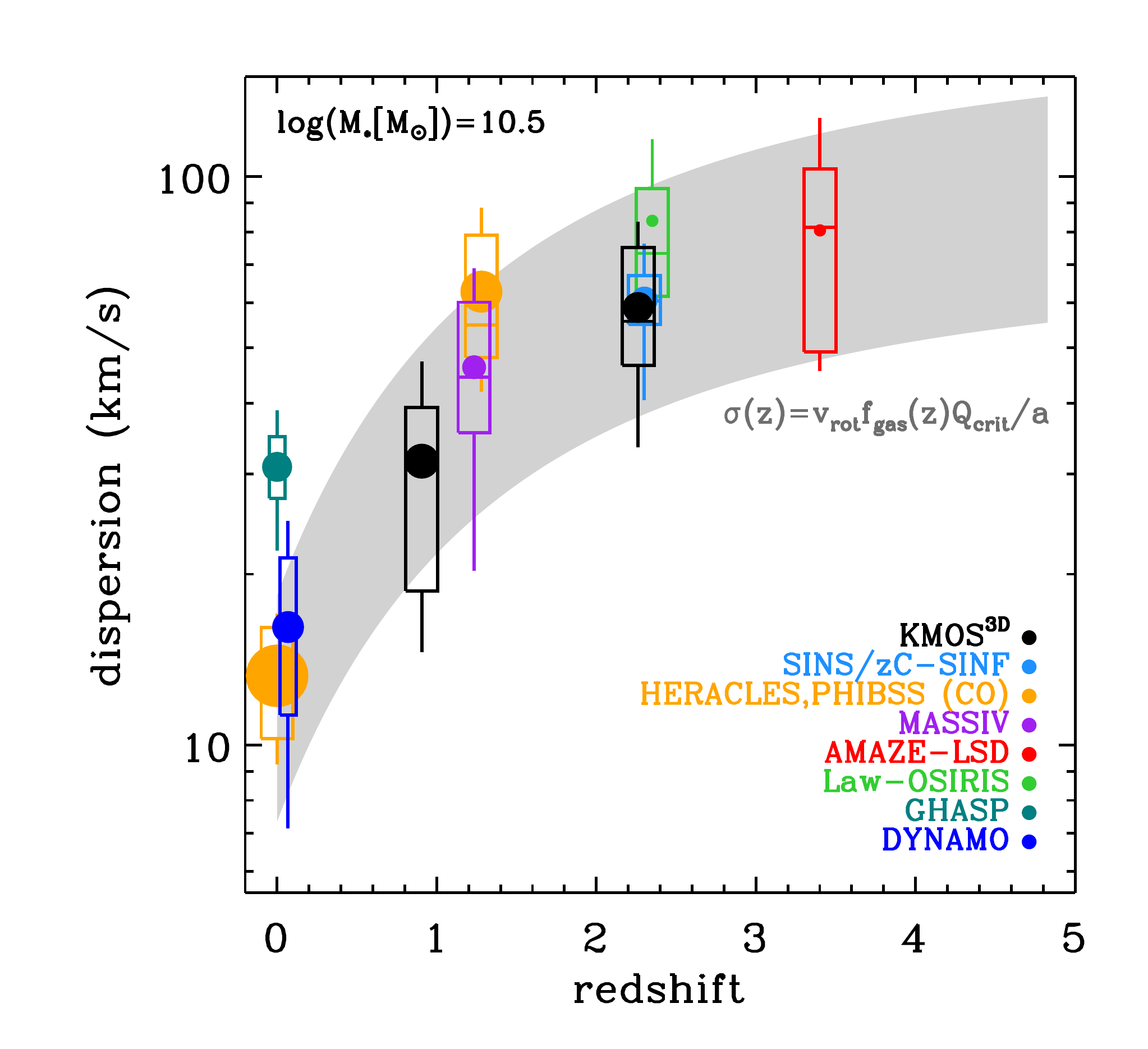}
\caption[...]{Same as Fig.~\ref{fig.dispevo} but showing the medians (horizontal lines) and means (circles) of samples that have high-resolution IFS data, their 50\% distributions (boxes) and 90\% distribution (vertical lines). The points representing each sample are sized to the average relative half-light size of the galaxies and adjusted using equations \ref{eq.fgas1}-\ref{eq.constants} to an average $\log(M_*)=10.5$. The grey band is described by ${\sigma_0} = {v_\mathrm{rot}  Q_\mathrm{crit}  f_\mathrm{gas}(z) } / \mathrm{a}$ where $f_\mathrm{gas}(z)$ is the gas fraction as a function of redshift as determined by equations~\ref{eq.fgas1}-\ref{eq.vsigma_fgas}, $Q_\mathrm{crit} =1$ and $\mathrm{a}=\sqrt{2}$ for a $\log(M_*)=10.5$ disk with constant rotational velocity.  The upper and lower boundaries of the curves are defined by $v_\mathrm{obs} = 250$ and 100 km s$^{-1}$ respectively.
 }
\label{fig.dispevo3}
\end{center}
\end{figure}

Figure~\ref{fig.dispevo3} shows the average velocity dispersions for disk galaxies only, taken from IFS surveys with good spatial and spectral resolution. The prediction for the evolution of dispersion for $\log{M_*}=10.5$ galaxies using the above assumptions is shown by the grey band, bounded by $v_\mathrm{obs}=100-250$ km s$^{-1}$ corresponding to the approximate spread of the peak of $v_\mathrm{rot}$ in Fig.~\ref{fig.disphisto}. The spread in $\sigma_0$ values for each survey, their 50\% distributions and 90\% distributions, are shown by the boxes and vertical lines respectively.

The observed dispersions and predicted evolution are in remarkably good agreement indicating that the evolution of measured velocity dispersions can be described by the evolution of key properties ($f_\mathrm{gas}$, $t_\mathrm{dep}$, sSFR) consistent with the equilibrium model. 
While correlations between $v/\sigma_0$, $\sigma_0$, $f_\mathrm{gas}$ and sSFR are uncertain due to a lack of dynamic range and large errors on individual measurements in Fig.~\ref{fig.disptrends}, by expanding to the wider redshift range and using the global scaling for $f_\mathrm{gas}$ (rather than inferring it for individual galaxies from other observed parameters) the influence of a more active and gas-rich environment on velocity dispersion is seen in Fig.~\ref{fig.dispevo3}.

Some caveats arise from assumptions made in the derivation for the adopted stellar mass, rotational velocity range, and critical Toomre parameter.  For instance, the sSFR is dependent on stellar mass (e.g. \citealt{2009ApJ...690..937D,2010ApJ...718.1001B,2014arXiv1407.1843W}) and the samples included in the analysis have average stellar masses ranging from $\log{M_*}=9.4-11.0$.  For a comparison at the same stellar mass, we adjust the average velocity dispersion for each sample to a reference stellar mass of $\log{M_*}=10.5$ using the ratio of equation~\ref{eq.fgas1} solved at the reference mass and the average mass of the sample. The average absolute adjustment is 5 \kms~with the largest adjustment being to the PHIBSS sample (25 \kms), which has an average stellar mass of $\log{M_*}=11.0$.

\begin{figure}  
\includegraphics[scale=0.42,  trim=0.5cm 0cm 0cm 0cm, clip ]{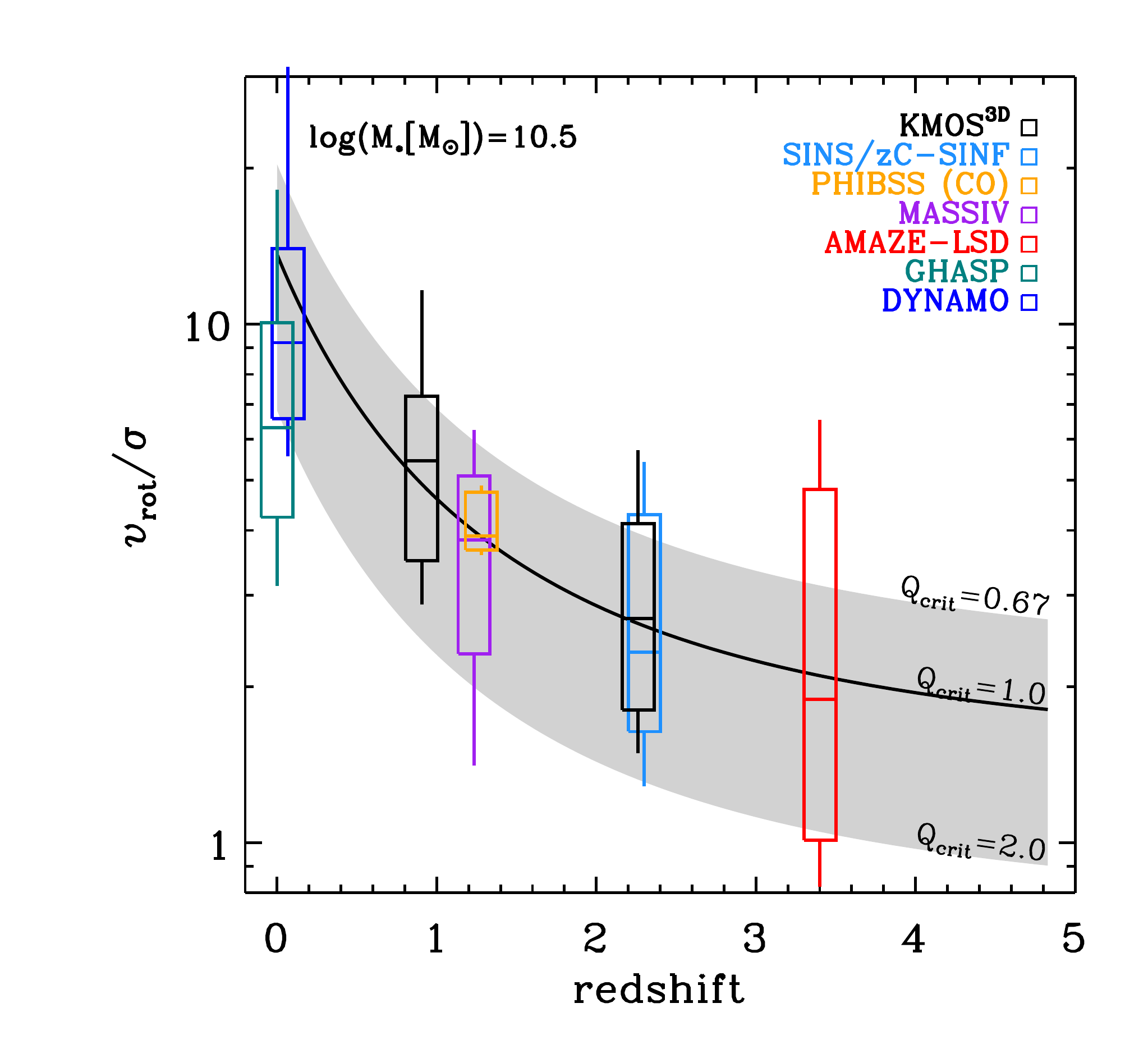}
\caption[...]{The medians (horizontal lines) of the disk stability diagnostic, $v_\mathrm{rot}/\sigma$, for IFS samples with boxes representing the central 50\% of the sample and the vertical lines representing the 90\% distribution. The values are adjusted to an average $\log(M_*)=10.5$.  The upper and lower boundaries of the curve are the predictions following equation~\ref{eq.disp_evo} where $Q_\mathrm{crit}=0.67, 2.0$ for a thick gas disk and composite disk. $Q_\mathrm{crit}=1$, the value adopted in this analysis, is shown by the solid line. The data are from the same surveys as described in Fig.~\ref{fig.dispevo} and Fig.~\ref{fig.dispevo3}. 
 }
\label{fig.dispevo4}
\end{figure}

The expected range of velocity dispersions at a given redshift can be widened by increasing the range of rotational velocities considered in equation~\ref{eq.disp_evo}. The boundary values of $100<(v_\mathrm{obs} [\mathrm{km~s}^{-1}])<250$ are used in Fig.~\ref{fig.dispevo3} to encompasses the peaks of the $z\sim1$ and $z\sim2$ histograms of rotational velocities of the \kmostd galaxies in Fig.~\ref{fig.disphisto} and are consistent with the other surveys considered with the exception of the sample of Law et al., which has average $v_\mathrm{rot}=50$ \kms.

We have made the assumption that all disk galaxies considered for this analysis are quasi-stable disks with $Q_\mathrm{crit}=1.0$, the value derived for a pure thin gas disk. However, the critical Toomre parameter is 0.67 for a thick gas disk and increases by factors $1-2$ for a stellar-plus-gas disk (e.g. \citealt{2007ApJ...660.1232K}). There are indications from our data and the literature that disk galaxies at $z\sim1-2$ are both thick (e.g. \citealt{2006ApJ...650..644E,2011ApJ...733..101G}) and composite (e.g. \citealt{2014arXiv1402.0866L}) thus we adopt $Q=1$ as an acceptable average value.  Increasing $Q_\mathrm{crit}$ to $Q=2$, as may be expected towards lower redshifts where stellar disks could play a more significant role in the stability of the system, would more than double the predictions for measured velocity dispersion. The resulting prediction would be consistent with the $z\sim0$ GHASP sample (Fig.~\ref{fig.dispevo}), which lies above the prediction in Fig.~\ref{fig.dispevo3}. 

To further test the validity of the assumption of $Q=1$ we use the observed $(v_\mathrm{rot}/\sigma)$ for disk galaxies, which is expected to be nearly constant over cosmic time in balance with $a/(f_\mathrm{gas}Q_\mathrm{crit})$ from equation~\ref{eq.vsigma_fgas} as shown in Fig.~\ref{fig.dispevo4}. Here $Q_\mathrm{crit}$ is varied between 0.67 and 2 to account for the difference between thin and thick gas disks and composite gas plus stellar disks.  All $z\gtrsim1$ samples are consistent with $Q=1$, although in most cases the full distribution of the sample encompasses a range of $Q$ values.
The $z=0$ surveys are in better agreement with $Q=2$. 
To extract more conclusive $Q_ \mathrm{crit}$ values, and/or a possible evolution in $Q_\mathrm{crit}$, observations with gas fractions measured on a galaxy-by-galaxy basis directly from molecular tracers are needed.  

\section{Conclusions}
\label{sec.conclusions}

This paper presents the design and first results of \kmostdns, a deep integral field spectroscopic survey targeting $>600$ $z=0.7-2.7$ galaxies, being carried out over the next $\sim5$ years with KMOS on the VLT.  \kmostd is designed to be a representative survey with a simple coherent selection from a comprehensive mass-selected parent catalog with redshifts from the near-infrared 3D-HST grism survey. The survey pushes well below the main sequence of star formation to characterise the internal dynamics and star formation of galaxies transiting from star forming to passive.

To date 246 galaxies have been observed for the \kmostd survey with observing times ranging from $2-20$ hours. A total of 191 galaxies with $M_*>4\times10^9$ \Msun~are detected in \halpha with KMOS; 90 galaxies at $z\sim1$ and 101 galaxies at $z\sim2$ with 70\% of the detected galaxies resolved. Detections cover $\gtrsim3$ dex in SFR and sSFR. Given the depth of the survey we detect \NII$\lambda6584$ in 77\% of the \halpha detected galaxies.

First results from the survey reveal that the main sequence of star-formation is dominated by rotating galaxies at both redshift regimes demonstrating the build up of size, central mass concentration, and ordered rotation when moving to higher galaxy stellar masses. We find  $93$\% of galaxies at $z\sim1$ and $74$\% at $z\sim2$ are rotationally-supported, as determined from a continuous velocity gradient and $v_\mathrm{rot}/\sigma_0>1$. We find a disk fraction of 58\% when applying the additional stricter criteria that the projected velocity dispersion distribution peaks on or near the kinematic center, the velocity gradient is measured along the photometric major axis (for inclined systems), and the closeness of the kinematic centroid to the center of the galaxy continuum. Galaxies well below the MS show rotational signatures while the few galaxies observed so far above the MS are compact with unresolved internal motions. Galaxies that are resolved but not rotating are found primarily at low $M_*$. We observe 11 galaxy `close pairs' (within 500 \kms~and $\sim12$ kpc) that have a variety of kinematic structure from rotating companions to chaotic motions and are found in all populated regions of the SFR$-M_*$ plane.

With the \kmostd data we confirm a factor of two decrease in ionized gas velocity dispersions from 50 km s$^{-1}$ at $z\sim2$ and 25 km s$^{-1}$ at $z\sim1$ using representative populations measured with consistent methods.  When these measurements are considered in the context of disk velocity dispersions from $z=0-4$, we report an evolution of ionized gas velocity dispersion that closely follows the evolution of specific star formation rate and gas fractions, consistent with the ``equilibrium''  or ``regulator'' model and providing evidence that disk turbulence is being set by the balance of gas fuelling and star formation as predicted by marginally stable disk theory.

The new multi-IFU KMOS instrument allows us to take the next major step in IFS surveys of distant galaxies, enabling sensitive observations of large samples across a broad range of redshifts with wider and more uniform coverage of galaxy parameter space. The first year data and results from our \kmostd survey and other surveys open up new avenues in investigating the early evolution of galaxies. As more data are collected over the next few years better constrains on the kinematic and star-forming properties should provide a much more complete picture of the processes driving the growth and star-formation shutdown of galaxies at the crucial $z\sim0.5-3$ epochs.

\acknowledgments{
We wish to thank the ESO staff, and in particular the staff at Paranal Observatory, for their helpful and enthusiastic support during the many observing runs over which the KMOS GTO were carried out. We thank the entire KMOS instrument and Commissioning team for their hard work, which allowed our observational program to be carried out so successfully. We also thank the software development team of \spark for all their work with us to get the most out of the data. This paper and the \kmostd survey have benefitted from many constructive, insightful, and enthusiastic discussions with many colleagues whom we are very grateful to, especially M. Franx, A. Renzini, K. Whitaker, K. Glazebrook, and R. Bassett. DJW and MF acknowledge the support of the Deutsche Forschungsgemeinschaft via Project ID 387/1-1. We thank the referee for a thorough reading and valuable comments.}

\bibliographystyle{apj}

\begin{thebibliography}{}
\expandafter\ifx\csname natexlab\endcsname\relax\def\natexlab#1{#1}\fi

\bibitem[{{Arribas} {et~al.}(2014){Arribas}, {Colina}, {Bellocchi}, {Maiolino},
  \& {Villar-Mart{\'{\i}}n}}]{2014A&A...568A..14A}
{Arribas}, S., {Colina}, L., {Bellocchi}, E., {Maiolino}, R., \&
  {Villar-Mart{\'{\i}}n}, M. 2014, \aap, 568, A14

\bibitem[{{Aumer} {et~al.}(2010){Aumer}, {Burkert}, {Johansson}, \&
  {Genzel}}]{2010ApJ...719.1230A}
{Aumer}, M., {Burkert}, A., {Johansson}, P.~H., \& {Genzel}, R. 2010, \apj,
  719, 1230

\bibitem[{{Bigiel} {et~al.}(2011){Bigiel}, {Leroy}, {Walter}, {Brinks}, {de
  Blok}, {Kramer}, {Rix}, {Schruba}, {Schuster}, {Usero}, \&
  {Wiesemeyer}}]{2011ApJ...730L..13B}
{Bigiel}, F., {Leroy}, A.~K., {Walter}, F., {et~al.} 2011, \apjl, 730, L13

\bibitem[{{Bluck} {et~al.}(2009){Bluck}, {Conselice}, {Bouwens}, {Daddi},
  {Dickinson}, {Papovich}, \& {Yan}}]{2009MNRAS.394L..51B}
{Bluck}, A.~F.~L., {Conselice}, C.~J., {Bouwens}, R.~J., {et~al.} 2009, \mnras,
  394, L51

\bibitem[{{Bongiorno} {et~al.}(2012){Bongiorno}, {Merloni}, {Brusa},
  {Magnelli}, {Salvato}, {Mignoli}, {Zamorani}, {Fiore}, {Rosario}, {Mainieri},
  {Hao}, {Comastri}, {Vignali}, {Balestra}, {Bardelli}, {Berta}, {Civano},
  {Kampczyk}, {Le Floc'h}, {Lusso}, {Lutz}, {Pozzetti}, {Pozzi}, {Riguccini},
  {Shankar}, \& {Silverman}}]{2012MNRAS.427.3103B}
{Bongiorno}, A., {Merloni}, A., {Brusa}, M., {et~al.} 2012, \mnras, 427, 3103

\bibitem[{{Bouch{\'e}} {et~al.}(2010){Bouch{\'e}}, {Dekel}, {Genzel}, {Genel},
  {Cresci}, {F{\"o}rster Schreiber}, {Shapiro}, {Davies}, \&
  {Tacconi}}]{2010ApJ...718.1001B}
{Bouch{\'e}}, N., {Dekel}, A., {Genzel}, R., {et~al.} 2010, \apj, 718, 1001

\bibitem[{{Bournaud} {et~al.}(2010){Bournaud}, {Elmegreen}, {Teyssier},
  {Block}, \& {Puerari}}]{2010MNRAS.409.1088B}
{Bournaud}, F., {Elmegreen}, B.~G., {Teyssier}, R., {Block}, D.~L., \&
  {Puerari}, I. 2010, \mnras, 409, 1088

\bibitem[{{Bournaud} {et~al.}(2011){Bournaud}, {Chapon}, {Teyssier}, {Powell},
  {Elmegreen}, {Elmegreen}, {Duc}, {Contini}, {Epinat}, \&
  {Shapiro}}]{2011ApJ...730....4B}
{Bournaud}, F., {Chapon}, D., {Teyssier}, R., {et~al.} 2011, \apj, 730, 4

\bibitem[{{Bouwens} {et~al.}(2012){Bouwens}, {Illingworth}, {Oesch}, {Franx},
  {Labb{\'e}}, {Trenti}, {van Dokkum}, {Carollo}, {Gonz{\'a}lez}, {Smit}, \&
  {Magee}}]{2012ApJ...754...83B}
{Bouwens}, R.~J., {Illingworth}, G.~D., {Oesch}, P.~A., {et~al.} 2012, \apj,
  754, 83

\bibitem[{{Bower} {et~al.}(2006){Bower}, {Benson}, {Malbon}, {Helly}, {Frenk},
  {Baugh}, {Cole}, \& {Lacey}}]{2006MNRAS.370..645B}
{Bower}, R.~G., {Benson}, A.~J., {Malbon}, R., {et~al.} 2006, \mnras, 370, 645

\bibitem[{Brammer {et~al.}(2012)Brammer, van Dokkum, Franx, Fumagalli, Patel,
  Rix, Skelton, Kriek, Nelson, Schmidt, Bezanson, da~Cunha, Erb, Fan,
  Schreiber, Illingworth, Labb{\'e}, Leja, Lundgren, Magee, Marchesini,
  McCarthy, Momcheva, Muzzin, Quadri, Steidel, Tal, Wake, Whitaker, \&
  Williams}]{brammer:2012:04}
Brammer, G., van Dokkum, P., Franx, M., {et~al.} 2012, 1204.2829

\bibitem[{{Brusa} {et~al.}(2009){Brusa}, {Fiore}, {Santini}, {Grazian},
  {Comastri}, {Zamorani}, {Hasinger}, {Merloni}, {Civano}, {Fontana}, \&
  {Mainieri}}]{2009A&A...507.1277B}
{Brusa}, M., {Fiore}, F., {Santini}, P., {et~al.} 2009, \aap, 507, 1277

\bibitem[{{Cacciato} {et~al.}(2012){Cacciato}, {Dekel}, \&
  {Genel}}]{2012MNRAS.421..818C}
{Cacciato}, M., {Dekel}, A., \& {Genel}, S. 2012, \mnras, 421, 818

\bibitem[{{Calzetti} {et~al.}(2000){Calzetti}, {Armus}, {Bohlin}, {Kinney},
  {Koornneef}, \& {Storchi-Bergmann}}]{2000ApJ...533..682C}
{Calzetti}, D., {Armus}, L., {Bohlin}, R.~C., {et~al.} 2000, \apj, 533, 682

\bibitem[{{Chabrier}(2003)}]{2003PASP..115..763C}
{Chabrier}, G. 2003, \pasp, 115, 763

\bibitem[{{Civano} {et~al.}(2012){Civano}, {Elvis}, {Brusa}, {Comastri},
  {Salvato}, {Zamorani}, {Aldcroft}, {Bongiorno}, {Capak}, {Cappelluti},
  {Cisternas}, {Fiore}, {Fruscione}, {Hao}, {Kartaltepe}, {Koekemoer}, {Gilli},
  {Impey}, {Lanzuisi}, {Lusso}, {Mainieri}, {Miyaji}, {Lilly}, {Masters},
  {Puccetti}, {Schawinski}, {Scoville}, {Silverman}, {Trump}, {Urry},
  {Vignali}, \& {Wright}}]{2012ApJS..201...30C}
{Civano}, F., {Elvis}, M., {Brusa}, M., {et~al.} 2012, \apjs, 201, 30

\bibitem[{{Combes} {et~al.}(2011){Combes}, {Garc{\'{\i}}a-Burillo}, {Braine},
  {Schinnerer}, {Walter}, \& {Colina}}]{2011A&A...528A.124C}
{Combes}, F., {Garc{\'{\i}}a-Burillo}, S., {Braine}, J., {et~al.} 2011, \aap,
  528, A124+

\bibitem[{{Contini} {et~al.}(2012){Contini}, {Garilli}, {Le F{\`e}vre},
  {Kissler-Patig}, {Amram}, {Epinat}, {Moultaka}, {Paioro}, {Queyrel}, {Tasca},
  {Tresse}, {Vergani}, {L{\'o}pez-Sanjuan}, \&
  {Perez-Montero}}]{2012A&A...539A..91C}
{Contini}, T., {Garilli}, B., {Le F{\`e}vre}, O., {et~al.} 2012, \aap, 539, A91

\bibitem[{{Cresci} {et~al.}(2009){Cresci}, {Hicks}, {Genzel}, {Schreiber},
  {Davies}, {Bouch{\'e}}, {Buschkamp}, {Genel}, {Shapiro}, {Tacconi},
  {Sommer-Larsen}, {Burkert}, {Eisenhauer}, {Gerhard}, {Lutz}, {Naab},
  {Sternberg}, {Cimatti}, {Daddi}, {Erb}, {Kurk}, {Lilly}, {Renzini},
  {Shapley}, {Steidel}, \& {Caputi}}]{2009ApJ...697..115C}
{Cresci}, G., {Hicks}, E.~K.~S., {Genzel}, R., {et~al.} 2009, \apj, 697, 115

\bibitem[{{Daddi} {et~al.}(2007{\natexlab{a}}){Daddi}, {Dickinson}, {Morrison},
  {Chary}, {Cimatti}, {Elbaz}, {Frayer}, {Renzini}, {Pope}, {Alexander},
  {Bauer}, {Giavalisco}, {Huynh}, {Kurk}, \& {Mignoli}}]{2007ApJ...670..156D}
{Daddi}, E., {Dickinson}, M., {Morrison}, G., {et~al.} 2007{\natexlab{a}},
  \apj, 670, 156

\bibitem[{{Daddi} {et~al.}(2007{\natexlab{b}}){Daddi}, {Alexander},
  {Dickinson}, {Gilli}, {Renzini}, {Elbaz}, {Cimatti}, {Chary}, {Frayer},
  {Bauer}, {Brandt}, {Giavalisco}, {Grogin}, {Huynh}, {Kurk}, {Mignoli},
  {Morrison}, {Pope}, \& {Ravindranath}}]{2007ApJ...670..173D}
{Daddi}, E., {Alexander}, D.~M., {Dickinson}, M., {et~al.} 2007{\natexlab{b}},
  \apj, 670, 173

\bibitem[{{Daddi} {et~al.}(2010){Daddi}, {Bournaud}, {Walter}, {Dannerbauer},
  {Carilli}, {Dickinson}, {Elbaz}, {Morrison}, {Riechers}, {Onodera}, {Salmi},
  {Krips}, \& {Stern}}]{2010ApJ...713..686D}
{Daddi}, E., {Bournaud}, F., {Walter}, F., {et~al.} 2010, \apj, 713, 686

\bibitem[{{Damen} {et~al.}(2009){Damen}, {Labb{\'e}}, {Franx}, {van Dokkum},
  {Taylor}, \& {Gawiser}}]{2009ApJ...690..937D}
{Damen}, M., {Labb{\'e}}, I., {Franx}, M., {et~al.} 2009, \apj, 690, 937

\bibitem[{{Dav{\'e}} {et~al.}(2012){Dav{\'e}}, {Finlator}, \&
  {Oppenheimer}}]{2012MNRAS.421...98D}
{Dav{\'e}}, R., {Finlator}, K., \& {Oppenheimer}, B.~D. 2012, \mnras, 421, 98

\bibitem[{{Davies} {et~al.}(2011){Davies}, {F{\"o}rster Schreiber}, {Cresci},
  {Genzel}, {Bouch{\'e}}, {Burkert}, {Buschkamp}, {Genel}, {Hicks}, {Kurk},
  {Lutz}, {Newman}, {Shapiro}, {Sternberg}, {Tacconi}, \&
  {Wuyts}}]{2011ApJ...741...69D}
{Davies}, R., {F{\"o}rster Schreiber}, N.~M., {Cresci}, G., {et~al.} 2011,
  \apj, 741, 69

\bibitem[{{Davies} {et~al.}(2013){Davies}, {Agudo Berbel}, {Wiezorrek},
  {Cirasuolo}, {Foerster Schreiber}, {Jung}, {Muschielok}, {Ott}, {Ramsay},
  {Schlichter}, {Sharples}, \& {Wegner}}]{2013arXiv1308.6679D}
{Davies}, R., {Agudo Berbel}, A., {Wiezorrek}, E., {et~al.} 2013, ArXiv
  e-prints, arXiv:1308.6679

\bibitem[{{Dekel} \& {Mandelker}(2014)}]{2014arXiv1402.2283D}
{Dekel}, A., \& {Mandelker}, N. 2014, ArXiv e-prints, arXiv:1402.2283

\bibitem[{{Dekel} {et~al.}(2009){Dekel}, {Sari}, \&
  {Ceverino}}]{2009ApJ...703..785D}
{Dekel}, A., {Sari}, R., \& {Ceverino}, D. 2009, \apj, 703, 785

\bibitem[{{Dib} {et~al.}(2006){Dib}, {Bell}, \& {Burkert}}]{Dib:2006fk}
{Dib}, S., {Bell}, E., \& {Burkert}, A. 2006, \apj, 638, 797

\bibitem[{{Dutton} {et~al.}(2010){Dutton}, {van den Bosch}, \&
  {Dekel}}]{2010MNRAS.405.1690D}
{Dutton}, A.~A., {van den Bosch}, F.~C., \& {Dekel}, A. 2010, \mnras, 405, 1690

\bibitem[{{Elbaz} {et~al.}(2007){Elbaz}, {Daddi}, {Le Borgne}, {Dickinson},
  {Alexander}, {Chary}, {Starck}, {Brandt}, {Kitzbichler}, {MacDonald},
  {Nonino}, {Popesso}, {Stern}, \& {Vanzella}}]{2007A&A...468...33E}
{Elbaz}, D., {Daddi}, E., {Le Borgne}, D., {et~al.} 2007, \aap, 468, 33

\bibitem[{{Elbaz} {et~al.}(2011){Elbaz}, {Dickinson}, {Hwang},
  {D{\'{\i}}az-Santos}, {Magdis}, {Magnelli}, {Le Borgne}, {Galliano},
  {Pannella}, {Chanial}, {Armus}, {Charmandaris}, {Daddi}, {Aussel}, {Popesso},
  {Kartaltepe}, {Altieri}, {Valtchanov}, {Coia}, {Dannerbauer}, {Dasyra},
  {Leiton}, {Mazzarella}, {Alexander}, {Buat}, {Burgarella}, {Chary}, {Gilli},
  {Ivison}, {Juneau}, {Le Floc'h}, {Lutz}, {Morrison}, {Mullaney}, {Murphy},
  {Pope}, {Scott}, {Brodwin}, {Calzetti}, {Cesarsky}, {Charlot}, {Dole},
  {Eisenhardt}, {Ferguson}, {F{\"o}rster Schreiber}, {Frayer}, {Giavalisco},
  {Huynh}, {Koekemoer}, {Papovich}, {Reddy}, {Surace}, {Teplitz}, {Yun}, \&
  {Wilson}}]{2011A&A...533A.119E}
{Elbaz}, D., {Dickinson}, M., {Hwang}, H.~S., {et~al.} 2011, \aap, 533, A119

\bibitem[{{Elmegreen} \& {Burkert}(2010)}]{2010ApJ...712..294E}
{Elmegreen}, B.~G., \& {Burkert}, A. 2010, \apj, 712, 294

\bibitem[{{Elmegreen} \& {Elmegreen}(2006)}]{2006ApJ...650..644E}
{Elmegreen}, B.~G., \& {Elmegreen}, D.~M. 2006, \apj, 650, 644

\bibitem[{{Epinat} {et~al.}(2010){Epinat}, {Amram}, {Balkowski}, \&
  {Marcelin}}]{2010MNRAS.401.2113E}
{Epinat}, B., {Amram}, P., {Balkowski}, C., \& {Marcelin}, M. 2010, \mnras,
  401, 2113

\bibitem[{{Epinat} {et~al.}(2009){Epinat}, {Contini}, {Le F{\`e}vre},
  {Vergani}, {Garilli}, {Amram}, {Queyrel}, {Tasca}, \&
  {Tresse}}]{2009A&A...504..789E}
{Epinat}, B., {Contini}, T., {Le F{\`e}vre}, O., {et~al.} 2009, \aap, 504, 789

\bibitem[{{Epinat} {et~al.}(2012){Epinat}, {Tasca}, {Amram}, {Contini}, {Le
  F{\`e}vre}, {Queyrel}, {Vergani}, {Garilli}, {Kissler-Patig}, {Moultaka},
  {Paioro}, {Tresse}, {Bournaud}, {L{\'o}pez-Sanjuan}, \&
  {Perret}}]{2012A&A...539A..92E}
{Epinat}, B., {Tasca}, L., {Amram}, P., {et~al.} 2012, \aap, 539, A92

\bibitem[{{Erb} {et~al.}(2004){Erb}, {Steidel}, {Shapley}, {Pettini}, \&
  {Adelberger}}]{2004ApJ...612..122E}
{Erb}, D.~K., {Steidel}, C.~C., {Shapley}, A.~E., {Pettini}, M., \&
  {Adelberger}, K.~L. 2004, \apj, 612, 122

\bibitem[{{Forbes} {et~al.}(2012){Forbes}, {Krumholz}, \&
  {Burkert}}]{2012ApJ...754...48F}
{Forbes}, J., {Krumholz}, M., \& {Burkert}, A. 2012, \apj, 754, 48

\bibitem[{{F{\"o}rster Schreiber} {et~al.}(2006){F{\"o}rster Schreiber},
  {Genzel}, {Lehnert}, {Bouch{\'e}}, {Verma}, {Erb}, {Shapley}, {Steidel},
  {Davies}, {Lutz}, {Nesvadba}, {Tacconi}, {Eisenhauer}, {Abuter}, {Gilbert},
  {Gillessen}, \& {Sternberg}}]{2006ApJ...645.1062F}
{F{\"o}rster Schreiber}, N.~M., {Genzel}, R., {Lehnert}, M.~D., {et~al.} 2006,
  \apj, 645, 1062

\bibitem[{{F{\"o}rster Schreiber} {et~al.}(2009){F{\"o}rster Schreiber},
  {Genzel}, {Bouch{\'e}}, {Cresci}, {Davies}, {Buschkamp}, {Shapiro},
  {Tacconi}, {Hicks}, {Genel}, {Shapley}, {Erb}, {Steidel}, {Lutz},
  {Eisenhauer}, {Gillessen}, {Sternberg}, {Renzini}, {Cimatti}, {Daddi},
  {Kurk}, {Lilly}, {Kong}, {Lehnert}, {Nesvadba}, {Verma}, {McCracken},
  {Arimoto}, {Mignoli}, \& {Onodera}}]{2009ApJ...706.1364F}
{F{\"o}rster Schreiber}, N.~M., {Genzel}, R., {Bouch{\'e}}, N., {et~al.} 2009,
  \apj, 706, 1364

\bibitem[{{F{\"o}rster Schreiber} {et~al.}(2014){F{\"o}rster Schreiber},
  {Genzel}, {Newman}, {Kurk}, {Lutz}, {Tacconi}, {Wuyts}, {Bandara}, {Burkert},
  {Buschkamp}, {Carollo}, {Cresci}, {Daddi}, {Davies}, {Eisenhauer}, {Hicks},
  {Lang}, {Lilly}, {Mainieri}, {Mancini}, {Naab}, {Peng}, {Renzini}, {Rosario},
  {Shapiro Griffin}, {Shapley}, {Sternberg}, {Tacchella}, {Vergani},
  {Wisnioski}, {Wuyts}, \& {Zamorani}}]{2014ApJ...787...38F}
{F{\"o}rster Schreiber}, N.~M., {Genzel}, R., {Newman}, S.~F., {et~al.} 2014,
  \apj, 787, 38

\bibitem[{{Franx} {et~al.}(1991){Franx}, {Illingworth}, \& {de
  Zeeuw}}]{1991ApJ...383..112F}
{Franx}, M., {Illingworth}, G., \& {de Zeeuw}, T. 1991, \apj, 383, 112

\bibitem[{{Freeman}(1970)}]{1970ApJ...160..811F}
{Freeman}, K.~C. 1970, \apj, 160, 811

\bibitem[{{Genel} {et~al.}(2010){Genel}, {Bouch{\'e}}, {Naab}, {Sternberg}, \&
  {Genzel}}]{2010ApJ...719..229G}
{Genel}, S., {Bouch{\'e}}, N., {Naab}, T., {Sternberg}, A., \& {Genzel}, R.
  2010, \apj, 719, 229

\bibitem[{{Genzel} {et~al.}(2006){Genzel}, {Tacconi}, {Eisenhauer},
  {F{\"o}rster Schreiber}, {Cimatti}, {Daddi}, {Bouch{\'e}}, {Davies},
  {Lehnert}, {Lutz}, {Nesvadba}, {Verma}, {Abuter}, {Shapiro}, {Sternberg},
  {Renzini}, {Kong}, {Arimoto}, \& {Mignoli}}]{2006Natur.442..786G}
{Genzel}, R., {Tacconi}, L.~J., {Eisenhauer}, F., {et~al.} 2006, \nat, 442, 786

\bibitem[{{Genzel} {et~al.}(2008){Genzel}, {Burkert}, {Bouch{\'e}}, {Cresci},
  {F{\"o}rster Schreiber}, {Shapley}, {Shapiro}, {Tacconi}, {Buschkamp},
  {Cimatti}, {Daddi}, {Davies}, {Eisenhauer}, {Erb}, {Genel}, {Gerhard},
  {Hicks}, {Lutz}, {Naab}, {Ott}, {Rabien}, {Renzini}, {Steidel}, {Sternberg},
  \& {Lilly}}]{2008ApJ...687...59G}
{Genzel}, R., {Burkert}, A., {Bouch{\'e}}, N., {et~al.} 2008, \apj, 687, 59

\bibitem[{{Genzel} {et~al.}(2011){Genzel}, {Newman}, {Jones}, {F{\"o}rster
  Schreiber}, {Shapiro}, {Genel}, {Lilly}, {Renzini}, {Tacconi}, {Bouch{\'e}},
  {Burkert}, {Cresci}, {Buschkamp}, {Carollo}, {Ceverino}, {Davies}, {Dekel},
  {Eisenhauer}, {Hicks}, {Kurk}, {Lutz}, {Mancini}, {Naab}, {Peng},
  {Sternberg}, {Vergani}, \& {Zamorani}}]{2011ApJ...733..101G}
{Genzel}, R., {Newman}, S., {Jones}, T., {et~al.} 2011, \apj, 733, 101

\bibitem[{{Genzel} {et~al.}(2014{\natexlab{a}}){Genzel}, {Tacconi}, {Lutz},
  {Saintonge}, {Berta}, {Magnelli}, {Combes}, {Garc{\'{\i}}a-Burillo}, {Neri},
  {Bolatto}, {Contini}, {Lilly}, {Boissier}, {Boone}, {Bouch{\'e}}, {Bournaud},
  {Burkert}, {Carollo}, {Colina}, {Cooper}, {Cox}, {Feruglio}, {F{\"o}rster
  Schreiber}, {Freundlich}, {Gracia-Carpio}, {Juneau}, {Kovac}, {Lippa},
  {Naab}, {Salome}, {Renzini}, {Sternberg}, {Walter}, {Weiner}, {Weiss}, \&
  {Wuyts}}]{2014arXiv1409.1171G}
{Genzel}, R., {Tacconi}, L.~J., {Lutz}, D., {et~al.} 2014{\natexlab{a}}, ArXiv
  e-prints, arXiv:1409.1171

\bibitem[{{Genzel} {et~al.}(2014{\natexlab{b}}){Genzel}, {F{\"o}rster
  Schreiber}, {Rosario}, {Lang}, {Lutz}, {Wisnioski}, {Wuyts}, {Wuyts},
  {Bandara}, {Bender}, {Berta}, {Kurk}, {Mendel}, {Tacconi}, {Wilman},
  {Beifiori}, {Brammer}, {Burkert}, {Buschkamp}, {Chan}, {Carollo}, {Davies},
  {Eisenhauer}, {Fabricius}, {Fossati}, {Kriek}, {Kulkarni}, {Lilly},
  {Mancini}, {Momcheva}, {Naab}, {Nelson}, {Renzini}, {Saglia}, {Sharples},
  {Sternberg}, {Tacchella}, \& {van Dokkum}}]{2014ApJ...796....7G}
{Genzel}, R., {F{\"o}rster Schreiber}, N.~M., {Rosario}, D., {et~al.}
  2014{\natexlab{b}}, \apj, 796, 7 

\bibitem[{{Genzel} {et~al.}(2014{\natexlab{c}}){Genzel}, {F{\"o}rster
  Schreiber}, {Lang}, {Tacchella}, {Tacconi}, {Wuyts}, {Bandara}, {Burkert},
  {Buschkamp}, {Carollo}, {Cresci}, {Davies}, {Eisenhauer}, {Hicks}, {Kurk},
  {Lilly}, {Lutz}, {Mancini}, {Naab}, {Newman}, {Peng}, {Renzini}, {Shapiro
  Griffin}, {Sternberg}, {Vergani}, {Wisnioski}, {Wuyts}, \&
  {Zamorani}}]{2014ApJ...785...75G}
{Genzel}, R., {F{\"o}rster Schreiber}, N.~M., {Lang}, P., {et~al.}
  2014{\natexlab{c}}, \apj, 785, 75

\bibitem[{{Glazebrook}(2013)}]{2013arXiv1305.2469G}
{Glazebrook}, K. 2013, ArXiv e-prints, arXiv:1305.2469

\bibitem[{{Gnerucci} {et~al.}(2011){Gnerucci}, {Marconi}, {Cresci}, {Maiolino},
  {Mannucci}, {Calura}, {Cimatti}, {Cocchia}, {Grazian}, {Matteucci}, {Nagao},
  {Pozzetti}, \& {Troncoso}}]{2011A&A...528A..88G}
{Gnerucci}, A., {Marconi}, A., {Cresci}, G., {et~al.} 2011, \aap, 528, A88

\bibitem[{{Gonz{\'a}lez} {et~al.}(2014){Gonz{\'a}lez}, {Bouwens},
  {Illingworth}, {Labb{\'e}}, {Oesch}, {Franx}, \&
  {Magee}}]{2014ApJ...781...34G}
{Gonz{\'a}lez}, V., {Bouwens}, R., {Illingworth}, G., {et~al.} 2014, \apj, 781,
  34

\bibitem[{{Green} {et~al.}(2010){Green}, {Glazebrook}, {McGregor}, {Abraham},
  {Poole}, {Damjanov}, {McCarthy}, {Colless}, \& {Sharp}}]{Green:2010fk}
{Green}, A.~W., {Glazebrook}, K., {McGregor}, P.~J., {et~al.} 2010, \nat, 467,
  684

\bibitem[{{Green} {et~al.}(2014){Green}, {Glazebrook}, {McGregor}, {Damjanov},
  {Wisnioski}, {Abraham}, {Colless}, {Sharp}, {Crain}, {Poole}, \&
  {McCarthy}}]{2014MNRAS.437.1070G}
---. 2014, \mnras, 437, 1070

\bibitem[{{Grogin} {et~al.}(2011){Grogin}, {Kocevski}, {Faber}, {Ferguson},
  {Koekemoer}, {Riess}, {Acquaviva}, {Alexander}, {Almaini}, {Ashby}, {Barden},
  {Bell}, {Bournaud}, {Brown}, {Caputi}, {Casertano}, {Cassata}, {Castellano},
  {Challis}, {Chary}, {Cheung}, {Cirasuolo}, {Conselice}, {Roshan Cooray},
  {Croton}, {Daddi}, {Dahlen}, {Dav{\'e}}, {de Mello}, {Dekel}, {Dickinson},
  {Dolch}, {Donley}, {Dunlop}, {Dutton}, {Elbaz}, {Fazio}, {Filippenko},
  {Finkelstein}, {Fontana}, {Gardner}, {Garnavich}, {Gawiser}, {Giavalisco},
  {Grazian}, {Guo}, {Hathi}, {H{\"a}ussler}, {Hopkins}, {Huang}, {Huang},
  {Jha}, {Kartaltepe}, {Kirshner}, {Koo}, {Lai}, {Lee}, {Li}, {Lotz}, {Lucas},
  {Madau}, {McCarthy}, {McGrath}, {McIntosh}, {McLure}, {Mobasher},
  {Moustakas}, {Mozena}, {Nandra}, {Newman}, {Niemi}, {Noeske}, {Papovich},
  {Pentericci}, {Pope}, {Primack}, {Rajan}, {Ravindranath}, {Reddy}, {Renzini},
  {Rix}, {Robaina}, {Rodney}, {Rosario}, {Rosati}, {Salimbeni}, {Scarlata},
  {Siana}, {Simard}, {Smidt}, {Somerville}, {Spinrad}, {Straughn}, {Strolger},
  {Telford}, {Teplitz}, {Trump}, {van der Wel}, {Villforth}, {Wechsler},
  {Weiner}, {Wiklind}, {Wild}, {Wilson}, {Wuyts}, {Yan}, \&
  {Yun}}]{2011ApJS..197...35G}
{Grogin}, N.~A., {Kocevski}, D.~D., {Faber}, S.~M., {et~al.} 2011, \apjs, 197,
  35

\bibitem[{{Guo} {et~al.}(2010){Guo}, {White}, {Li}, \&
  {Boylan-Kolchin}}]{2010MNRAS.404.1111G}
{Guo}, Q., {White}, S., {Li}, C., \& {Boylan-Kolchin}, M. 2010, \mnras, 404,
  1111

\bibitem[{Hainline {et~al.}(2012)Hainline, Shapley, Greene, Steidel, Reddy, \&
  Erb}]{hainline:2012:06}
Hainline, K.~N., Shapley, A.~E., Greene, J.~E., {et~al.} 2012, 1206.3308

\bibitem[{{Harrison} {et~al.}(2012){Harrison}, {Alexander}, {Swinbank},
  {Smail}, {Alaghband-Zadeh}, {Bauer}, {Chapman}, {Del Moro}, {Hickox},
  {Ivison}, {Men{\'e}ndez-Delmestre}, {Mullaney}, \&
  {Nesvadba}}]{2012MNRAS.426.1073H}
{Harrison}, C.~M., {Alexander}, D.~M., {Swinbank}, A.~M., {et~al.} 2012,
  \mnras, 426, 1073

\bibitem[{{Hopkins} {et~al.}(2010){Hopkins}, {Bundy}, {Croton}, {Hernquist},
  {Keres}, {Khochfar}, {Stewart}, {Wetzel}, \& {Younger}}]{2010ApJ...715..202H}
{Hopkins}, P.~F., {Bundy}, K., {Croton}, D., {et~al.} 2010, \apj, 715, 202

\bibitem[{{Hung} {et~al.}(2013){Hung}, {Sanders}, {Casey}, {Lee}, {Barnes},
  {Capak}, {Kartaltepe}, {Koss}, {Larson}, {Le Floc'h}, {Lockhart}, {Man},
  {Mann}, {Riguccini}, {Scoville}, \& {Symeonidis}}]{2013ApJ...778..129H}
{Hung}, C.-L., {Sanders}, D.~B., {Casey}, C.~M., {et~al.} 2013, \apj, 778, 129

\bibitem[{{Immeli} {et~al.}(2004){Immeli}, {Samland}, {Gerhard}, \&
  {Westera}}]{2004A&A...413..547I}
{Immeli}, A., {Samland}, M., {Gerhard}, O., \& {Westera}, P. 2004, \aap, 413,
  547

\bibitem[{{Kartaltepe} {et~al.}(2012){Kartaltepe}, {Dickinson}, {Alexander},
  {Bell}, {Dahlen}, {Elbaz}, {Faber}, {Lotz}, {McIntosh}, {Wiklind}, {Altieri},
  {Aussel}, {Bethermin}, {Bournaud}, {Charmandaris}, {Conselice}, {Cooray},
  {Dannerbauer}, {Dav{\'e}}, {Dunlop}, {Dekel}, {Ferguson}, {Grogin}, {Hwang},
  {Ivison}, {Kocevski}, {Koekemoer}, {Koo}, {Lai}, {Leiton}, {Lucas}, {Lutz},
  {Magdis}, {Magnelli}, {Morrison}, {Mozena}, {Mullaney}, {Newman}, {Pope},
  {Popesso}, {van der Wel}, {Weiner}, \& {Wuyts}}]{2012ApJ...757...23K}
{Kartaltepe}, J.~S., {Dickinson}, M., {Alexander}, D.~M., {et~al.} 2012, \apj,
  757, 23

\bibitem[{{Kassin} {et~al.}(2012){Kassin}, {Weiner}, {Faber}, {Gardner},
  {Willmer}, {Coil}, {Cooper}, {Devriendt}, {Dutton}, {Guhathakurta}, {Koo},
  {Metevier}, {Noeske}, \& {Primack}}]{2012ApJ...758..106K}
{Kassin}, S.~A., {Weiner}, B.~J., {Faber}, S.~M., {et~al.} 2012, \apj, 758, 106

\bibitem[{{Kim} \& {Ostriker}(2007)}]{2007ApJ...660.1232K}
{Kim}, W.-T., \& {Ostriker}, E.~C. 2007, \apj, 660, 1232

\bibitem[{{Koekemoer} {et~al.}(2011){Koekemoer}, {Faber}, {Ferguson}, {Grogin},
  {Kocevski}, {Koo}, {Lai}, {Lotz}, {Lucas}, {McGrath}, {Ogaz}, {Rajan},
  {Riess}, {Rodney}, {Strolger}, {Casertano}, {Castellano}, {Dahlen},
  {Dickinson}, {Dolch}, {Fontana}, {Giavalisco}, {Grazian}, {Guo}, {Hathi},
  {Huang}, {van der Wel}, {Yan}, {Acquaviva}, {Alexander}, {Almaini}, {Ashby},
  {Barden}, {Bell}, {Bournaud}, {Brown}, {Caputi}, {Cassata}, {Challis},
  {Chary}, {Cheung}, {Cirasuolo}, {Conselice}, {Roshan Cooray}, {Croton},
  {Daddi}, {Dav{\'e}}, {de Mello}, {de Ravel}, {Dekel}, {Donley}, {Dunlop},
  {Dutton}, {Elbaz}, {Fazio}, {Filippenko}, {Finkelstein}, {Frazer}, {Gardner},
  {Garnavich}, {Gawiser}, {Gruetzbauch}, {Hartley}, {H{\"a}ussler},
  {Herrington}, {Hopkins}, {Huang}, {Jha}, {Johnson}, {Kartaltepe},
  {Khostovan}, {Kirshner}, {Lani}, {Lee}, {Li}, {Madau}, {McCarthy},
  {McIntosh}, {McLure}, {McPartland}, {Mobasher}, {Moreira}, {Mortlock},
  {Moustakas}, {Mozena}, {Nandra}, {Newman}, {Nielsen}, {Niemi}, {Noeske},
  {Papovich}, {Pentericci}, {Pope}, {Primack}, {Ravindranath}, {Reddy},
  {Renzini}, {Rix}, {Robaina}, {Rosario}, {Rosati}, {Salimbeni}, {Scarlata},
  {Siana}, {Simard}, {Smidt}, {Snyder}, {Somerville}, {Spinrad}, {Straughn},
  {Telford}, {Teplitz}, {Trump}, {Vargas}, {Villforth}, {Wagner}, {Wandro},
  {Wechsler}, {Weiner}, {Wiklind}, {Wild}, {Wilson}, {Wuyts}, \&
  {Yun}}]{2011ApJS..197...36K}
{Koekemoer}, A.~M., {Faber}, S.~M., {Ferguson}, H.~C., {et~al.} 2011, \apjs,
  197, 36

\bibitem[{{Krajnovi{\'c}} {et~al.}(2006){Krajnovi{\'c}}, {Cappellari}, {de
  Zeeuw}, \& {Copin}}]{2006MNRAS.366..787K}
{Krajnovi{\'c}}, D., {Cappellari}, M., {de Zeeuw}, P.~T., \& {Copin}, Y. 2006,
  \mnras, 366, 787

\bibitem[{{Lackner} {et~al.}(2014){Lackner}, {Silverman}, {Salvato},
  {Kampczyk}, {Kartaltepe}, {Sanders}, {Capak}, {Civano}, {Ilbert}, {Jahnke},
  {Koekemoer}, {Lee}, {Le Fevre}, {Liu}, {Scoville}, {Sheth}, \&
  {Toft}}]{2014arXiv1406.2327L}
{Lackner}, C.~N., {Silverman}, J.~D., {Salvato}, M., {et~al.} 2014, ArXiv
  e-prints, arXiv:1406.2327

\bibitem[{{Lang} {et~al.}(2014){Lang}, {Wuyts}, {Somerville}, {Forster
  Schreiber}, {Genzel}, {Bell}, {Brammer}, {Dekel}, {Faber}, {Ferguson},
  {Grogin}, {Kocevski}, {Koekemoer}, {Lutz}, {McGrath}, {Momcheva}, {Nelson},
  {Primack}, {Rosario}, {Skelton}, {Tacconi}, {van Dokkum}, \&
  {Whitaker}}]{2014arXiv1402.0866L}
{Lang}, P., {Wuyts}, S., {Somerville}, R., {et~al.} 2014, ArXiv e-prints,
  arXiv:1402.0866

\bibitem[{{Law} {et~al.}(2007){Law}, {Steidel}, {Erb}, {Larkin}, {Pettini},
  {Shapley}, \& {Wright}}]{2007ApJ...669..929L}
{Law}, D.~R., {Steidel}, C.~C., {Erb}, D.~K., {et~al.} 2007, \apj, 669, 929

\bibitem[{{Law} {et~al.}(2009){Law}, {Steidel}, {Erb}, {Larkin}, {Pettini},
  {Shapley}, \& {Wright}}]{2009ApJ...697.2057L}
---. 2009, \apj, 697, 2057

\bibitem[{{Law} {et~al.}(2012{\natexlab{a}}){Law}, {Steidel}, {Shapley},
  {Nagy}, {Reddy}, \& {Erb}}]{2012ApJ...759...29L}
{Law}, D.~R., {Steidel}, C.~C., {Shapley}, A.~E., {et~al.} 2012{\natexlab{a}},
  \apj, 759, 29

\bibitem[{{Law} {et~al.}(2012{\natexlab{b}}){Law}, {Steidel}, {Shapley},
  {Nagy}, {Reddy}, \& {Erb}}]{2012ApJ...745...85L}
---. 2012{\natexlab{b}}, \apj, 745, 85

\bibitem[{{Le F{\`e}vre} {et~al.}(2000){Le F{\`e}vre}, {Abraham}, {Lilly},
  {Ellis}, {Brinchmann}, {Schade}, {Tresse}, {Colless}, {Crampton},
  {Glazebrook}, {Hammer}, \& {Broadhurst}}]{2000MNRAS.311..565L}
{Le F{\`e}vre}, O., {Abraham}, R., {Lilly}, S.~J., {et~al.} 2000, \mnras, 311,
  565

\bibitem[{{Leroy} {et~al.}(2008){Leroy}, {Walter}, {Brinks}, {Bigiel}, {de
  Blok}, {Madore}, \& {Thornley}}]{2008AJ....136.2782L}
{Leroy}, A.~K., {Walter}, F., {Brinks}, E., {et~al.} 2008, \aj, 136, 2782

\bibitem[{{Leroy} {et~al.}(2009){Leroy}, {Walter}, {Bigiel}, {Usero}, {Weiss},
  {Brinks}, {de Blok}, {Kennicutt}, {Schuster}, {Kramer}, {Wiesemeyer}, \&
  {Roussel}}]{2009AJ....137.4670L}
{Leroy}, A.~K., {Walter}, F., {Bigiel}, F., {et~al.} 2009, \aj, 137, 4670

\bibitem[{{Lilly} {et~al.}(2013){Lilly}, {Carollo}, {Pipino}, {Renzini}, \&
  {Peng}}]{2013ApJ...772..119L}
{Lilly}, S.~J., {Carollo}, C.~M., {Pipino}, A., {Renzini}, A., \& {Peng}, Y.
  2013, \apj, 772, 119

\bibitem[{{Lotz} {et~al.}(2008){Lotz}, {Davis}, {Faber}, {Guhathakurta},
  {Gwyn}, {Huang}, {Koo}, {Le Floc'h}, {Lin}, {Newman}, {Noeske}, {Papovich},
  {Willmer}, {Coil}, {Conselice}, {Cooper}, {Hopkins}, {Metevier}, {Primack},
  {Rieke}, \& {Weiner}}]{2008ApJ...672..177L}
{Lotz}, J.~M., {Davis}, M., {Faber}, S.~M., {et~al.} 2008, \apj, 672, 177

\bibitem[{{Lutz} {et~al.}(2011){Lutz}, {Poglitsch}, {Altieri}, {Andreani},
  {Aussel}, {Berta}, {Bongiovanni}, {Brisbin}, {Cava}, {Cepa}, {Cimatti},
  {Daddi}, {Dominguez-Sanchez}, {Elbaz}, {F{\"o}rster Schreiber}, {Genzel},
  {Grazian}, {Gruppioni}, {Harwit}, {Le Floc'h}, {Magdis}, {Magnelli},
  {Maiolino}, {Nordon}, {P{\'e}rez Garc{\'{\i}}a}, {Popesso}, {Pozzi},
  {Riguccini}, {Rodighiero}, {Saintonge}, {Sanchez Portal}, {Santini}, {Shao},
  {Sturm}, {Tacconi}, {Valtchanov}, {Wetzstein}, \& {Wieprecht}}]{lutz:2011aa}
{Lutz}, D., {Poglitsch}, A., {Altieri}, B., {et~al.} 2011, \aap, 532, A90

\bibitem[{{Magnelli} {et~al.}(2013){Magnelli}, {Popesso}, {Berta}, {Pozzi},
  {Elbaz}, {Lutz}, {Dickinson}, {Altieri}, {Andreani}, {Aussel},
  {B{\'e}thermin}, {Bongiovanni}, {Cepa}, {Charmandaris}, {Chary}, {Cimatti},
  {Daddi}, {F{\"o}rster Schreiber}, {Genzel}, {Gruppioni}, {Harwit}, {Hwang},
  {Ivison}, {Magdis}, {Maiolino}, {Murphy}, {Nordon}, {Pannella}, {P{\'e}rez
  Garc{\'{\i}}a}, {Poglitsch}, {Rosario}, {Sanchez-Portal}, {Santini}, {Scott},
  {Sturm}, {Tacconi}, \& {Valtchanov}}]{2013A&A...553A.132M}
{Magnelli}, B., {Popesso}, P., {Berta}, S., {et~al.} 2013, \aap, 553, A132

\bibitem[{{Martig} {et~al.}(2009){Martig}, {Bournaud}, {Teyssier}, \&
  {Dekel}}]{2009ApJ...707..250M}
{Martig}, M., {Bournaud}, F., {Teyssier}, R., \& {Dekel}, A. 2009, \apj, 707,
  250

\bibitem[{{Mullaney} {et~al.}(2012){Mullaney}, {Daddi}, {B{\'e}thermin},
  {Elbaz}, {Juneau}, {Pannella}, {Sargent}, {Alexander}, \&
  {Hickox}}]{2012ApJ...753L..30M}
{Mullaney}, J.~R., {Daddi}, E., {B{\'e}thermin}, M., {et~al.} 2012, \apjl, 753,
  L30

\bibitem[{{Nelson} {et~al.}(2013){Nelson}, {van Dokkum}, {Momcheva}, {Brammer},
  {Lundgren}, {Skelton}, {Whitaker}, {Da Cunha}, {F{\"o}rster Schreiber},
  {Franx}, {Fumagalli}, {Kriek}, {Labbe}, {Leja}, {Patel}, {Rix}, {Schmidt},
  {van der Wel}, \& {Wuyts}}]{2013ApJ...763L..16N}
{Nelson}, E.~J., {van Dokkum}, P.~G., {Momcheva}, I., {et~al.} 2013, \apjl,
  763, L16

\bibitem[{{Nestor} {et~al.}(2011){Nestor}, {Johnson}, {Wild}, {M{\'e}nard},
  {Turnshek}, {Rao}, \& {Pettini}}]{2011MNRAS.412.1559N}
{Nestor}, D.~B., {Johnson}, B.~D., {Wild}, V., {et~al.} 2011, \mnras, 412, 1559

\bibitem[{{Nesvadba} {et~al.}(2008){Nesvadba}, {Lehnert}, {De Breuck},
  {Gilbert}, \& {van Breugel}}]{2008A&A...491..407N}
{Nesvadba}, N.~P.~H., {Lehnert}, M.~D., {De Breuck}, C., {Gilbert}, A.~M., \&
  {van Breugel}, W. 2008, \aap, 491, 407


\bibitem[{{Newman} {et~al.}(2012){Newman}, {Shapiro Griffin}, {Genzel},
  {Davies}, {F{\"o}rster-Schreiber}, {Tacconi}, {Kurk}, {Wuyts}, {Genel},
  {Lilly}, {Renzini}, {Bouch{\'e}}, {Burkert}, {Cresci}, {Buschkamp},
  {Carollo}, {Eisenhauer}, {Hicks}, {Lutz}, {Mancini}, {Naab}, {Peng}, \&
  {Vergani}}]{2012ApJ...752..111N}
{Newman}, S.~F., {Shapiro Griffin}, K., {Genzel}, R., {et~al.} 2012, \apj, 752,
  111

\bibitem[{{Newman} {et~al.}(2013){Newman}, {Genzel}, {F{\"o}rster Schreiber},
  {Shapiro Griffin}, {Mancini}, {Lilly}, {Renzini}, {Bouch{\'e}}, {Burkert},
  {Buschkamp}, {Carollo}, {Cresci}, {Davies}, {Eisenhauer}, {Genel}, {Hicks},
  {Kurk}, {Lutz}, {Naab}, {Peng}, {Sternberg}, {Tacconi}, {Wuyts}, {Zamorani},
  \& {Vergani}}]{2013ApJ...767..104N}
{Newman}, S.~F., {Genzel}, R., {F{\"o}rster Schreiber}, N.~M., {et~al.} 2013,
  \apj, 767, 104

\bibitem[{{Noeske} {et~al.}(2007){Noeske}, {Weiner}, {Faber}, {Papovich},
  {Koo}, {Somerville}, {Bundy}, {Conselice}, {Newman}, {Schiminovich}, {Le
  Floc'h}, {Coil}, {Rieke}, {Lotz}, {Primack}, {Barmby}, {Cooper}, {Davis},
  {Ellis}, {Fazio}, {Guhathakurta}, {Huang}, {Kassin}, {Martin}, {Phillips},
  {Rich}, {Small}, {Willmer}, \& {Wilson}}]{2007ApJ...660L..43N}
{Noeske}, K.~G., {Weiner}, B.~J., {Faber}, S.~M., {et~al.} 2007, \apjl, 660,
  L43

\bibitem[{{Peng} {et~al.}(2010){Peng}, {Lilly}, {Kova{\v c}}, {Bolzonella},
  {Pozzetti}, {Renzini}, {Zamorani}, {Ilbert}, {Knobel}, {Iovino}, {Maier},
  {Cucciati}, {Tasca}, {Carollo}, {Silverman}, {Kampczyk}, {de Ravel},
  {Sanders}, {Scoville}, {Contini}, {Mainieri}, {Scodeggio}, {Kneib}, {Le
  F{\`e}vre}, {Bardelli}, {Bongiorno}, {Caputi}, {Coppa}, {de la Torre},
  {Franzetti}, {Garilli}, {Lamareille}, {Le Borgne}, {Le Brun}, {Mignoli},
  {Perez Montero}, {Pello}, {Ricciardelli}, {Tanaka}, {Tresse}, {Vergani},
  {Welikala}, {Zucca}, {Oesch}, {Abbas}, {Barnes}, {Bordoloi}, {Bottini},
  {Cappi}, {Cassata}, {Cimatti}, {Fumana}, {Hasinger}, {Koekemoer},
  {Leauthaud}, {Maccagni}, {Marinoni}, {McCracken}, {Memeo}, {Meneux}, {Nair},
  {Porciani}, {Presotto}, \& {Scaramella}}]{2010ApJ...721..193P}
{Peng}, Y.-j., {Lilly}, S.~J., {Kova{\v c}}, K., {et~al.} 2010, \apj, 721, 193

\bibitem[{{Powell} {et~al.}(2013){Powell}, {Bournaud}, {Chapon}, \&
  {Teyssier}}]{2013MNRAS.434.1028P}
{Powell}, L.~C., {Bournaud}, F., {Chapon}, D., \& {Teyssier}, R. 2013, \mnras,
  434, 1028

\bibitem[{{Reddy} {et~al.}(2005){Reddy}, {Erb}, {Steidel}, {Shapley},
  {Adelberger}, \& {Pettini}}]{2005ApJ...633..748R}
{Reddy}, N.~A., {Erb}, D.~K., {Steidel}, C.~C., {et~al.} 2005, \apj, 633, 748

\bibitem[{{Reshetnikov} {et~al.}(2003){Reshetnikov}, {Dettmar}, \&
  {Combes}}]{2003A&A...399..879R}
{Reshetnikov}, V.~P., {Dettmar}, R.-J., \& {Combes}, F. 2003, \aap, 399, 879

\bibitem[{{Rodighiero} {et~al.}(2011){Rodighiero}, {Daddi}, {Baronchelli},
  {Cimatti}, {Renzini}, {Aussel}, {Popesso}, {Lutz}, {Andreani}, {Berta},
  {Cava}, {Elbaz}, {Feltre}, {Fontana}, {F{\"o}rster Schreiber},
  {Franceschini}, {Genzel}, {Grazian}, {Gruppioni}, {Ilbert}, {Le Floch},
  {Magdis}, {Magliocchetti}, {Magnelli}, {Maiolino}, {McCracken}, {Nordon},
  {Poglitsch}, {Santini}, {Pozzi}, {Riguccini}, {Tacconi}, {Wuyts}, \&
  {Zamorani}}]{2011ApJ...739L..40R}
{Rodighiero}, G., {Daddi}, E., {Baronchelli}, I., {et~al.} 2011, \apjl, 739,
  L40

\bibitem[{{Rosario} {et~al.}(2012){Rosario}, {Santini}, {Lutz}, {Shao},
  {Maiolino}, {Alexander}, {Altieri}, {Andreani}, {Aussel}, {Bauer}, {Berta},
  {Bongiovanni}, {Brandt}, {Brusa}, {Cepa}, {Cimatti}, {Cox}, {Daddi}, {Elbaz},
  {Fontana}, {F{\"o}rster Schreiber}, {Genzel}, {Grazian}, {Le Floch},
  {Magnelli}, {Mainieri}, {Netzer}, {Nordon}, {P{\'e}rez Garcia}, {Poglitsch},
  {Popesso}, {Pozzi}, {Riguccini}, {Rodighiero}, {Salvato}, {Sanchez-Portal},
  {Sturm}, {Tacconi}, {Valtchanov}, \& {Wuyts}}]{2012A&A...545A..45R}
{Rosario}, D.~J., {Santini}, P., {Lutz}, D., {et~al.} 2012, \aap, 545, A45

\bibitem[{{Saintonge} {et~al.}(2011){Saintonge}, {Kauffmann}, {Wang}, {Kramer},
  {Tacconi}, {Buchbender}, {Catinella}, {Graci{\'a}-Carpio}, {Cortese},
  {Fabello}, {Fu}, {Genzel}, {Giovanelli}, {Guo}, {Haynes}, {Heckman},
  {Krumholz}, {Lemonias}, {Li}, {Moran}, {Rodriguez-Fernandez}, {Schiminovich},
  {Schuster}, \& {Sievers}}]{2011MNRAS.415...61S}
{Saintonge}, A., {Kauffmann}, G., {Wang}, J., {et~al.} 2011, \mnras, 415, 61

\bibitem[{{Saintonge} {et~al.}(2012){Saintonge}, {Tacconi}, {Fabello}, {Wang},
  {Catinella}, {Genzel}, {Graci{\'a}-Carpio}, {Kramer}, {Moran}, {Heckman},
  {Schiminovich}, {Schuster}, \& {Wuyts}}]{2012ApJ...758...73S}
{Saintonge}, A., {Tacconi}, L.~J., {Fabello}, S., {et~al.} 2012, \apj, 758, 73

\bibitem[{{Saintonge} {et~al.}(2013){Saintonge}, {Lutz}, {Genzel}, {Magnelli},
  {Nordon}, {Tacconi}, {Baker}, {Bandara}, {Berta}, {F{\"o}rster Schreiber},
  {Poglitsch}, {Sturm}, {Wuyts}, \& {Wuyts}}]{2013ApJ...778....2S}
{Saintonge}, A., {Lutz}, D., {Genzel}, R., {et~al.} 2013, \apj, 778, 2

\bibitem[{{Schaye} {et~al.}(2014){Schaye}, {Crain}, {Bower}, {Furlong},
  {Schaller}, {Theuns}, {Dalla Vecchia}, {Frenk}, {McCarthy}, {Helly},
  {Jenkins}, {Rosas-Guevara}, {White}, {Baes}, {Booth}, {Camps}, {Navarro},
  {Qu}, {Rahmati}, {Sawala}, {Thomas}, \& {Trayford}}]{2014arXiv1407.7040S}
{Schaye}, J., {Crain}, R.~A., {Bower}, R.~G., {et~al.} 2014, ArXiv e-prints,
  arXiv:1407.7040

\bibitem[{{Schmidt} {et~al.}(2013){Schmidt}, {Rix}, {da Cunha}, {Brammer},
  {Cox}, {van Dokkum}, {F{\"o}rster Schreiber}, {Franx}, {Fumagalli},
  {Jonsson}, {Lundgren}, {Maseda}, {Momcheva}, {Nelson}, {Skelton}, {van der
  Wel}, \& {Whitaker}}]{2013MNRAS.432..285S}
{Schmidt}, K.~B., {Rix}, H.-W., {da Cunha}, E., {et~al.} 2013, \mnras, 432, 285

\bibitem[{{Shapiro} {et~al.}(2008){Shapiro}, {Genzel}, {F{\"o}rster Schreiber},
  {Tacconi}, {Bouch{\'e}}, {Cresci}, {Davies}, {Eisenhauer}, {Johansson},
  {Krajnovi{\'c}}, {Lutz}, {Naab}, {Arimoto}, {Arribas}, {Cimatti}, {Colina},
  {Daddi}, {Daigle}, {Erb}, {Hernandez}, {Kong}, {Mignoli}, {Onodera},
  {Renzini}, {Shapley}, \& {Steidel}}]{2008ApJ...682..231S}
{Shapiro}, K.~L., {Genzel}, R., {F{\"o}rster Schreiber}, N.~M., {et~al.} 2008,
  \apj, 682, 231

\bibitem[{{Shapiro} {et~al.}(2009){Shapiro}, {Genzel}, {Quataert}, {F{\"o}rster
  Schreiber}, {Davies}, {Tacconi}, {Armus}, {Bouch{\'e}}, {Buschkamp},
  {Cimatti}, {Cresci}, {Daddi}, {Eisenhauer}, {Erb}, {Genel}, {Hicks}, {Lilly},
  {Lutz}, {Renzini}, {Shapley}, {Steidel}, \& {Sternberg}}]{Shapiro:2009sj}
{Shapiro}, K.~L., {Genzel}, R., {Quataert}, E., {et~al.} 2009, \apj, 701, 955

\bibitem[{{Sharples} {et~al.}(2013){Sharples}, {Bender}, {Agudo Berbel},
  {Bezawada}, {Castillo}, {Cirasuolo}, {Davidson}, {Davies}, {Dubbeldam},
  {Fairley}, {Finger}, {F{\"o}rster Schreiber}, {Gonte}, {Hess}, {Jung},
  {Lewis}, {Lizon}, {Muschielok}, {Pasquini}, {Pirard}, {Popovic}, {Ramsay},
  {Rees}, {Richter}, {Riquelme}, {Rodrigues}, {Saviane}, {Schlichter},
  {Schmidtobreick}, {Segovia}, {Smette}, {Szeifert}, {van Kesteren}, {Wegner},
  \& {Wiezorrek}}]{2013Msngr.151...21S}
{Sharples}, R., {Bender}, R., {Agudo Berbel}, A., {et~al.} 2013, The Messenger,
  151, 21

\bibitem[{{Sharples} {et~al.}(2004){Sharples}, {Bender}, {Lehnert}, {Ramsay
  Howat}, {Bremer}, {Davies}, {Genzel}, {Hofmann}, {Ivison}, {Saglia}, \&
  {Thatte}}]{2004SPIE.5492.1179S}
{Sharples}, R.~M., {Bender}, R., {Lehnert}, M.~D., {et~al.} 2004, in Society of
  Photo-Optical Instrumentation Engineers (SPIE) Conference Series, Vol. 5492,
  Society of Photo-Optical Instrumentation Engineers (SPIE) Conference Series,
  ed. {A.~F.~M.~Moorwood \& M.~Iye}, 1179--1186

\bibitem[{{Skelton} {et~al.}(2014){Skelton}, {Whitaker}, {Momcheva}, {Brammer},
  {van Dokkum}, {Labbe}, {Franx}, {van der Wel}, {Bezanson}, {Da Cunha},
  {Fumagalli}, {Foerster Schreiber}, {Kriek}, {Leja}, {Lundgren}, {Magee},
  {Marchesini}, {Maseda}, {Nelson}, {Oesch}, {Pacifici}, {Patel}, {Price},
  {Rix}, {Tal}, {Wake}, \& {Wuyts}}]{2014arXiv1403.3689S}
{Skelton}, R.~E., {Whitaker}, K.~E., {Momcheva}, I.~G., {et~al.} 2014, ArXiv
  e-prints, arXiv:1403.3689

\bibitem[{{Sobral} {et~al.}(2013){Sobral}, {Swinbank}, {Stott}, {Matthee},
  {Bower}, {Smail}, {Best}, {Geach}, \& {Sharples}}]{2013ApJ...779..139S}
{Sobral}, D., {Swinbank}, A.~M., {Stott}, J.~P., {et~al.} 2013, \apj, 779, 139

\bibitem[{{Sparre} {et~al.}(2014){Sparre}, {Hayward}, {Springel},
  {Vogelsberger}, {Genel}, {Torrey}, {Nelson}, {Sijacki}, \&
  {Hernquist}}]{2014arXiv1409.0009S}
{Sparre}, M., {Hayward}, C.~C., {Springel}, V., {et~al.} 2014, ArXiv e-prints,
  arXiv:1409.0009

\bibitem[{{Stark} {et~al.}(2013){Stark}, {Schenker}, {Ellis}, {Robertson},
  {McLure}, \& {Dunlop}}]{2013ApJ...763..129S}
{Stark}, D.~P., {Schenker}, M.~A., {Ellis}, R., {et~al.} 2013, \apj, 763, 129

\bibitem[{{Steidel} {et~al.}(2010){Steidel}, {Erb}, {Shapley}, {Pettini},
  {Reddy}, {Bogosavljevi{\'c}}, {Rudie}, \& {Rakic}}]{2010ApJ...717..289S}
{Steidel}, C.~C., {Erb}, D.~K., {Shapley}, A.~E., {et~al.} 2010, \apj, 717, 289

\bibitem[{{Stott} {et~al.}(2013){Stott}, {Sobral}, {Smail}, {Bower}, {Best}, \&
  {Geach}}]{2013MNRAS.430.1158S}
{Stott}, J.~P., {Sobral}, D., {Smail}, I., {et~al.} 2013, \mnras, 430, 1158

\bibitem[{{Stott} {et~al.}(2014){Stott}, {Sobral}, {Swinbank}, {Smail},
  {Bower}, {Best}, {Sharples}, {Geach}, \& {Matthee}}]{2014MNRAS.443.2695S}
{Stott}, J.~P., {Sobral}, D., {Swinbank}, A.~M., {et~al.} 2014, \mnras, 443,
  2695

\bibitem[{{Swinbank} {et~al.}(2012){Swinbank}, {Smail}, {Sobral}, {Theuns},
  {Best}, \& {Geach}}]{2012ApJ...760..130S}
{Swinbank}, A.~M., {Smail}, I., {Sobral}, D., {et~al.} 2012, \apj, 760, 130

\bibitem[{{Tacchella} {et al.}(2014){Tacchella}, {Lang}, {Carollo}}]
{Tacchella}, S., {Lang}, P., {Carollo}, C. M., {et~al.} 2014, \apj, {submitted}

\bibitem[{{Tacconi} {et~al.}(2010){Tacconi}, {Genzel}, {Neri}, {Cox}, {Cooper},
  {Shapiro}, {Bolatto}, {Bouch{\'e}}, {Bournaud}, {Burkert}, {Combes},
  {Comerford}, {Davis}, {Schreiber}, {Garcia-Burillo}, {Gracia-Carpio}, {Lutz},
  {Naab}, {Omont}, {Shapley}, {Sternberg}, \& {Weiner}}]{2010Natur.463..781T}
{Tacconi}, L.~J., {Genzel}, R., {Neri}, R., {et~al.} 2010, \nat, 463, 781

\bibitem[{{Tacconi} {et~al.}(2013){Tacconi}, {Neri}, {Genzel}, {Combes},
  {Bolatto}, {Cooper}, {Wuyts}, {Bournaud}, {Burkert}, {Comerford}, {Cox},
  {Davis}, {F{\"o}rster Schreiber}, {Garc{\'{\i}}a-Burillo}, {Gracia-Carpio},
  {Lutz}, {Naab}, {Newman}, {Omont}, {Saintonge}, {Shapiro Griffin}, {Shapley},
  {Sternberg}, \& {Weiner}}]{2013ApJ...768...74T}
{Tacconi}, L.~J., {Neri}, R., {Genzel}, R., {et~al.} 2013, \apj, 768, 74

\bibitem[{{Toomre}(1964)}]{1964ApJ...139.1217T}
{Toomre}, A. 1964, \apj, 139, 1217

\bibitem[{{Ueda} {et~al.}(2008){Ueda}, {Watson}, {Stewart}, {Akiyama},
  {Schwope}, {Lamer}, {Ebrero}, {Carrera}, {Sekiguchi}, {Yamada}, {Simpson},
  {Hasinger}, \& {Mateos}}]{2008ApJS..179..124U}
{Ueda}, Y., {Watson}, M.~G., {Stewart}, I.~M., {et~al.} 2008, \apjs, 179, 124

\bibitem[{{van der Kruit} \& {Allen}(1978)}]{1978ARA&A..16..103V}
{van der Kruit}, P.~C., \& {Allen}, R.~J. 1978, \araa, 16, 103

\bibitem[{{van der Wel} {et~al.}(2012){van der Wel}, {Bell}, {H{\"a}ussler},
  {McGrath}, {Chang}, {Guo}, {McIntosh}, {Rix}, {Barden}, {Cheung}, {Faber},
  {Ferguson}, {Galametz}, {Grogin}, {Hartley}, {Kartaltepe}, {Kocevski},
  {Koekemoer}, {Lotz}, {Mozena}, {Peth}, \& {Peng}}]{2012ApJS..203...24V}
{van der Wel}, A., {Bell}, E.~F., {H{\"a}ussler}, B., {et~al.} 2012, \apjs,
  203, 24

\bibitem[{{van der Wel} {et~al.}(2014{\natexlab{a}}){van der Wel}, {Franx},
  {van Dokkum}, {Skelton}, {Momcheva}, {Whitaker}, {Brammer}, {Bell}, {Rix},
  {Wuyts}, {Ferguson}, {Holden}, {Barro}, {Koekemoer}, {Chang}, {McGrath},
  {H{\"a}ussler}, {Dekel}, {Behroozi}, {Fumagalli}, {Leja}, {Lundgren},
  {Maseda}, {Nelson}, {Wake}, {Patel}, {Labb{\'e}}, {Faber}, {Grogin}, \&
  {Kocevski}}]{2014ApJ...788...28V}
{van der Wel}, A., {Franx}, M., {van Dokkum}, P.~G., {et~al.}
  2014{\natexlab{a}}, \apj, 788, 28

\bibitem[{{van der Wel} {et~al.}(2014{\natexlab{b}}){van der Wel}, {Chang},
  {Bell}, {Holden}, {Ferguson}, {Giavalisco}, {Rix}, {Skelton}, {Whitaker},
  {Momcheva}, {Brammer}, {Kassin}, {Martig}, {Dekel}, {Ceverino}, {Koo},
  {Mozena}, {van Dokkum}, {Franx}, {Faber}, \& {Primack}}]{2014arXiv1407.4233V}
{van der Wel}, A., {Chang}, Y.-Y., {Bell}, E.~F., {et~al.} 2014{\natexlab{b}},
  ArXiv e-prints, arXiv:1407.4233

\bibitem[{{Vergani} {et~al.}(2012){Vergani}, {Epinat}, {Contini}, {Tasca},
  {Tresse}, {Amram}, {Garilli}, {Kissler-Patig}, {Le F{\`e}vre}, {Moultaka},
  {Paioro}, {Queyrel}, \& {L{\'o}pez-Sanjuan}}]{2012A&A...546A.118V}
{Vergani}, D., {Epinat}, B., {Contini}, T., {et~al.} 2012, \aap, 546, A118

\bibitem[{{Vogelsberger} {et~al.}(2013){Vogelsberger}, {Genel}, {Sijacki},
  {Torrey}, {Springel}, \& {Hernquist}}]{2013MNRAS.436.3031V}
{Vogelsberger}, M., {Genel}, S., {Sijacki}, D., {et~al.} 2013, \mnras, 436,
  3031

\bibitem[{{Wegner} \& {Muschielok}(2008)}]{2008SPIE.7019E..27W}
{Wegner}, M., \& {Muschielok}, B. 2008, in Society of Photo-Optical
  Instrumentation Engineers (SPIE) Conference Series, Vol. 7019, Society of
  Photo-Optical Instrumentation Engineers (SPIE) Conference Series

\bibitem[{{Weiner} {et~al.}(2006){Weiner}, {Willmer}, {Faber}, {Melbourne},
  {Kassin}, {Phillips}, {Harker}, {Metevier}, {Vogt}, \&
  {Koo}}]{2006ApJ...653.1027W}
{Weiner}, B.~J., {Willmer}, C.~N.~A., {Faber}, S.~M., {et~al.} 2006, \apj, 653,
  1027

\bibitem[{{Weiner} {et~al.}(2009){Weiner}, {Coil}, {Prochaska}, {Newman},
  {Cooper}, {Bundy}, {Conselice}, {Dutton}, {Faber}, {Koo}, {Lotz}, {Rieke}, \&
  {Rubin}}]{2009ApJ...692..187W}
{Weiner}, B.~J., {Coil}, A.~L., {Prochaska}, J.~X., {et~al.} 2009, \apj, 692,
  187

\bibitem[{{Whitaker} {et~al.}(2012{\natexlab{a}}){Whitaker}, {Kriek}, {van
  Dokkum}, {Bezanson}, {Brammer}, {Franx}, \&
  {Labb{\'e}}}]{2012ApJ...745..179W}
{Whitaker}, K.~E., {Kriek}, M., {van Dokkum}, P.~G., {et~al.}
  2012{\natexlab{a}}, \apj, 745, 179

\bibitem[{{Whitaker} {et~al.}(2012{\natexlab{b}}){Whitaker}, {van Dokkum},
  {Brammer}, \& {Franx}}]{2012ApJ...754L..29W}
{Whitaker}, K.~E., {van Dokkum}, P.~G., {Brammer}, G., \& {Franx}, M.
  2012{\natexlab{b}}, \apjl, 754, L29

\bibitem[{{Whitaker} {et~al.}(2013){Whitaker}, {van Dokkum}, {Brammer},
  {Momcheva}, {Skelton}, {Franx}, {Kriek}, {Labb{\'e}}, {Fumagalli},
  {Lundgren}, {Nelson}, {Patel}, \& {Rix}}]{2013ApJ...770L..39W}
{Whitaker}, K.~E., {van Dokkum}, P.~G., {Brammer}, G., {et~al.} 2013, \apjl,
  770, L39

\bibitem[{{Whitaker} {et~al.}(2014){Whitaker}, {Franx}, {Leja}, {van Dokkum},
  {Henry}, {Skelton}, {Fumagalli}, {Momcheva}, {Brammer}, {Labbe}, {Nelson}, \&
  {Rigby}}]{2014arXiv1407.1843W}
{Whitaker}, K.~E., {Franx}, M., {Leja}, J., {et~al.} 2014, ArXiv e-prints,
  arXiv:1407.1843

\bibitem[{{Wisnioski} {et~al.}(2011){Wisnioski}, {Glazebrook}, {Blake},
  {Wyder}, {Martin}, {Poole}, {Sharp}, {Couch}, {Kacprzak}, {Brough},
  {Colless}, {Contreras}, {Croom}, {Croton}, {Davis}, {Drinkwater}, {Forster},
  {Gilbank}, {Gladders}, {Jelliffe}, {Jurek}, {Li}, {Madore}, {Pimbblet},
  {Pracy}, {Woods}, \& {Yee}}]{2011MNRAS.417.2601W}
{Wisnioski}, E., {Glazebrook}, K., {Blake}, C., {et~al.} 2011, \mnras, 417,
  2601

\bibitem[{{Wright} {et~al.}(2009){Wright}, {Larkin}, {Law}, {Steidel},
  {Shapley}, \& {Erb}}]{2009ApJ...699..421W}
{Wright}, S.~A., {Larkin}, J.~E., {Law}, D.~R., {et~al.} 2009, \apj, 699, 421

\bibitem[{{Wuyts} {et~al.}(2014){Wuyts}, {Kurk}, {F{\"o}rster Schreiber},
  {Genzel}, {Wisnioski}, {Bandara}, {Wuyts}, {Beifiori}, {Bender}, {Brammer},
  {Burkert}, {Buschkamp}, {Carollo}, {Chan}, {Davies}, {Eisenhauer}, {Fossati},
  {Kulkarni}, {Lang}, {Lilly}, {Lutz}, {Mancini}, {Mendel}, {Momcheva}, {Naab},
  {Nelson}, {Renzini}, {Rosario}, {Saglia}, {Seitz}, {Sharples}, {Sternberg},
  {Tacchella}, {Tacconi}, {van Dokkum}, \& {Wilman}}]{2014ApJ...789L..40W}
{Wuyts}, E., {Kurk}, J., {F{\"o}rster Schreiber}, N.~M., {et~al.} 2014, \apjl,
  789, L40

\bibitem[{{Wuyts} {et~al.}(2011{\natexlab{a}}){Wuyts}, {F{\"o}rster Schreiber},
  {van der Wel}, {Magnelli}, {Guo}, {Genzel}, {Lutz}, {Aussel}, {Barro},
  {Berta}, {Cava}, {Graci{\'a}-Carpio}, {Hathi}, {Huang}, {Kocevski},
  {Koekemoer}, {Lee}, {Le Floc'h}, {McGrath}, {Nordon}, {Popesso}, {Pozzi},
  {Riguccini}, {Rodighiero}, {Saintonge}, \& {Tacconi}}]{2011ApJ...742...96W}
{Wuyts}, S., {F{\"o}rster Schreiber}, N.~M., {van der Wel}, A., {et~al.}
  2011{\natexlab{a}}, \apj, 742, 96

\bibitem[{{Wuyts} {et~al.}(2011{\natexlab{b}}){Wuyts}, {F{\"o}rster Schreiber},
  {Lutz}, {Nordon}, {Berta}, {Altieri}, {Andreani}, {Aussel}, {Bongiovanni},
  {Cepa}, {Cimatti}, {Daddi}, {Elbaz}, {Genzel}, {Koekemoer}, {Magnelli},
  {Maiolino}, {McGrath}, {P{\'e}rez Garc{\'{\i}}a}, {Poglitsch}, {Popesso},
  {Pozzi}, {Sanchez-Portal}, {Sturm}, {Tacconi}, \&
  {Valtchanov}}]{2011ApJ...738..106W}
{Wuyts}, S., {F{\"o}rster Schreiber}, N.~M., {Lutz}, D., {et~al.}
  2011{\natexlab{b}}, \apj, 738, 106

\bibitem[{{Wuyts} {et~al.}(2012){Wuyts}, {F{\"o}rster Schreiber}, {Genzel},
  {Guo}, {Barro}, {Bell}, {Dekel}, {Faber}, {Ferguson}, {Giavalisco}, {Grogin},
  {Hathi}, {Huang}, {Kocevski}, {Koekemoer}, {Koo}, {Lotz}, {Lutz}, {McGrath},
  {Newman}, {Rosario}, {Saintonge}, {Tacconi}, {Weiner}, \& {van der
  Wel}}]{2012ApJ...753..114W}
{Wuyts}, S., {F{\"o}rster Schreiber}, N.~M., {Genzel}, R., {et~al.} 2012, \apj,
  753, 114

\bibitem[{{Wuyts} {et~al.}(2013){Wuyts}, {F{\"o}rster Schreiber}, {Nelson},
  {van Dokkum}, {Brammer}, {Chang}, {Faber}, {Ferguson}, {Franx}, {Fumagalli},
  {Genzel}, {Grogin}, {Kocevski}, {Koekemoer}, {Lundgren}, {Lutz}, {McGrath},
  {Momcheva}, {Rosario}, {Skelton}, {Tacconi}, {van der Wel}, \&
  {Whitaker}}]{2013ApJ...779..135W}
{Wuyts}, S., {F{\"o}rster Schreiber}, N.~M., {Nelson}, E.~J., {et~al.} 2013,
  \apj, 779, 135

\bibitem[{{Xue} {et~al.}(2011){Xue}, {Luo}, {Brandt}, {Bauer}, {Lehmer},
  {Broos}, {Schneider}, {Alexander}, {Brusa}, {Comastri}, {Fabian}, {Gilli},
  {Hasinger}, {Hornschemeier}, {Koekemoer}, {Liu}, {Mainieri}, {Paolillo},
  {Rafferty}, {Rosati}, {Shemmer}, {Silverman}, {Smail}, {Tozzi}, \&
  {Vignali}}]{2011ApJS..195...10X}
{Xue}, Y.~Q., {Luo}, B., {Brandt}, W.~N., {et~al.} 2011, \apjs, 195, 10

\end{thebibliography}

\clearpage
\clearpage
\vspace*{15cm}


\begin{appendix}
\label{app.a2}
\section{ High S/N disk sample in first year data}
We present the observed-frame $IJH$ images, \halpha emission maps, velocity and velocity dispersion fields of the high-S/N disk sample at $z\sim1$ and $z\sim2$ from the first year \kmostd data. Velocity and velocity dispersion axis profiles extracted along the kinematic axis in apertures equivalent to the PSF are shown for each galaxy. We fit a Freeman exponential disk model to the velocity axis profile, given by
\begin{equation}
v_\mathrm{obs}(r)  = \frac{r}{r_0}\sqrt{\pi \mathrm{G} \Sigma_0 r_0 (I_0K_0 - I_1K_1)},
\label{eq.Vexpdisk}
\end{equation}
where $\Sigma_0$ is the central disk surface density, $r_0$ is the exponential radius and $I_i$ and $K_i$ are $i$-th order modified Bessel functions \citep{1970ApJ...160..811F}. The maximal velocity of this model is reached at $\sim0.88\sqrt{\pi \mathrm{G} \Sigma_0 r_0}$ at a turnover radius of $\sim2.15r_0$. Additionally, we fit the approximation for the velocity dispersion in a thick disk ($\sigma_{02}$; \citealt{2008ApJ...687...59G,2009ApJ...697..115C}) to the dispersion profile with a constant level of intrinsic dispersion, $\sigma_{01}$, added in quadrature as described by; 
\begin{equation}
\sigma_{02}(r) = \frac{v_\mathrm{rot}(r)h_z}{r},
\label{eq.sig02}
\end{equation} 
\begin{equation}
\sigma_\mathrm{mod}(r) = \sqrt{\sigma_{01}^2 + \sigma_{02}(r)^2}.
\label{eq.sigmod}
\end{equation} 
The models are convolved with the average PSF of the pointing prior to fitting.  In practice for this model, the constant scale height, $h_z$, determines the normalization of the $v_\mathrm{rot}/r$ curve in the $\sigma_{02}$ term while $\sigma_{01}$ determines the value that the model flattens to at radii beyond the turnover radius. In the cases where $\sigma_{01}>0$ the measured velocity dispersion $\sigma_0$ is $\sigma_0\approx\sigma_{01}$. When $\sigma_{01}$ is equal to zero, $\sigma_\mathrm{mod}$ approaches zero at infinity. 

\vspace{0.5cm}
\begin{figure*}[h]
\includegraphics[scale=0.8,  trim=1.5cm 0cm 1.0cm 0cm, clip, angle=90 ]{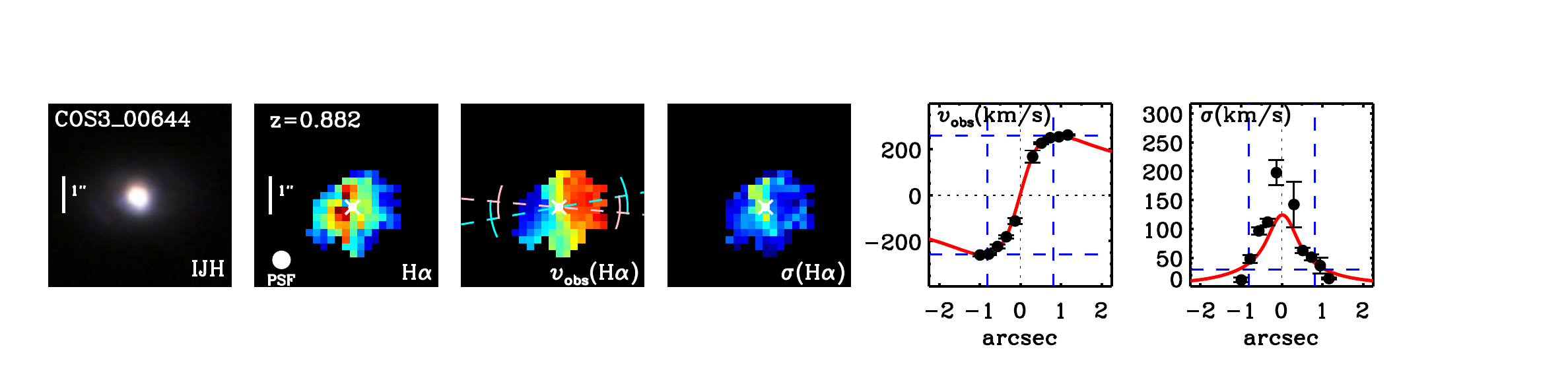}
\includegraphics[scale=0.8,  trim=1.5cm 0cm 1.0cm 0cm, clip, angle=90 ]{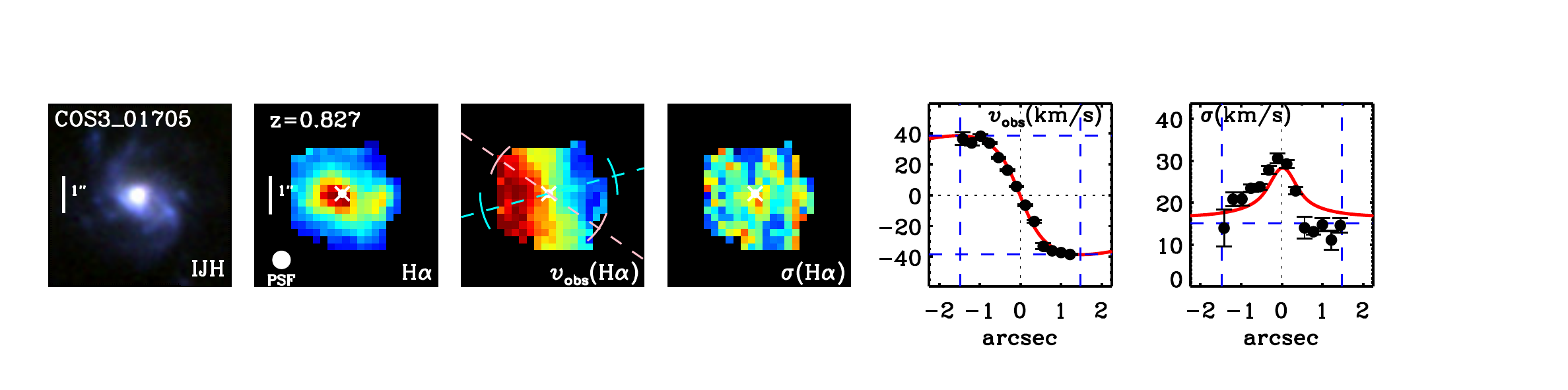}
\includegraphics[scale=0.8,  trim=0.0cm 0cm 1.0cm 0cm, clip, angle=90 ]{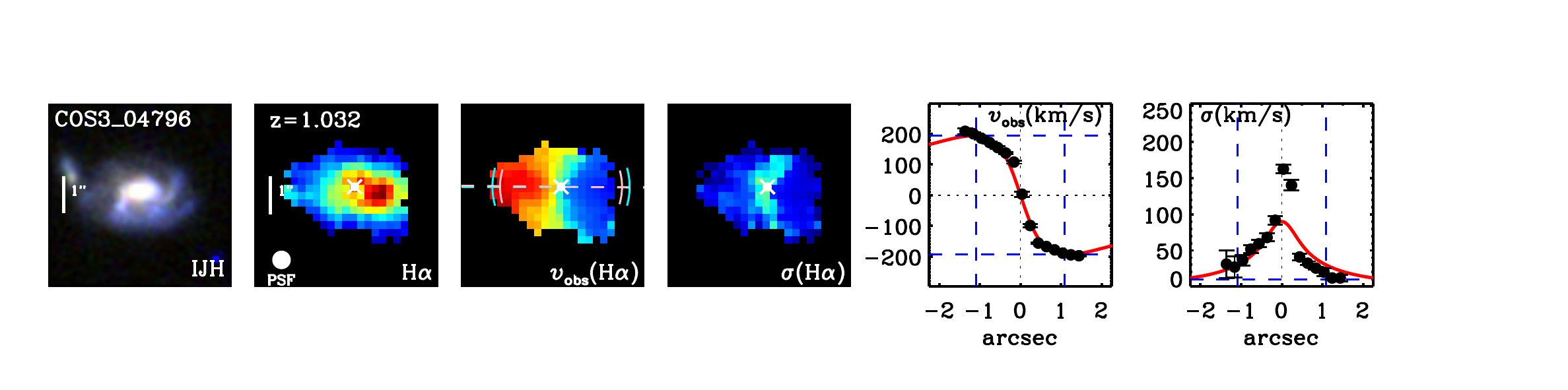}
\caption{HST images, kinematic maps and axis profiles for the high S/N disk galaxies in \kmostd first year data. From left to right for an individual galaxy: Observed-frame $IJH$ color composite image from CANDELS HST imaging, \halpha emission map from KMOS, corresponding velocity field, and velocity dispersion field, followed by the velocity and velocity dispersion axis profiles.  North is up and East is left for all galaxy images and maps. The axis profiles are extracted along the kinematic PA as denoted by the light blue line over plotted on the velocity map. The photometric PA, as determined by F160W HST images, is shown by the pink line. The blue arcs correspond to $\pm18^\circ$, the average misalignment between photometric and kinematic PAs, while the pink arcs correspond to $\pm3\sigma$ error on the photometric PA. The white circle in the \halpha image represents the FWHM of the PSF. Exponential disk models fit to the axis profiles, black data points, are shown by the red curves. The velocity fields are scaled by minimum and maximum $v(r)$ from the corresponding axis profile and velocity dispersion fields are scaled by the minimum pixel dispersion and the maximum dispersion from the dispersion axis profile.}
\label{afig.disks}
\end{figure*}

\begin{figure*}
\includegraphics[scale=0.8,  trim=1.5cm 0cm 1.0cm 0cm, clip, angle=90 ]{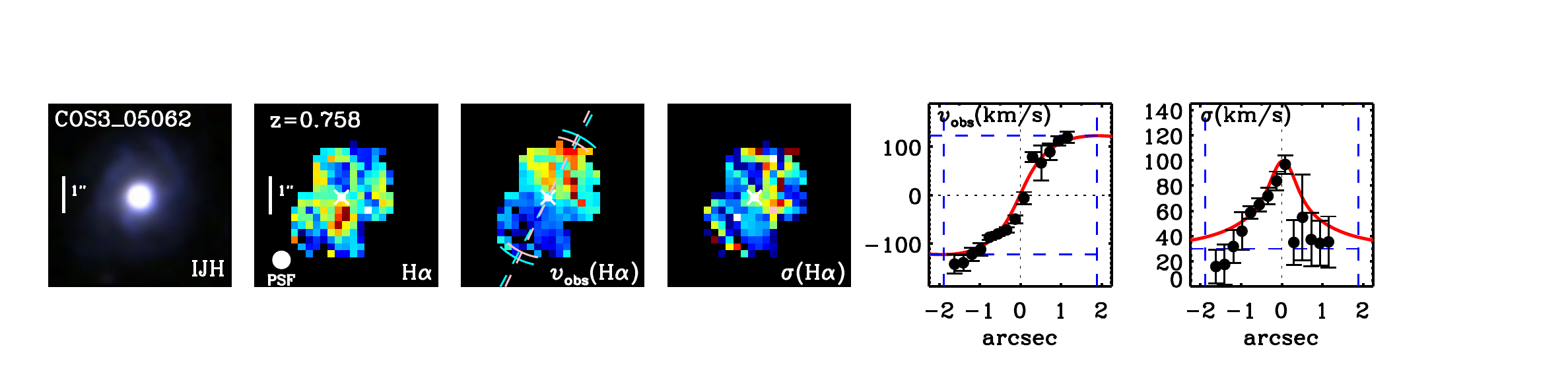}
\includegraphics[scale=0.8,  trim=1.5cm 0cm 1.0cm 0cm, clip, angle=90 ]{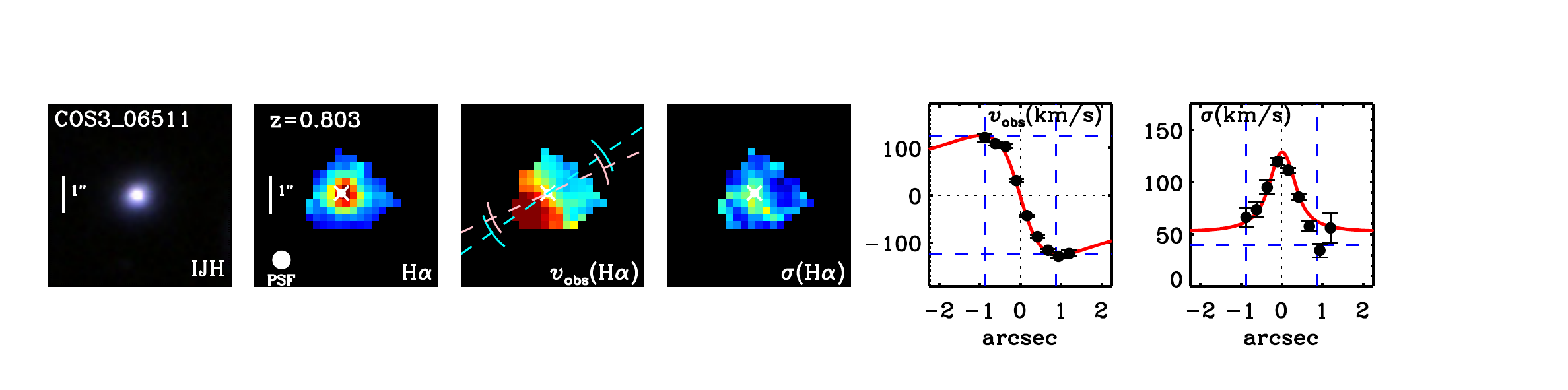}
\includegraphics[scale=0.8,  trim=1.5cm 0cm 1.0cm 0cm, clip, angle=90 ]{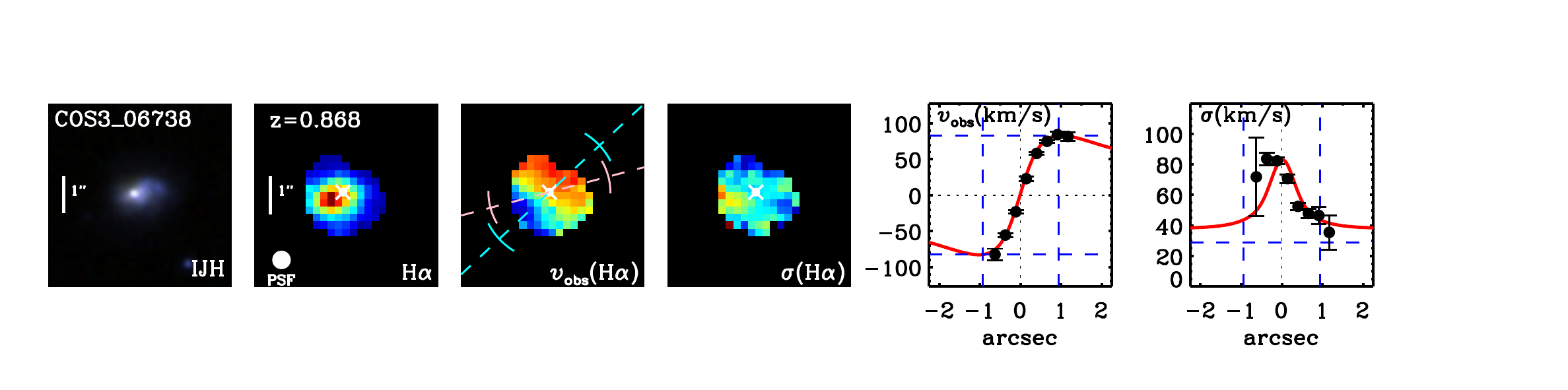}
\includegraphics[scale=0.8,  trim=1.5cm 0cm 1.0cm 0cm, clip, angle=90 ]{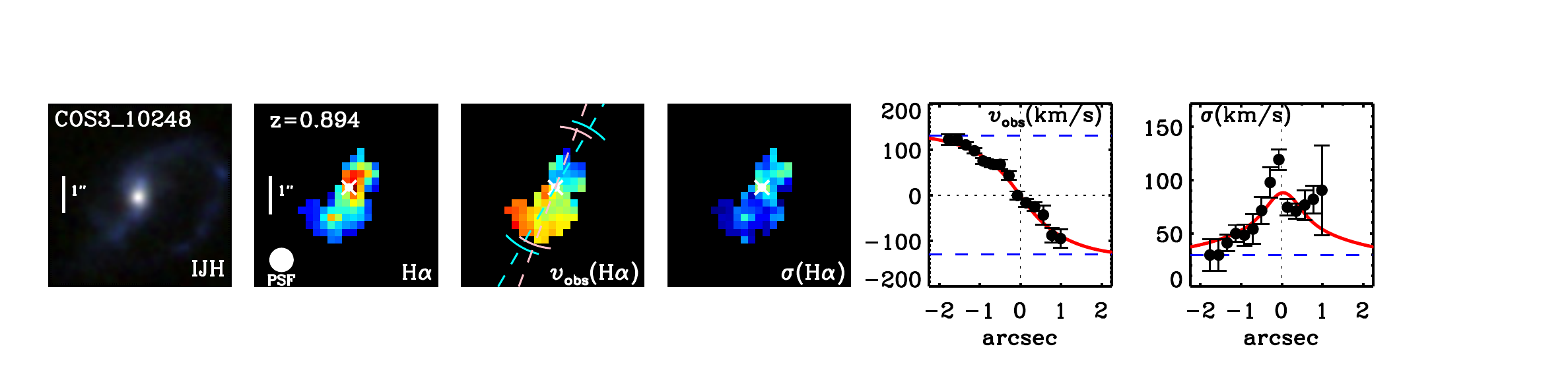}
\includegraphics[scale=0.8,  trim=1.5cm 0cm 1.0cm 0cm, clip, angle=90 ]{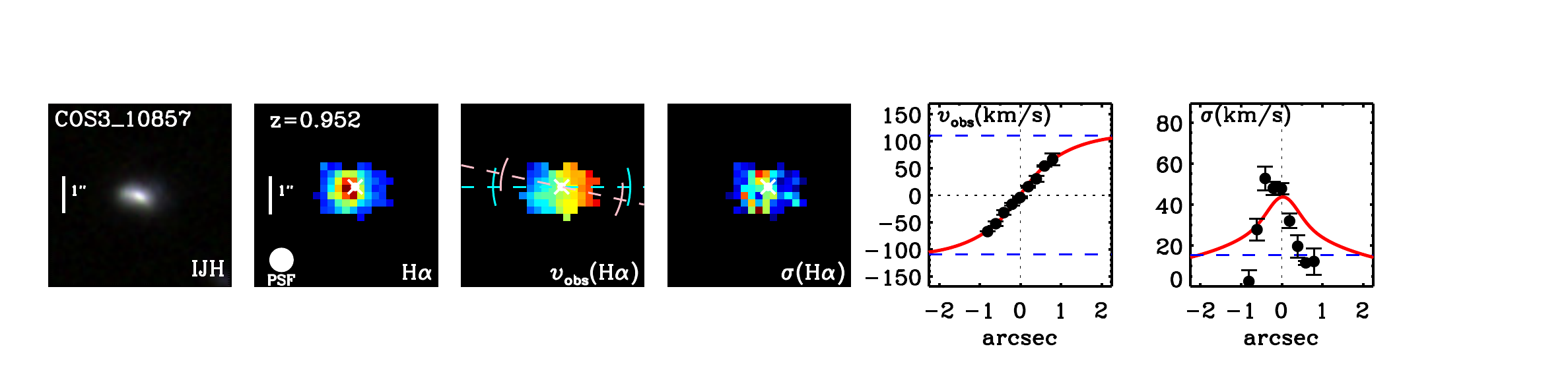}
\includegraphics[scale=0.8,  trim=1.5cm 0cm 1.0cm 0cm, clip, angle=90 ]{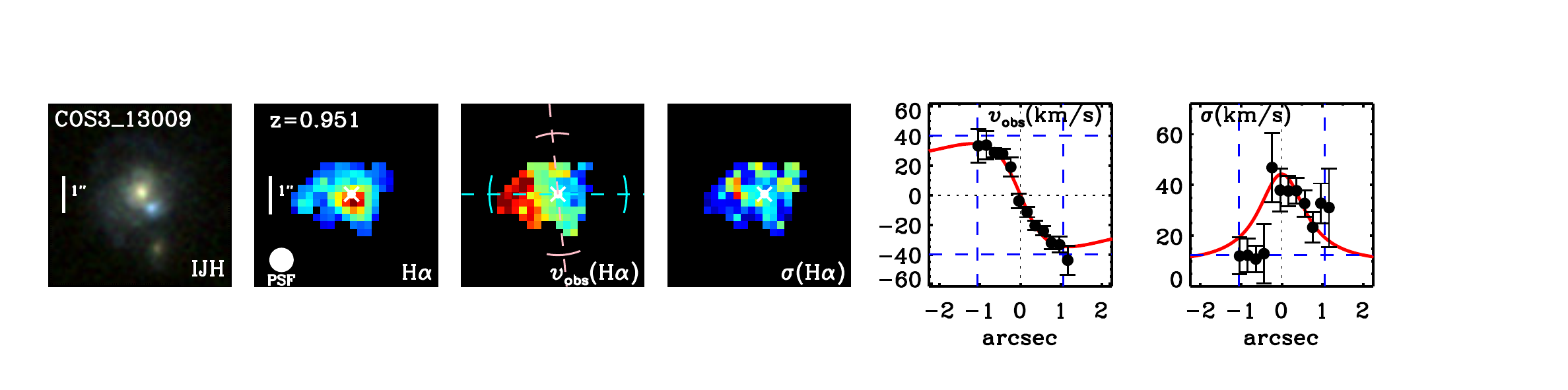}
\includegraphics[scale=0.8,  trim=1.5cm 0cm 1.0cm 0cm, clip, angle=90 ]{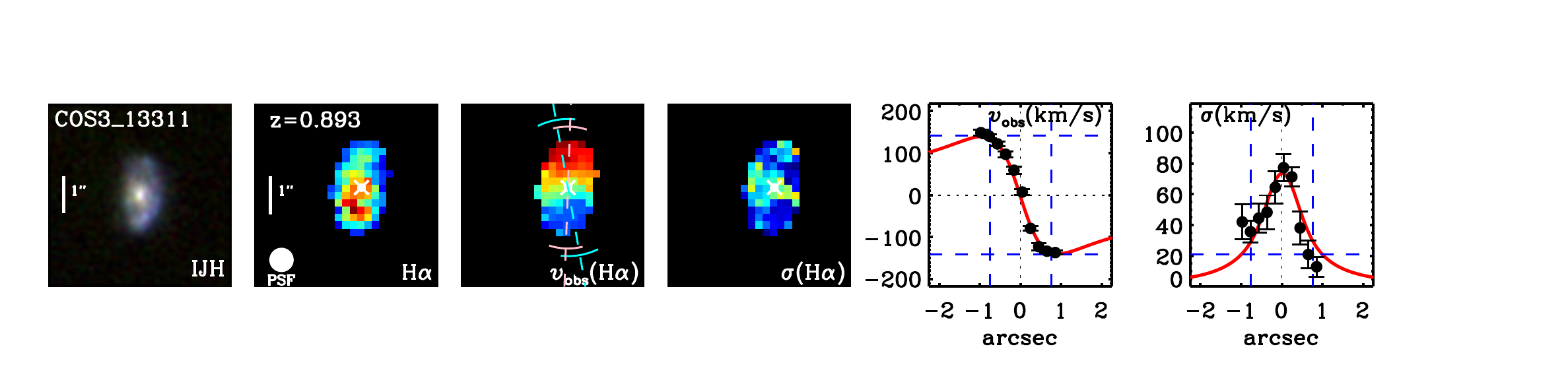}
\includegraphics[scale=0.8,  trim=0.5cm 0cm 1.0cm 0cm, clip, angle=90 ]{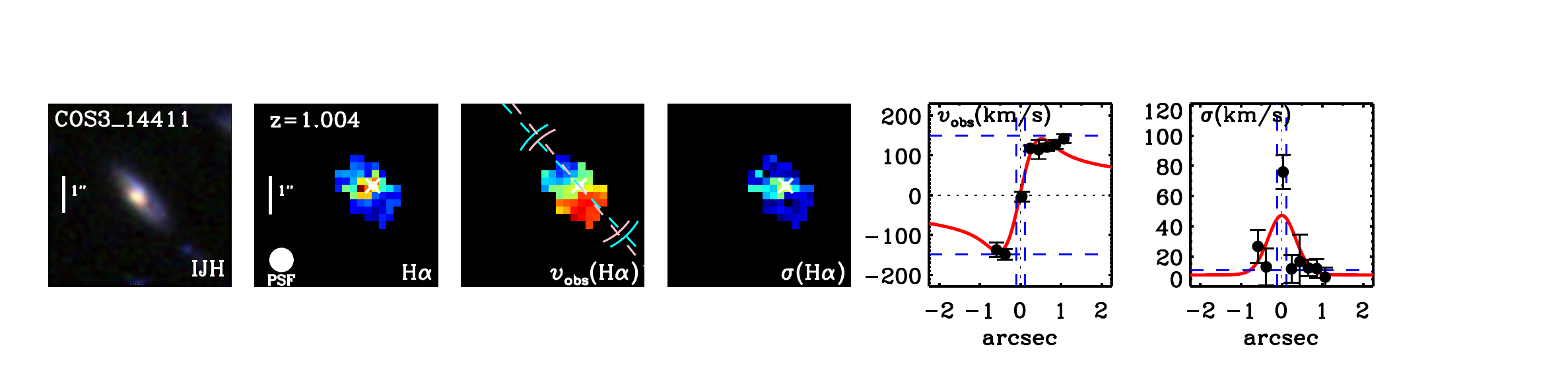}
\caption{\textit{cont.}; Kinematic maps and axis profiles for the high S/N disk galaxies in \kmostd first year data}
\end{figure*}
\addtocounter{figure}{-1}

\addtocounter{figure}{-1}
\begin{figure*}
\includegraphics[scale=0.8,  trim=1.5cm 0cm 1.0cm 0cm, clip, angle=90 ]{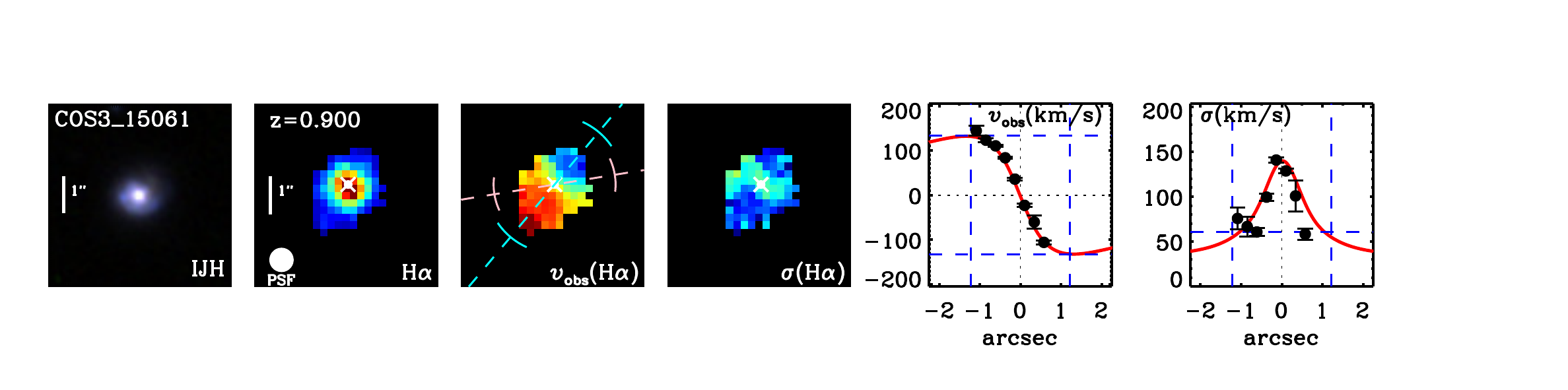}
\includegraphics[scale=0.8,  trim=1.5cm 0cm 1.0cm 0cm, clip, angle=90 ]{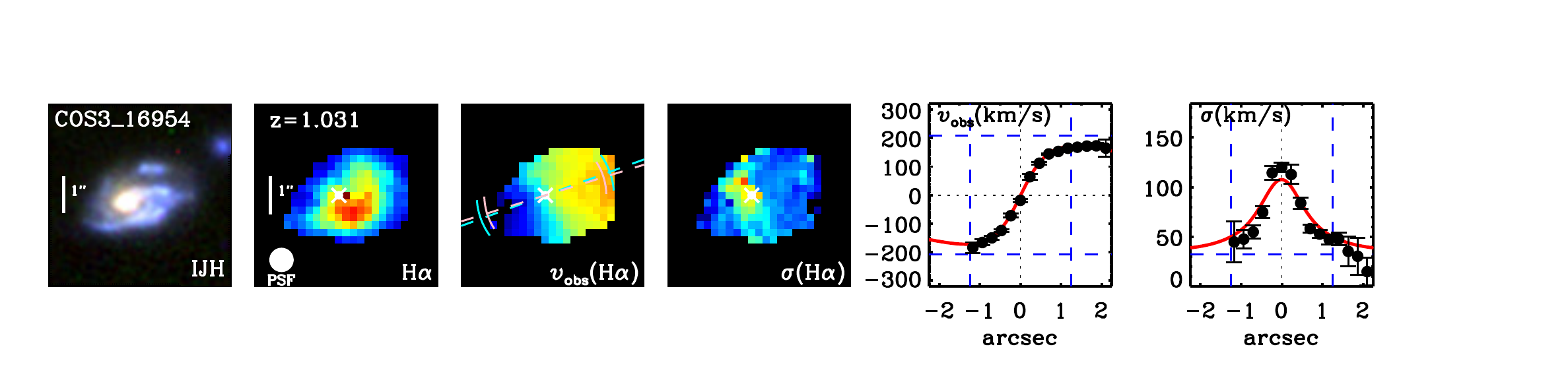}
\includegraphics[scale=0.8,  trim=1.5cm 0cm 1.0cm 0cm, clip, angle=90 ]{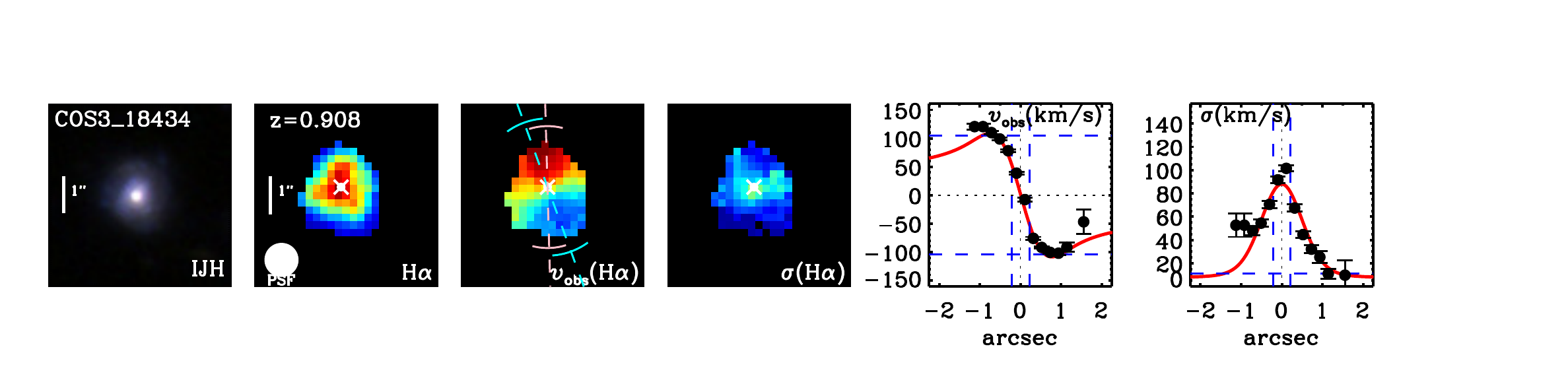}
\includegraphics[scale=0.8,  trim=1.5cm 0cm 1.0cm 0cm, clip, angle=90 ]{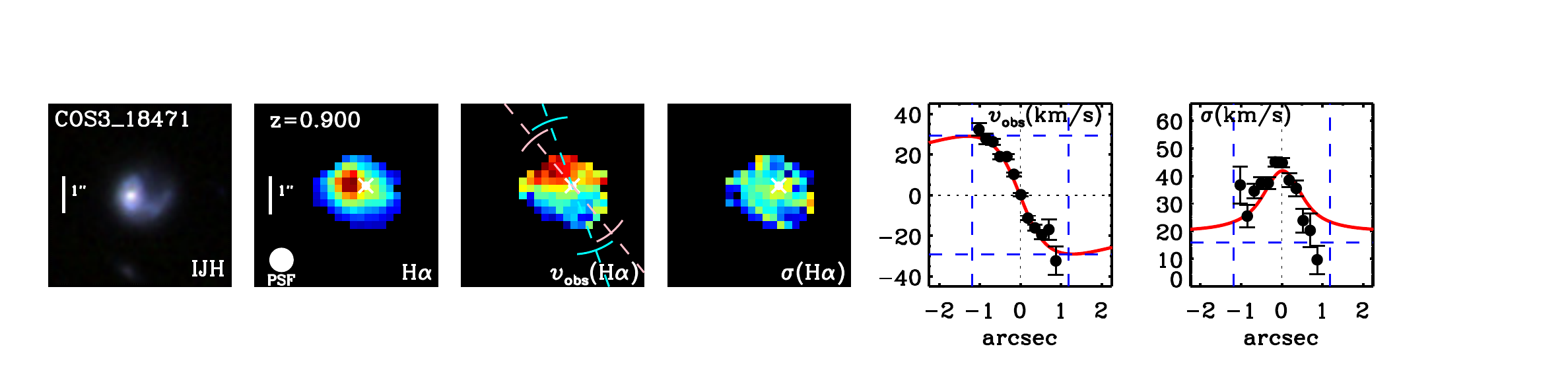}
\includegraphics[scale=0.8,  trim=1.5cm 0cm 1.0cm 0cm, clip, angle=90 ]{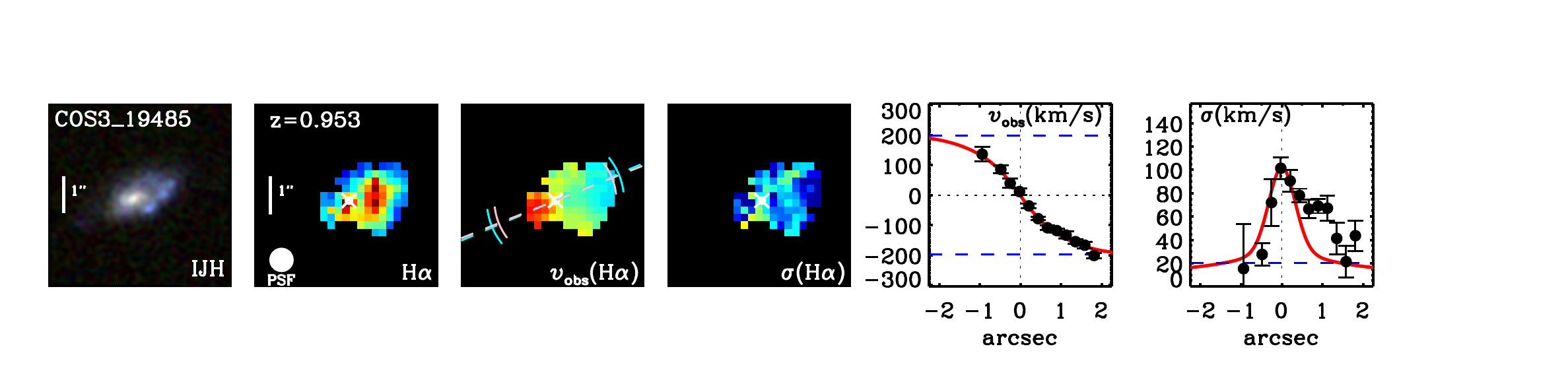}
\includegraphics[scale=0.8,  trim=1.5cm 0cm 1.0cm 0cm, clip, angle=90 ]{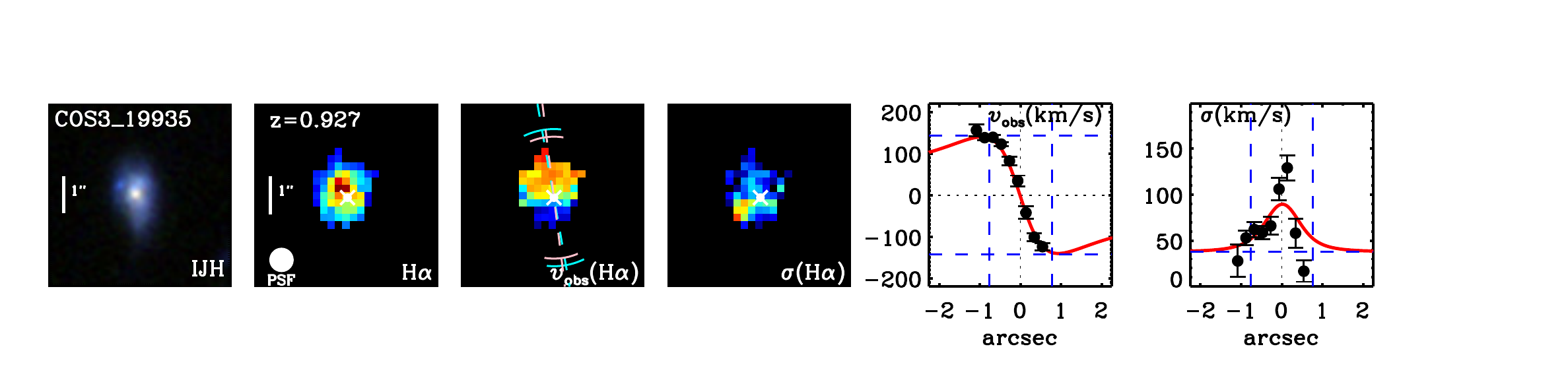}
\includegraphics[scale=0.8,  trim=1.5cm 0cm 1.0cm 0cm, clip, angle=90 ]{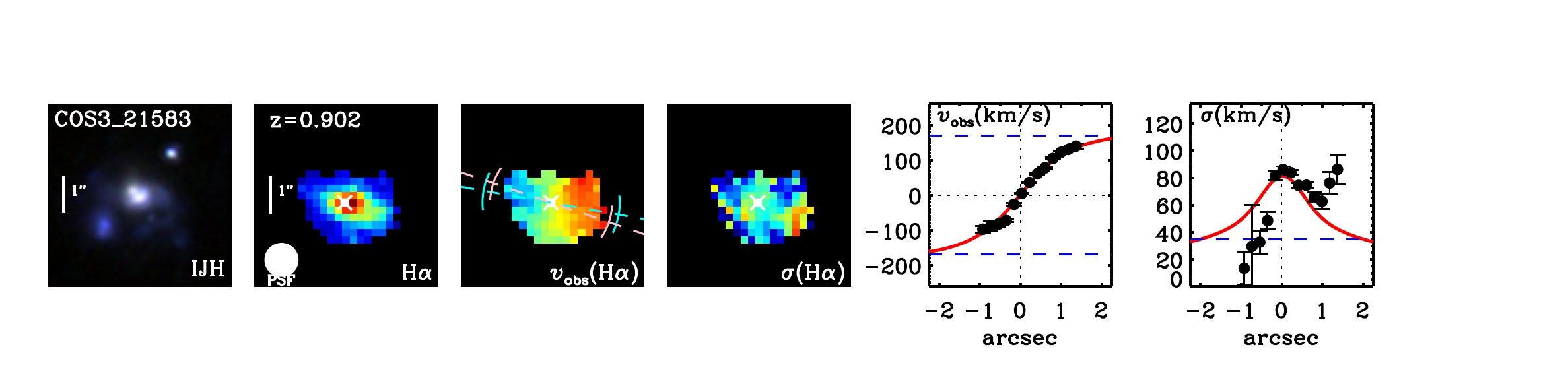}
\includegraphics[scale=0.8,  trim=0.5cm 0cm 1.0cm 0cm, clip, angle=90 ]{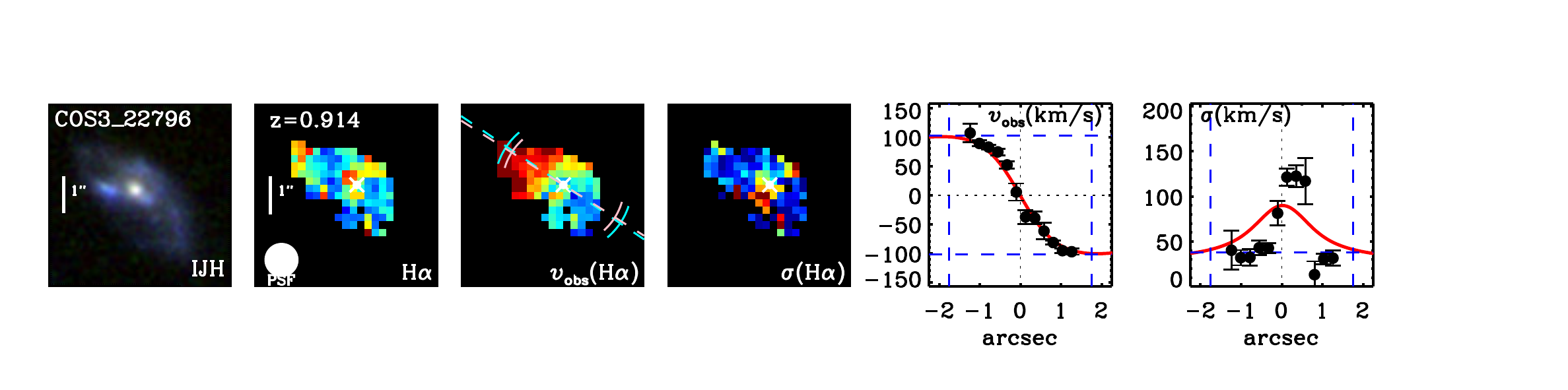}
\caption{\textit{cont.}; Kinematic maps and axis profiles for the high S/N disk galaxies in \kmostd first year data}
\end{figure*}
\addtocounter{figure}{-1}

\begin{figure*}
\begin{center}
\includegraphics[scale=0.8,  trim=1.5cm 0cm 1.0cm 0cm, clip, angle=90 ]{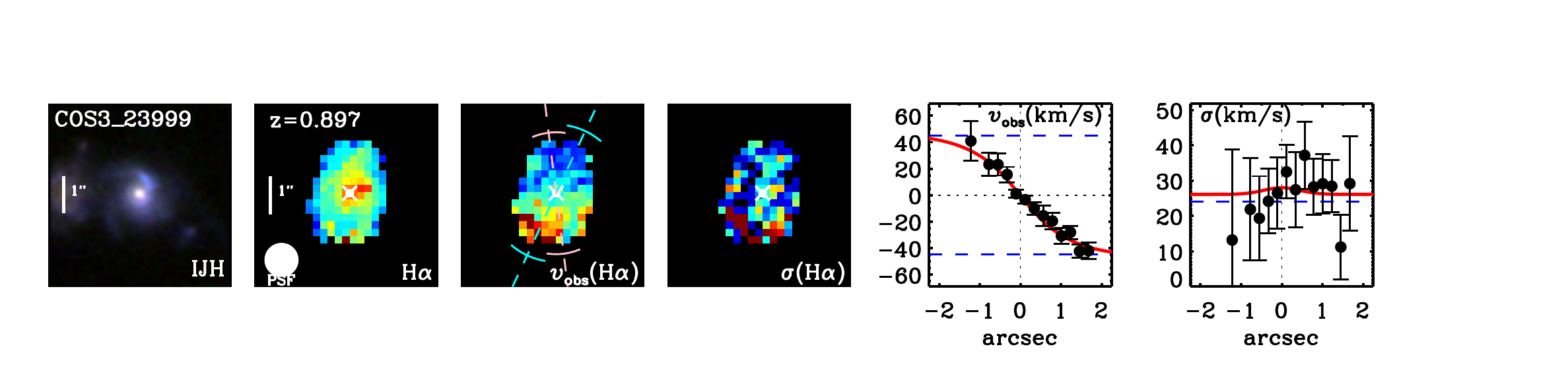}
\includegraphics[scale=0.8,  trim=1.5cm 0cm 1.0cm 0cm, clip, angle=90 ]{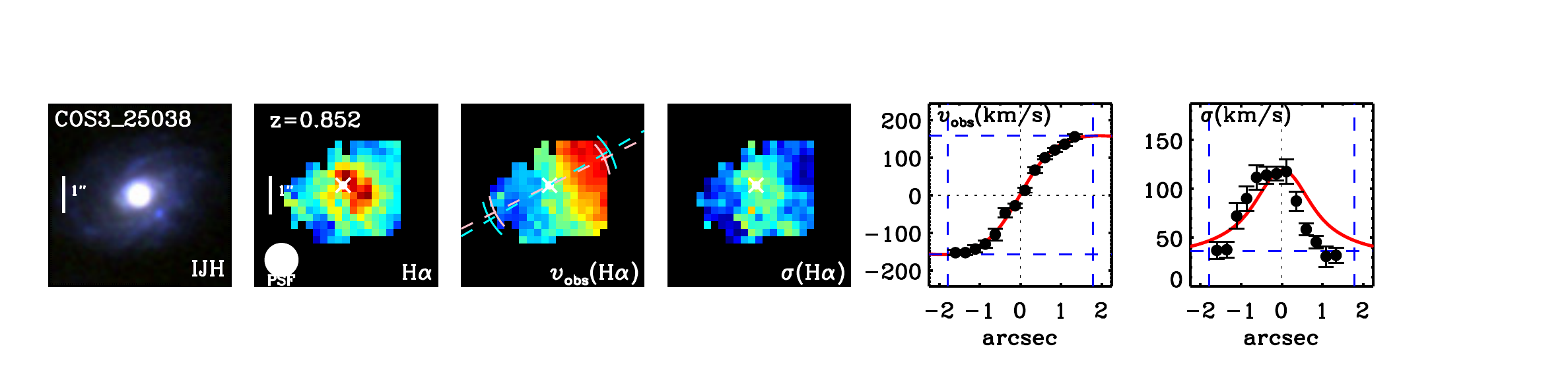}
\includegraphics[scale=0.8,  trim=1.5cm 0cm 1.0cm 0cm, clip, angle=90 ]{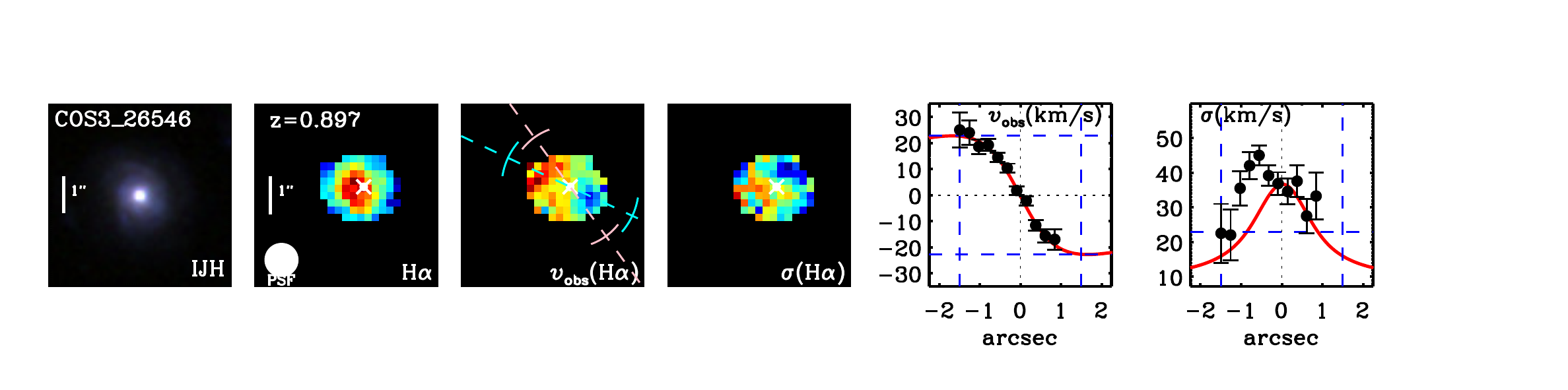}
\includegraphics[scale=0.8,  trim=1.5cm 0cm 1.0cm 0cm, clip, angle=90 ]{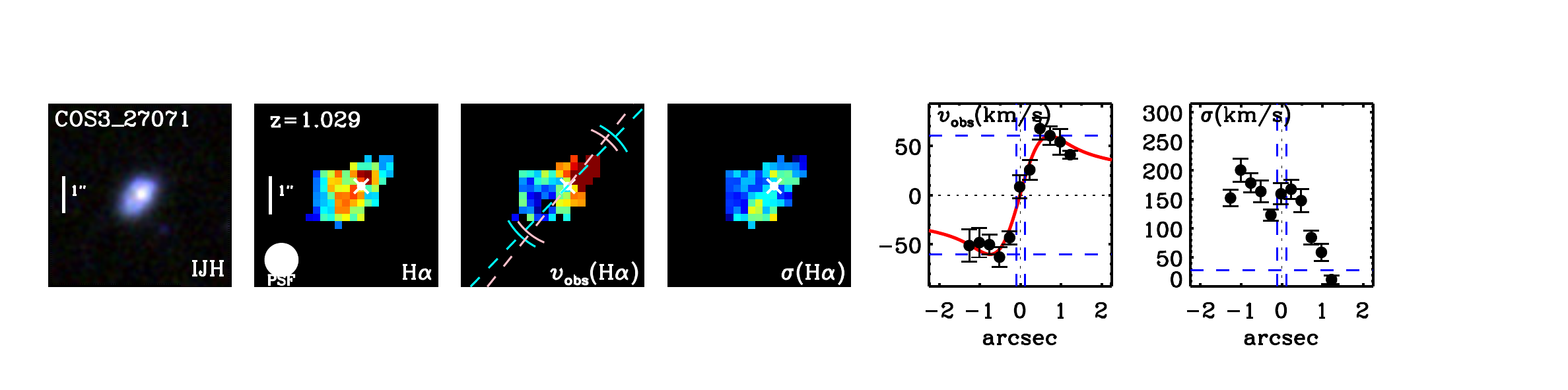}
\includegraphics[scale=0.8,  trim=1.5cm 0cm 1.0cm 0cm, clip, angle=90 ]{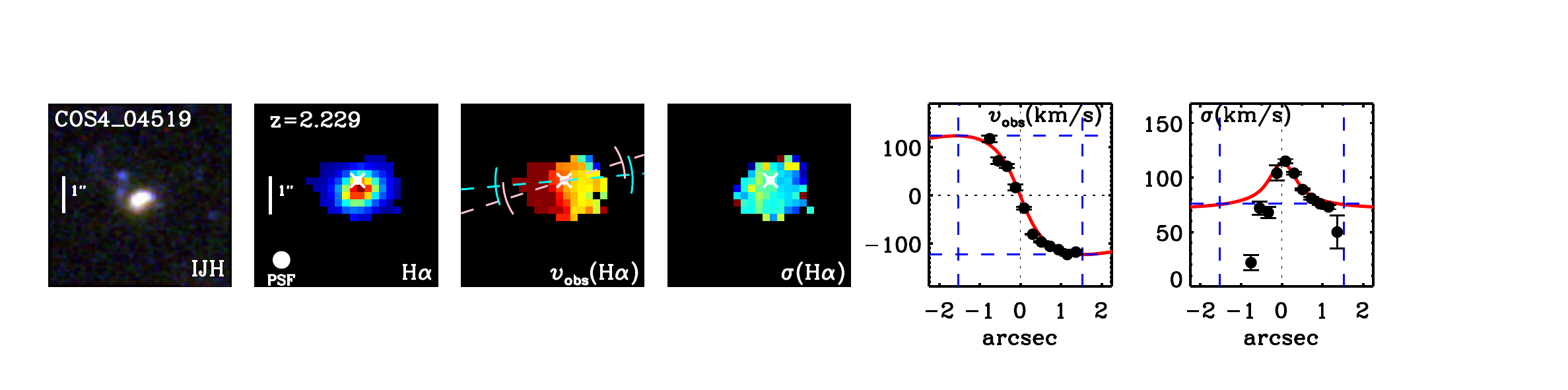}
\includegraphics[scale=0.8,  trim=1.5cm 0cm 1.0cm 0cm, clip, angle=90 ]{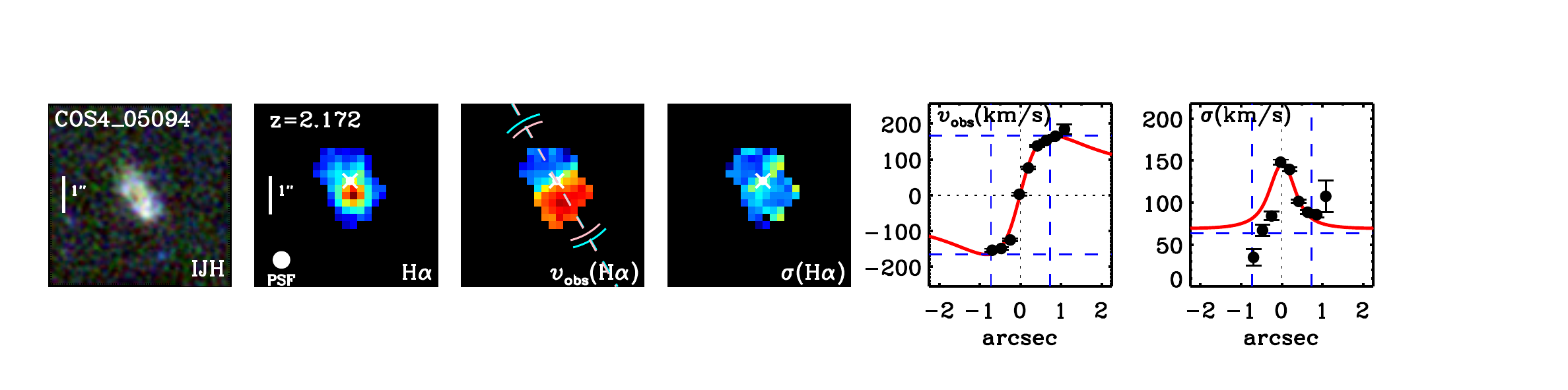}
\includegraphics[scale=0.8,  trim=1.5cm 0cm 1.0cm 0cm, clip, angle=90 ]{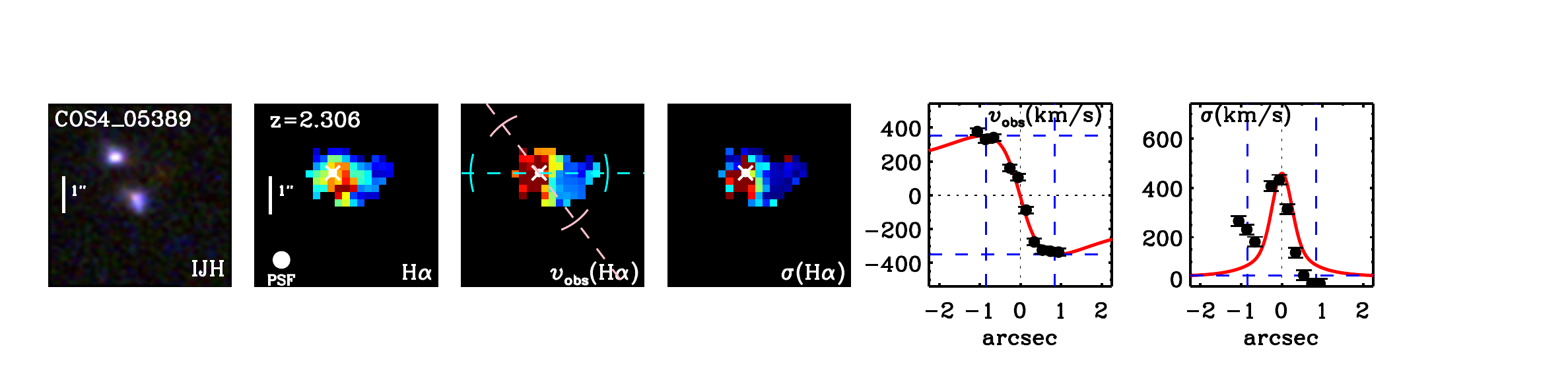}
\includegraphics[scale=0.8,  trim=0.5cm 0cm 1.0cm 0cm, clip, angle=90 ]{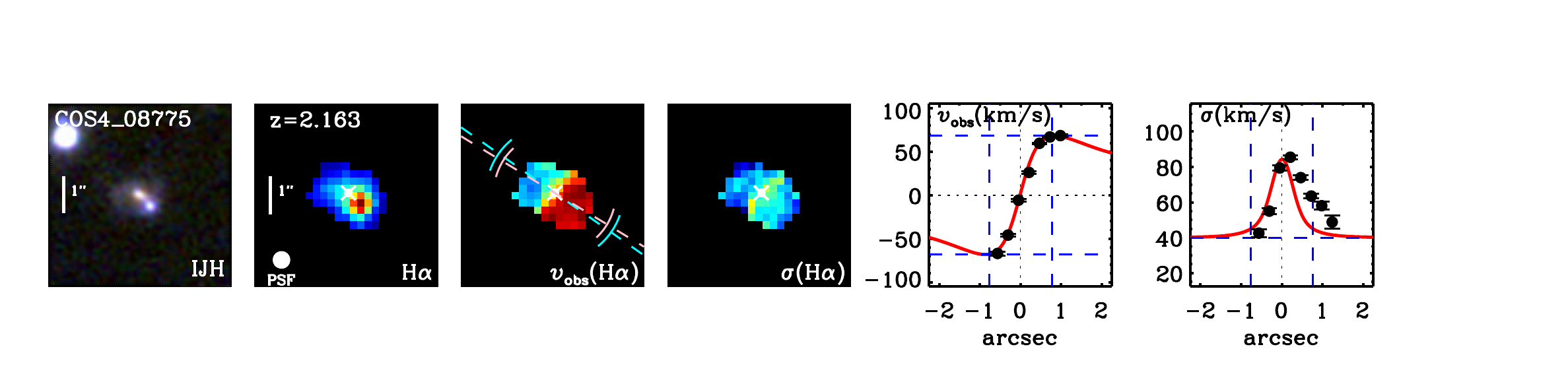}
\caption{\textit{cont.}; Kinematic maps and axis profiles for the high S/N disk galaxies in \kmostd first year data}
\end{center}
\end{figure*}
\addtocounter{figure}{-1}
\begin{figure*}
\begin{center}
\includegraphics[scale=0.8,  trim=1.5cm 0cm 1.0cm 0cm, clip, angle=90 ]{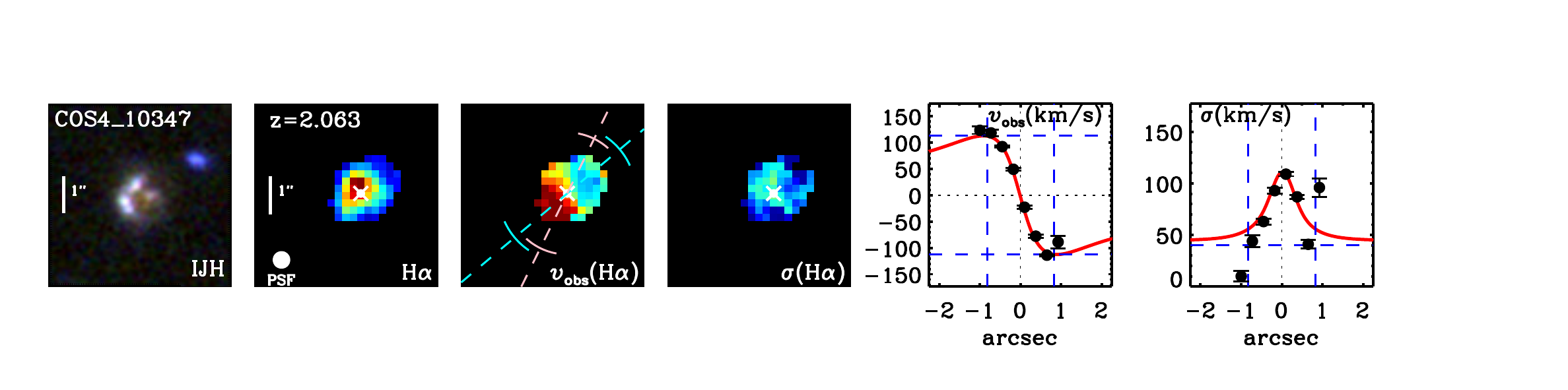}
\includegraphics[scale=0.8,  trim=1.5cm 0cm 1.0cm 0cm, clip, angle=90 ]{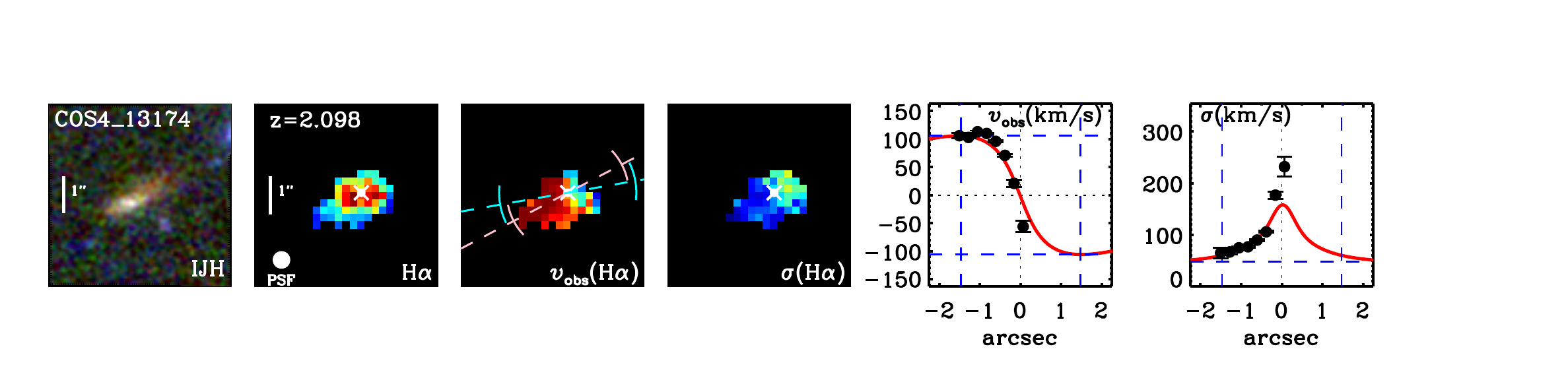}
\includegraphics[scale=0.8,  trim=1.5cm 0cm 1.0cm 0cm, clip, angle=90 ]{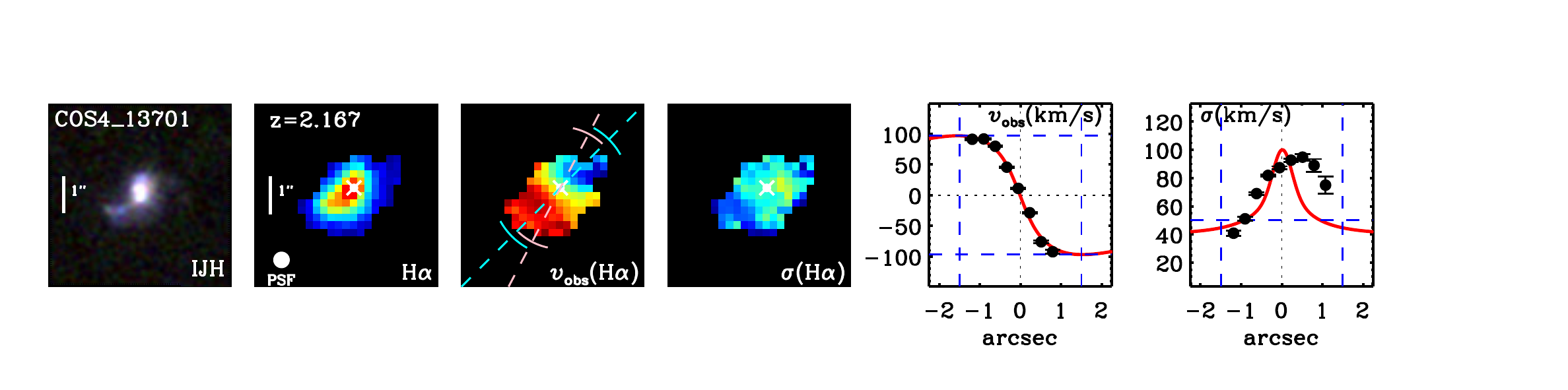}
\includegraphics[scale=0.8,  trim=1.5cm 0cm 1.0cm 0cm, clip, angle=90 ]{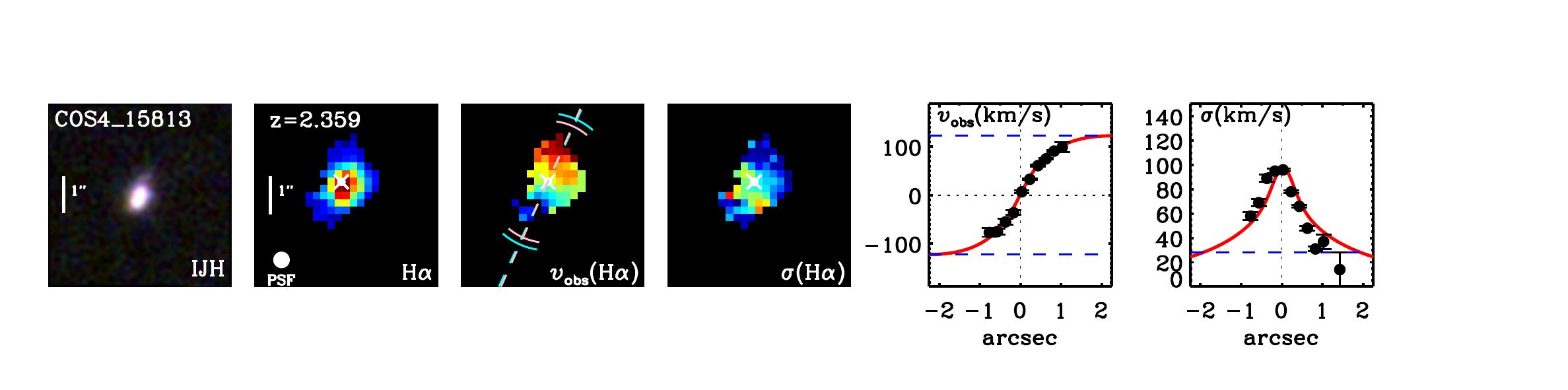}
\includegraphics[scale=0.8,  trim=1.5cm 0cm 1.0cm 0cm, clip, angle=90 ]{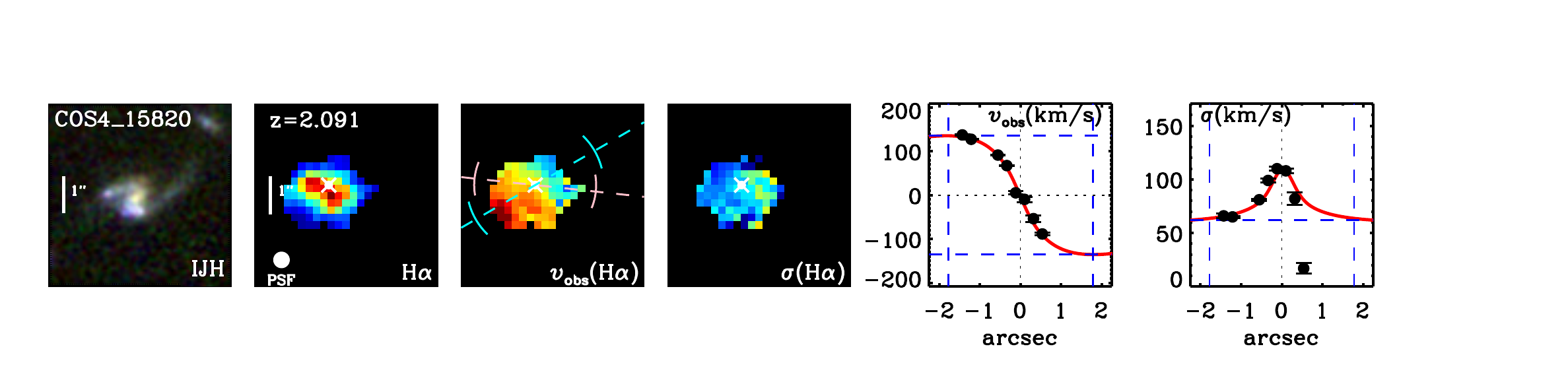}
\includegraphics[scale=0.8,  trim=1.5cm 0cm 1.0cm 0cm, clip, angle=90 ]{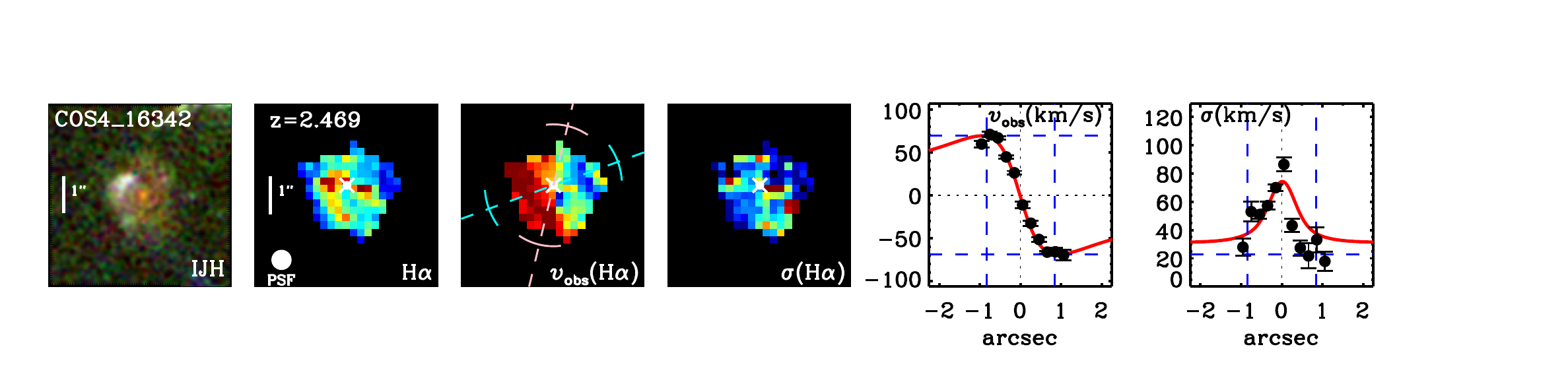}
\includegraphics[scale=0.8,  trim=1.5cm 0cm 1.0cm 0cm, clip, angle=90 ]{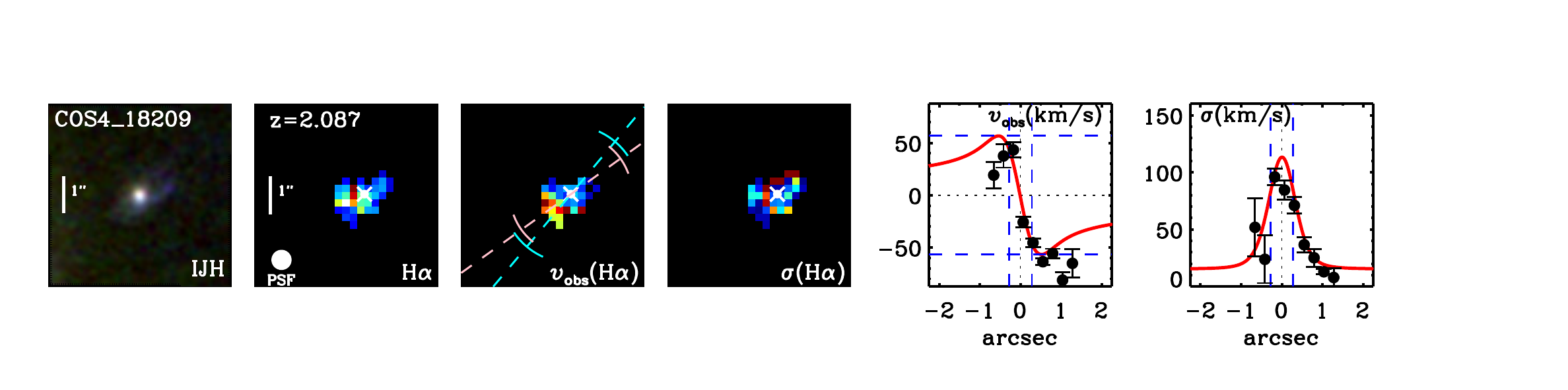}
\includegraphics[scale=0.8,  trim=0.5cm 0cm 1.0cm 0cm, clip, angle=90 ]{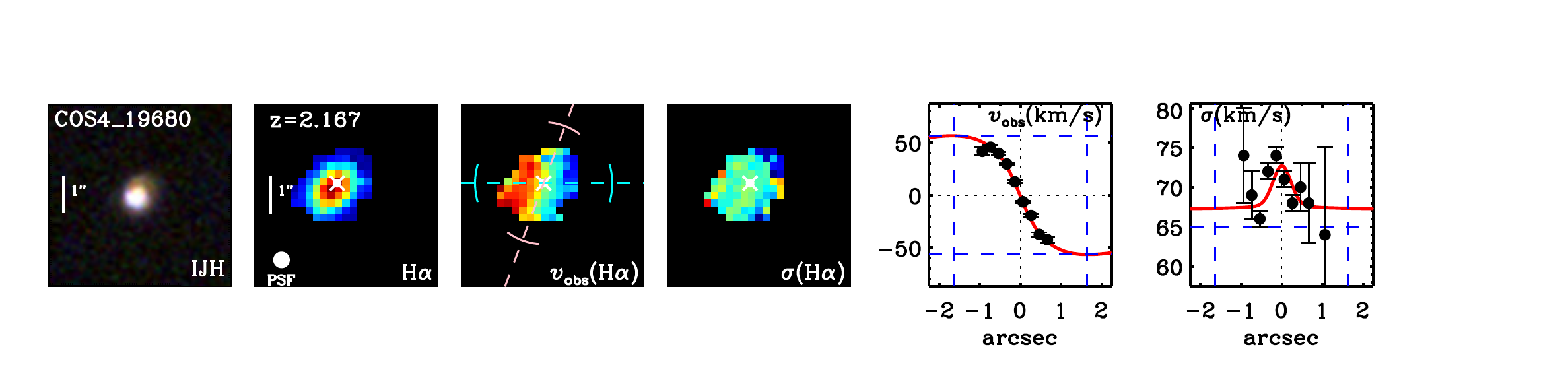}
\end{center}
\caption{\textit{cont.}; Kinematic maps and axis profiles for the high S/N disk galaxies in \kmostd first year data}
\end{figure*}
\addtocounter{figure}{-1}

\clearpage
\begin{figure*}
\begin{center}
\includegraphics[scale=0.8,  trim=1.5cm 0cm 1.0cm 0cm, clip, angle=90 ]{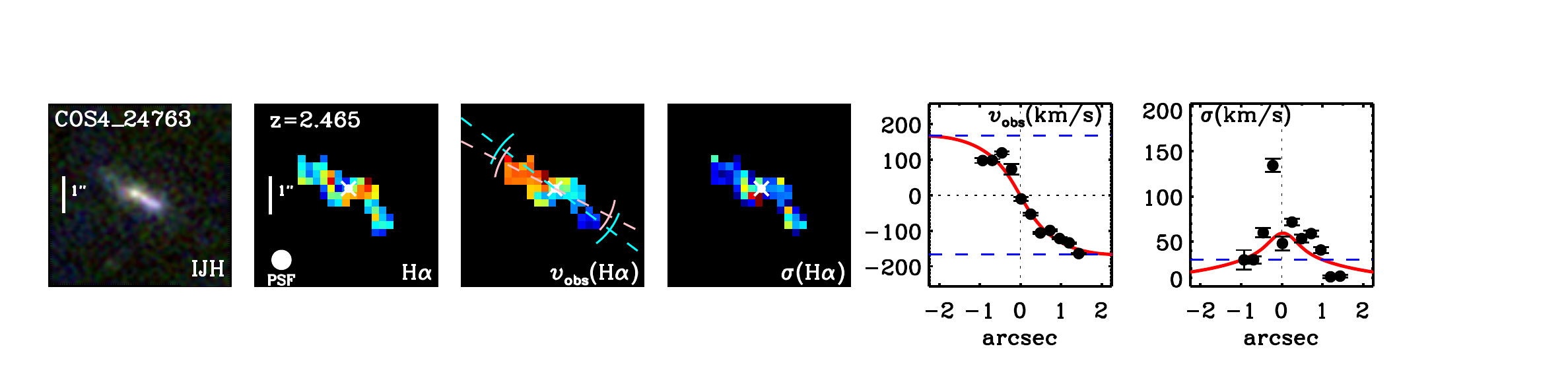}
\includegraphics[scale=0.8,  trim=1.5cm 0cm 1.0cm 0cm, clip, angle=90 ]{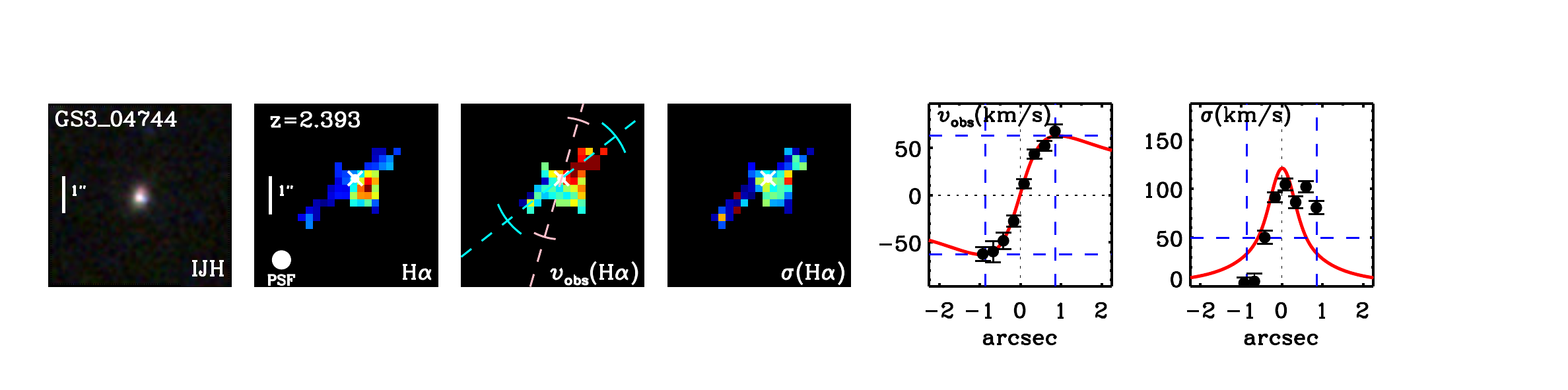}
\includegraphics[scale=0.8,  trim=1.5cm 0cm 1.0cm 0cm, clip, angle=90 ]{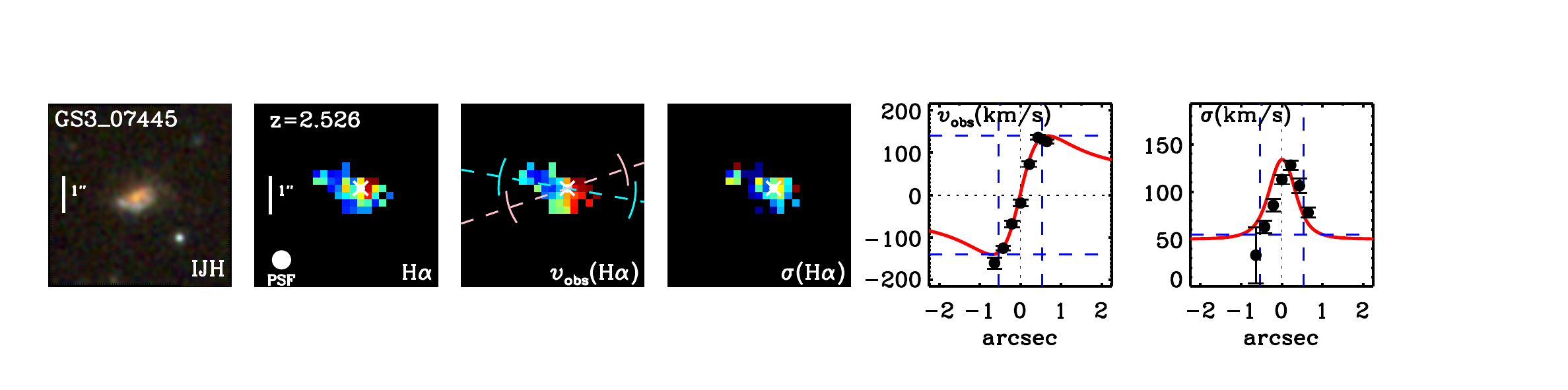}
\includegraphics[scale=0.8,  trim=1.5cm 0cm 1.0cm 0cm, clip, angle=90 ]{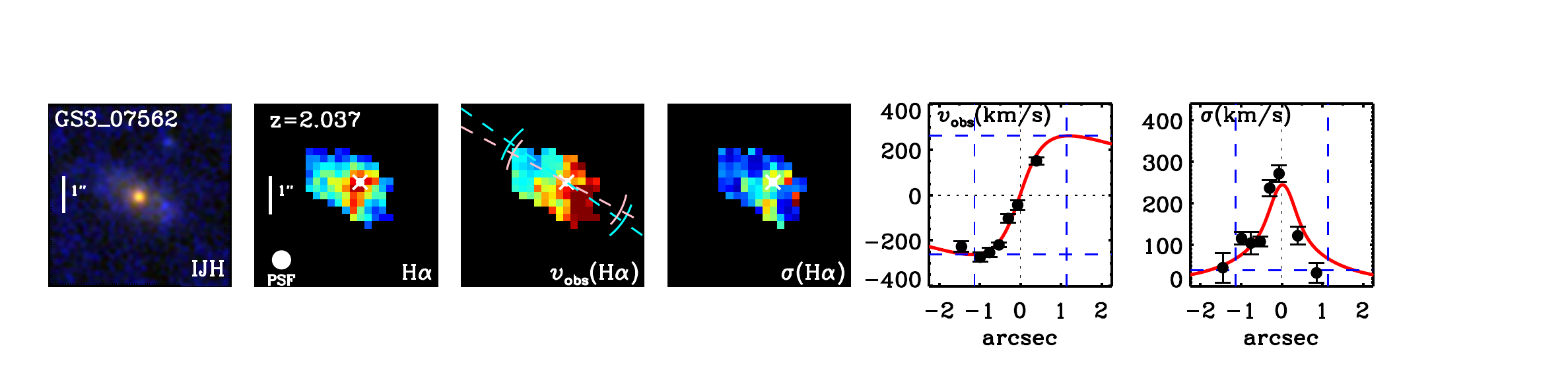}
\includegraphics[scale=0.8,  trim=1.5cm 0cm 1.0cm 0cm, clip, angle=90 ]{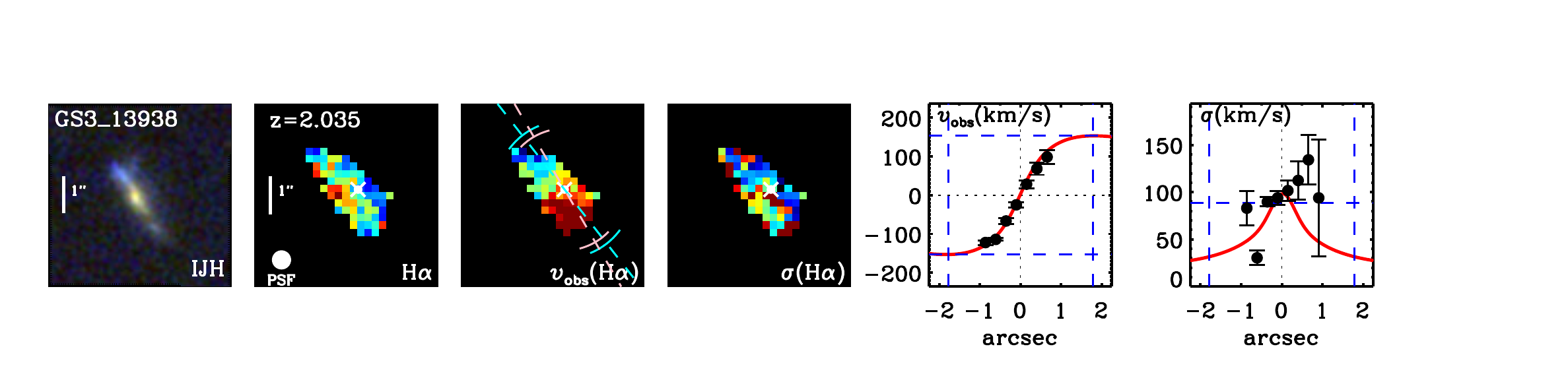}
\includegraphics[scale=0.8,  trim=1.5cm 0cm 1.0cm 0cm, clip, angle=90 ]{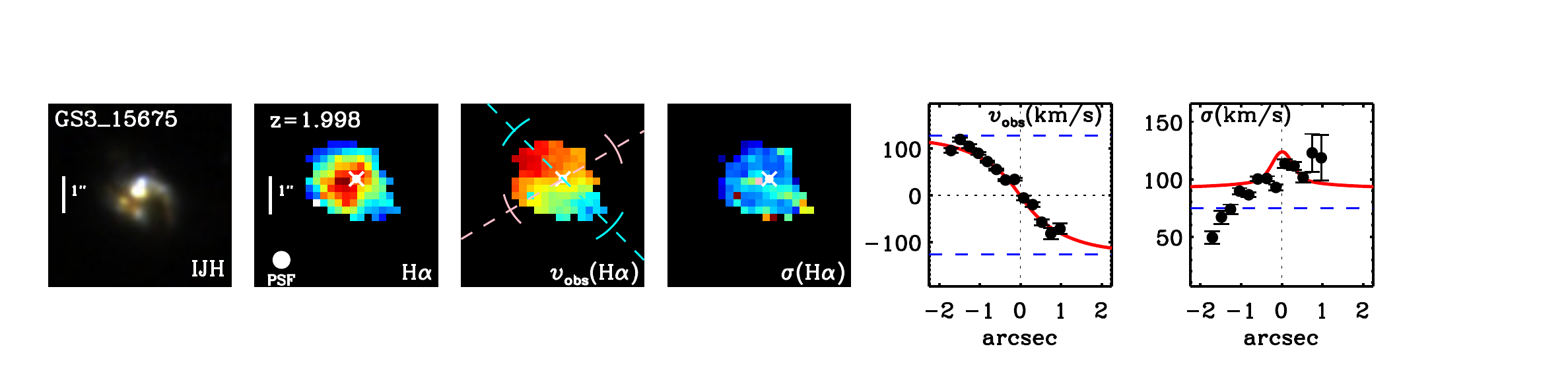}
\includegraphics[scale=0.8,  trim=1.5cm 0cm 1.0cm 0cm, clip, angle=90 ]{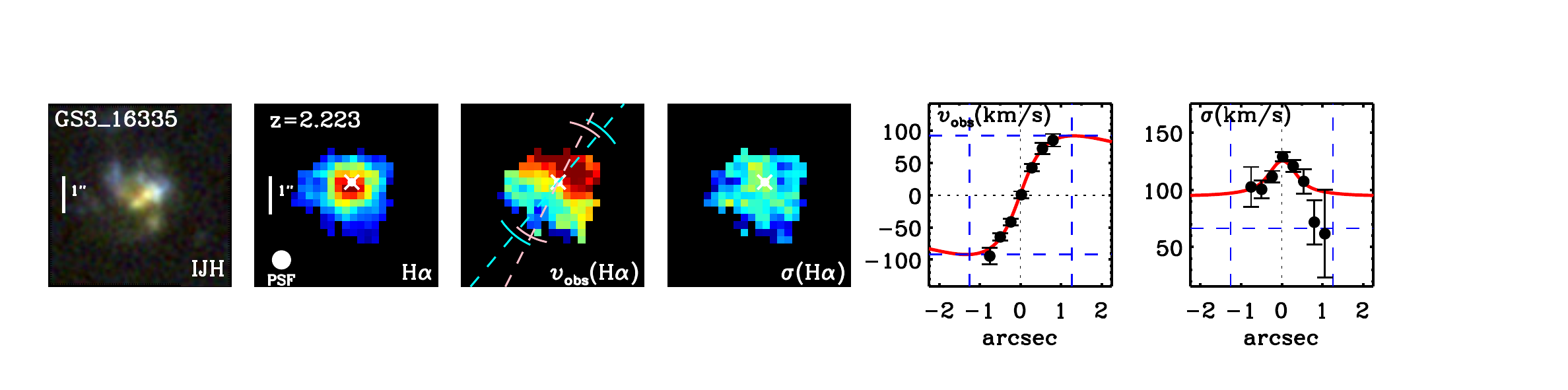}
\includegraphics[scale=0.8,  trim=0.5cm 0cm 1.0cm 0cm, clip, angle=90 ]{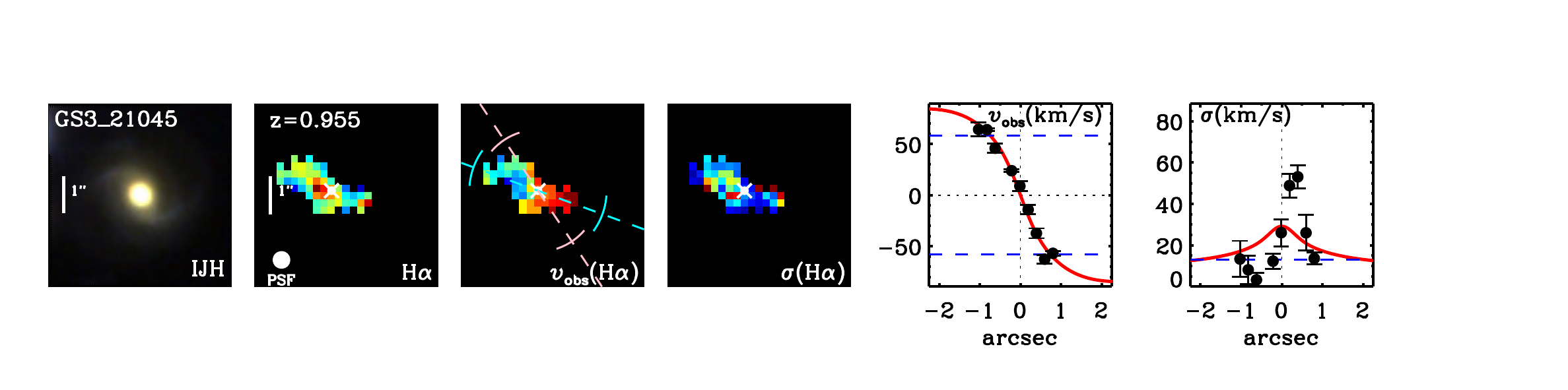}
\end{center}
\caption{\textit{cont.}; Kinematic maps and axis profiles for the high S/N disk galaxies in \kmostd first year data}
\end{figure*}
\addtocounter{figure}{-1}

\clearpage
\begin{figure*}
\begin{center}
\includegraphics[scale=0.8,  trim=1.5cm 0cm 1.0cm 0cm, clip, angle=90 ]{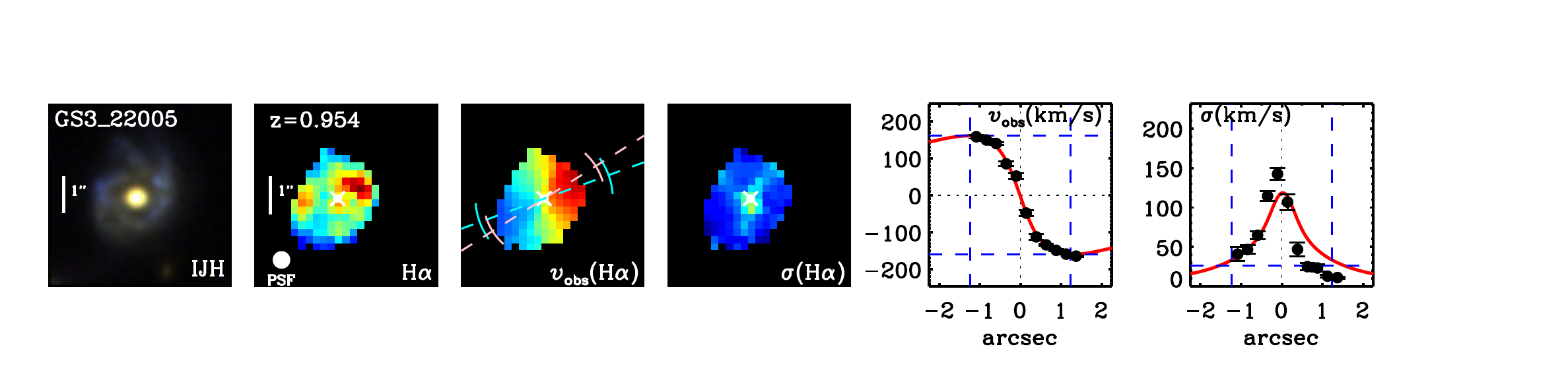}
\includegraphics[scale=0.8,  trim=1.5cm 0cm 1.0cm 0cm, clip, angle=90 ]{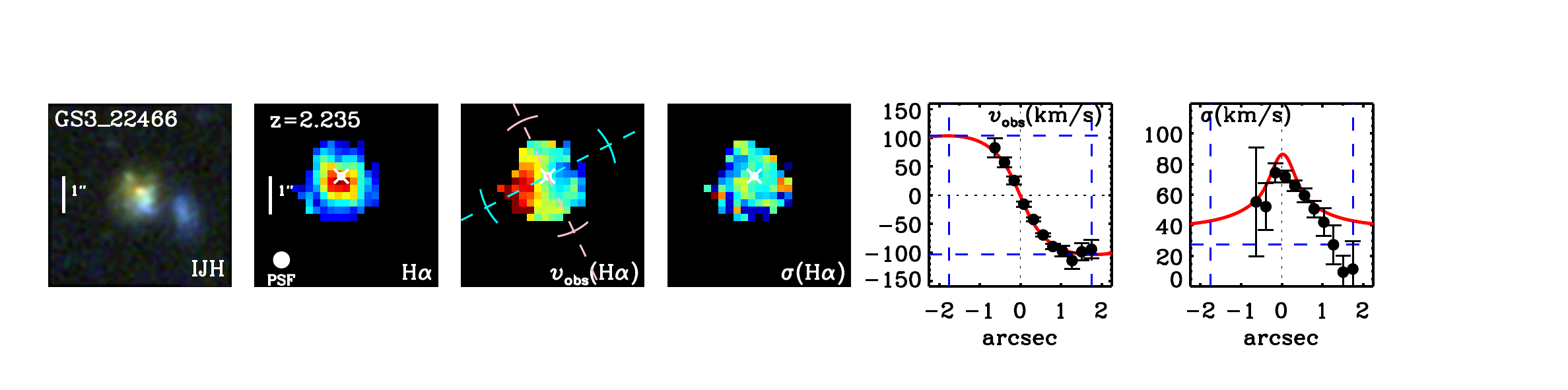}
\includegraphics[scale=0.8,  trim=1.5cm 0cm 1.0cm 0cm, clip, angle=90 ]{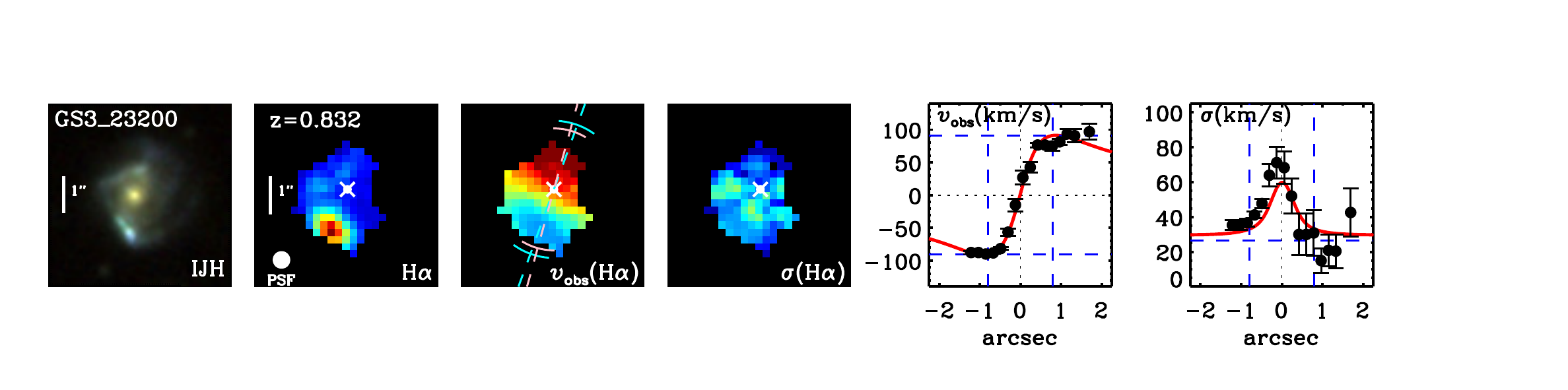}
\includegraphics[scale=0.8,  trim=1.5cm 0cm 1.0cm 0cm, clip, angle=90 ]{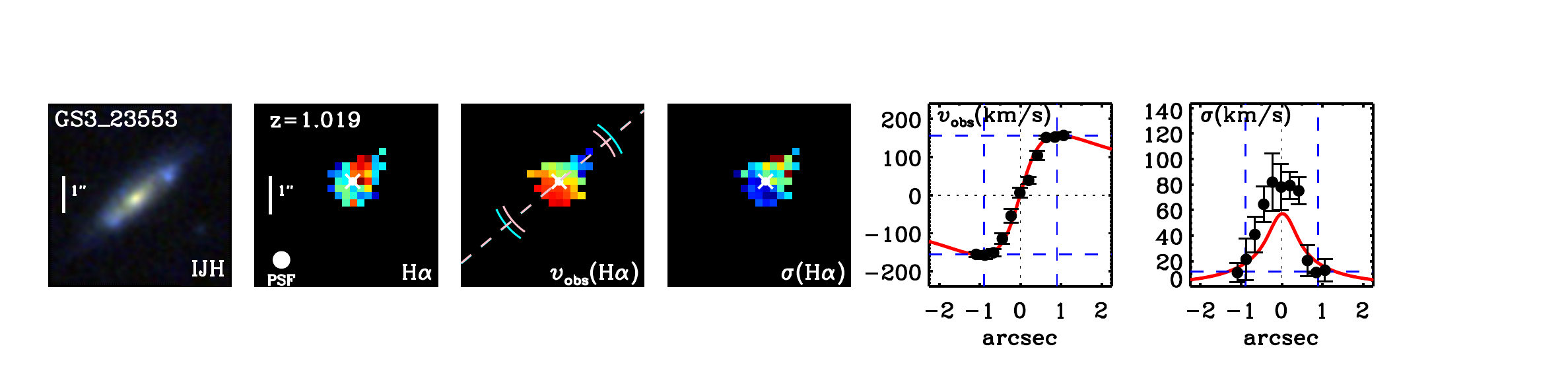}
\includegraphics[scale=0.8,  trim=1.5cm 0cm 1.0cm 0cm, clip, angle=90 ]{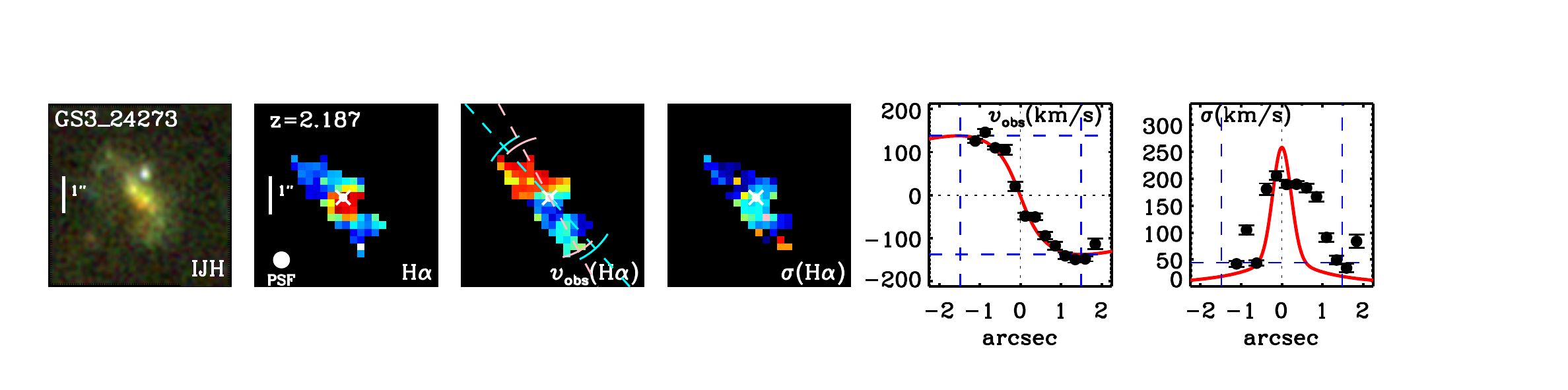}
\includegraphics[scale=0.8,  trim=1.5cm 0cm 1.0cm 0cm, clip, angle=90 ]{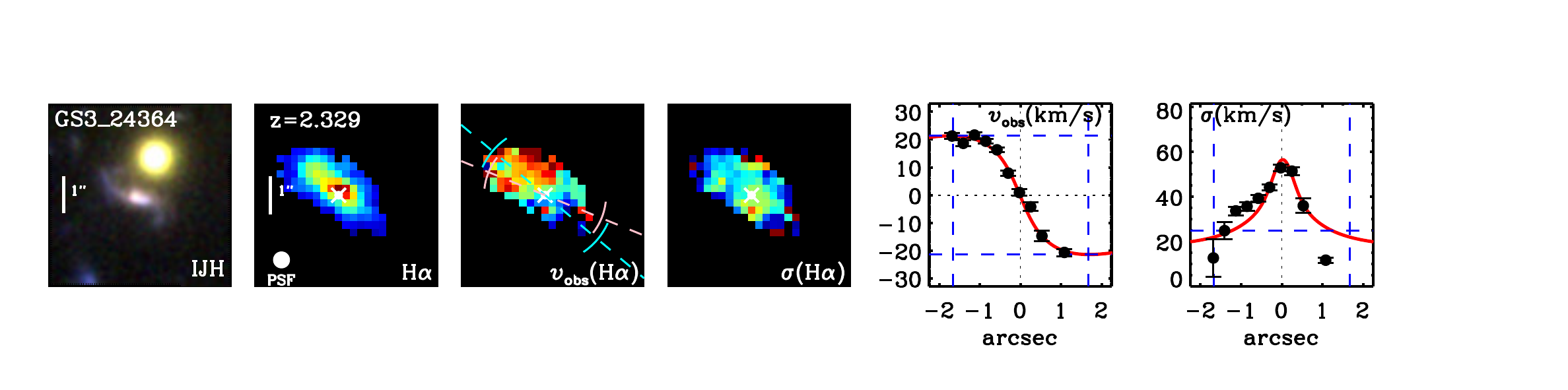}
\includegraphics[scale=0.8,  trim=1.5cm 0cm 1.0cm 0cm, clip, angle=90 ]{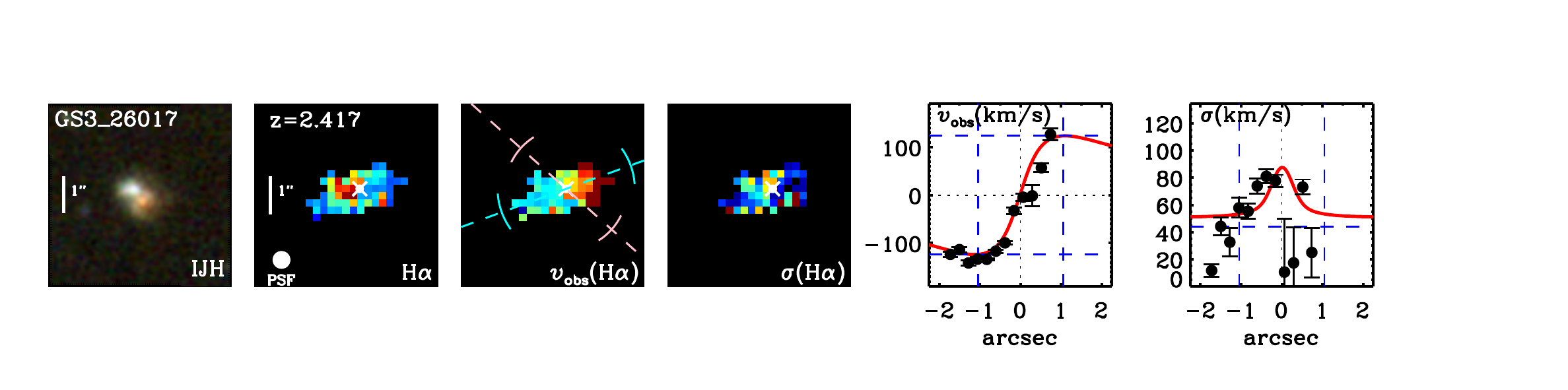}
\includegraphics[scale=0.8,  trim=0.5cm 0cm 1.0cm 0cm, clip, angle=90 ]{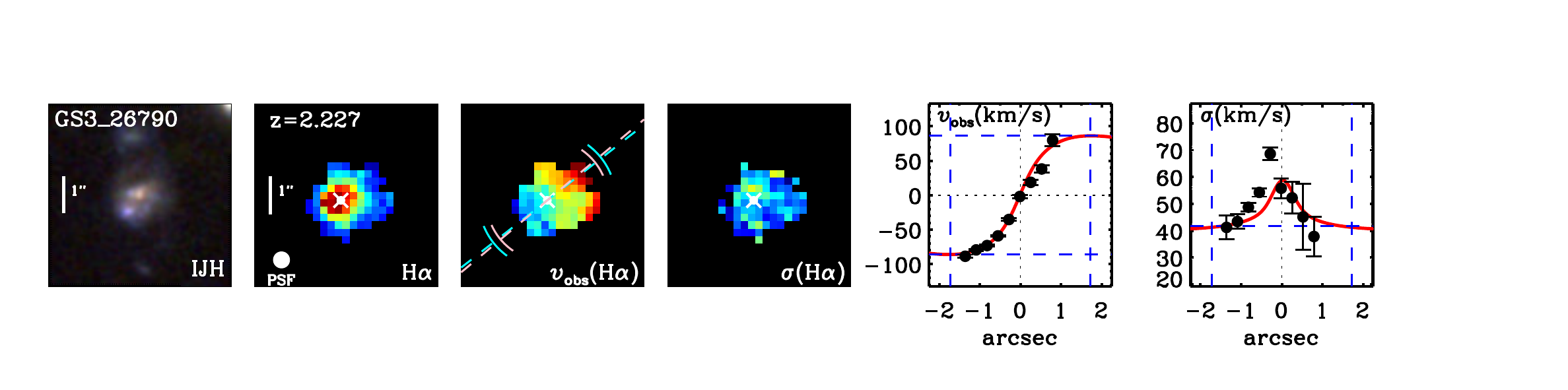}
\caption{\textit{cont.}; Kinematic maps and axis profiles for the high S/N disk galaxies in \kmostd first year data}
\end{center}
\end{figure*}
\addtocounter{figure}{-1}

\begin{figure*}
\begin{center}
\includegraphics[scale=0.8,  trim=1.5cm 0cm 1.0cm 0cm, clip, angle=90 ]{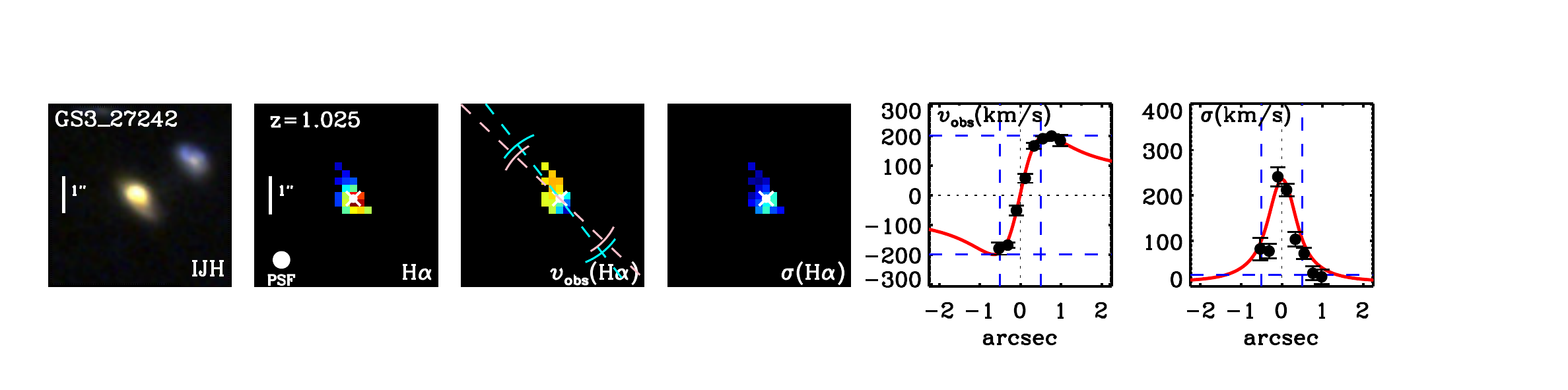}
\includegraphics[scale=0.8,  trim=1.5cm 0cm 1.0cm 0cm, clip, angle=90 ]{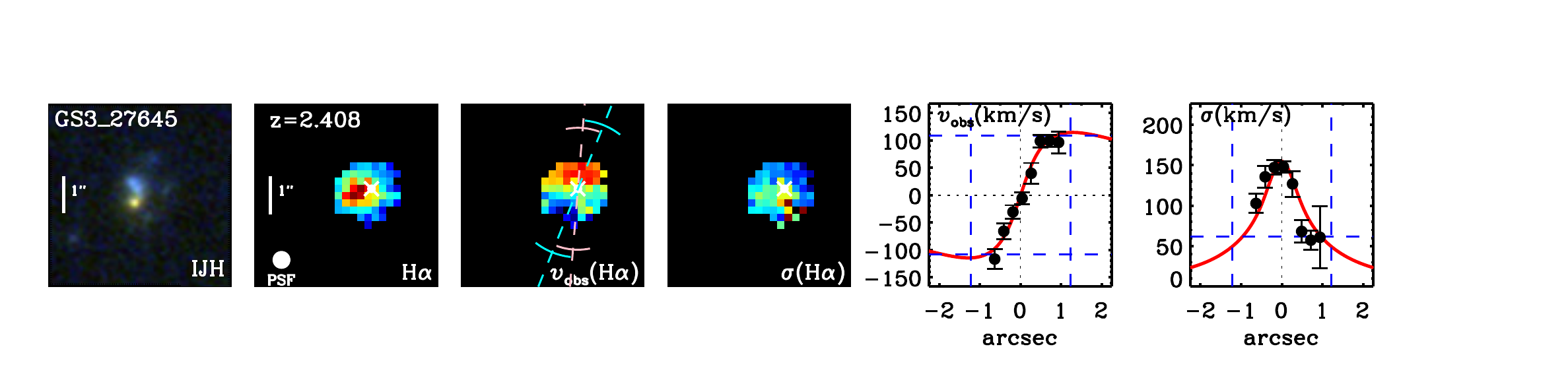}
\includegraphics[scale=0.8,  trim=1.5cm 0cm 1.0cm 0cm, clip, angle=90 ]{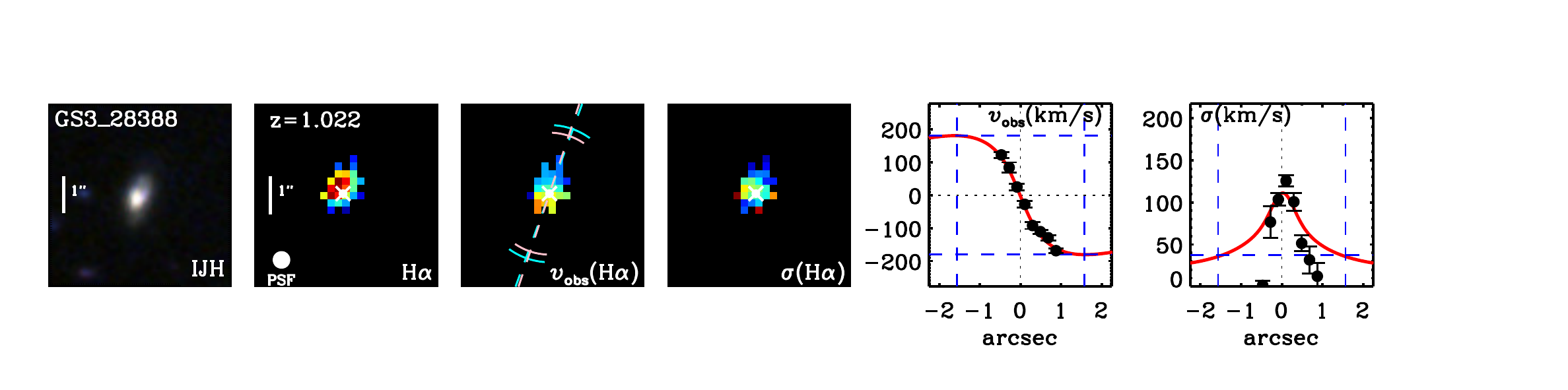}
\includegraphics[scale=0.8,  trim=1.5cm 0cm 1.0cm 0cm, clip, angle=90 ]{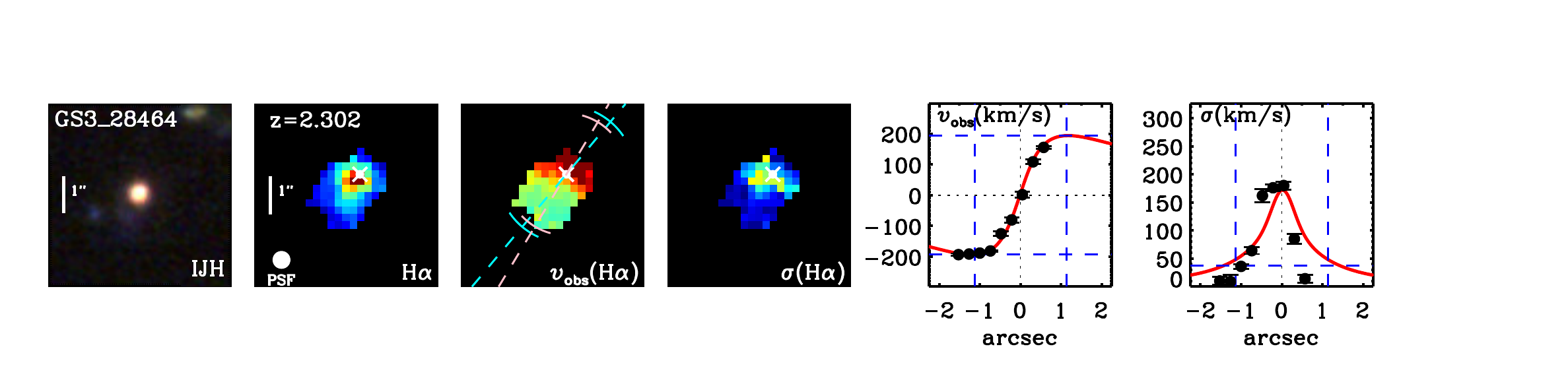}
\includegraphics[scale=0.8,  trim=1.5cm 0cm 1.0cm 0cm, clip, angle=90 ]{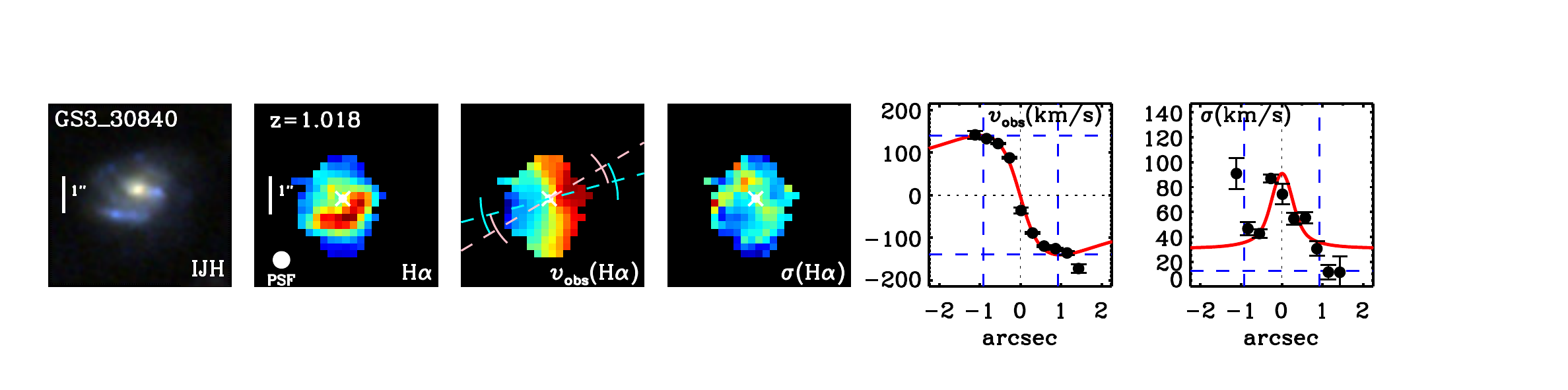}
\includegraphics[scale=0.8,  trim=1.5cm 0cm 1.0cm 0cm, clip, angle=90 ]{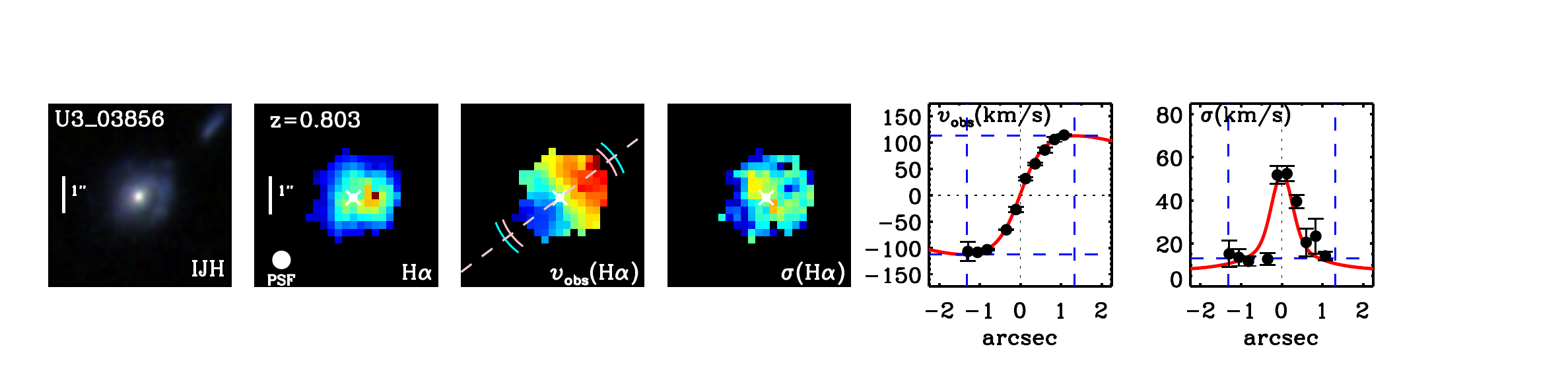}
\includegraphics[scale=0.8,  trim=1.5cm 0cm 1.0cm 0cm, clip, angle=90 ]{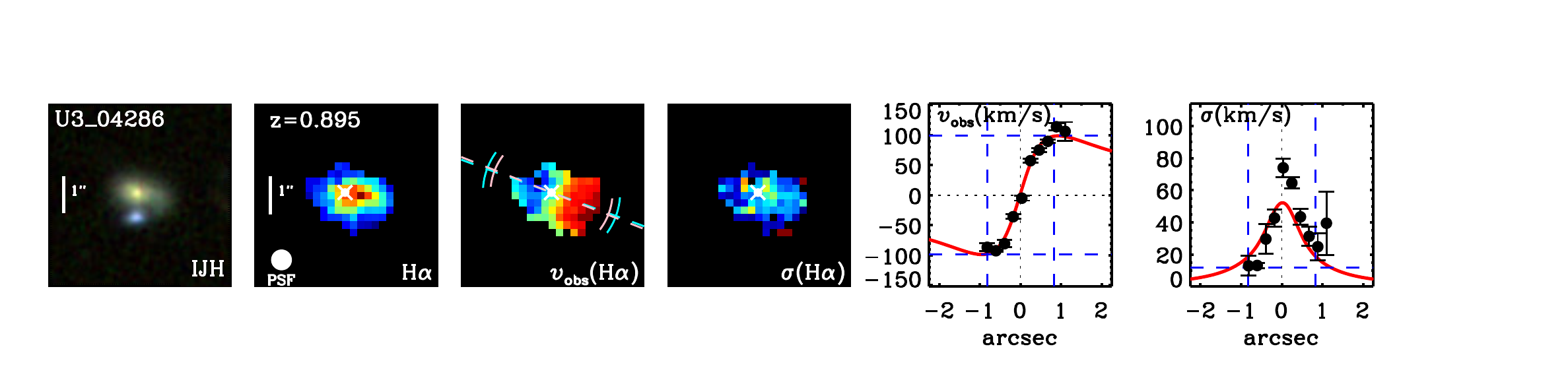}
\includegraphics[scale=0.8,  trim=0.5cm 0cm 1.0cm 0cm, clip, angle=90 ]{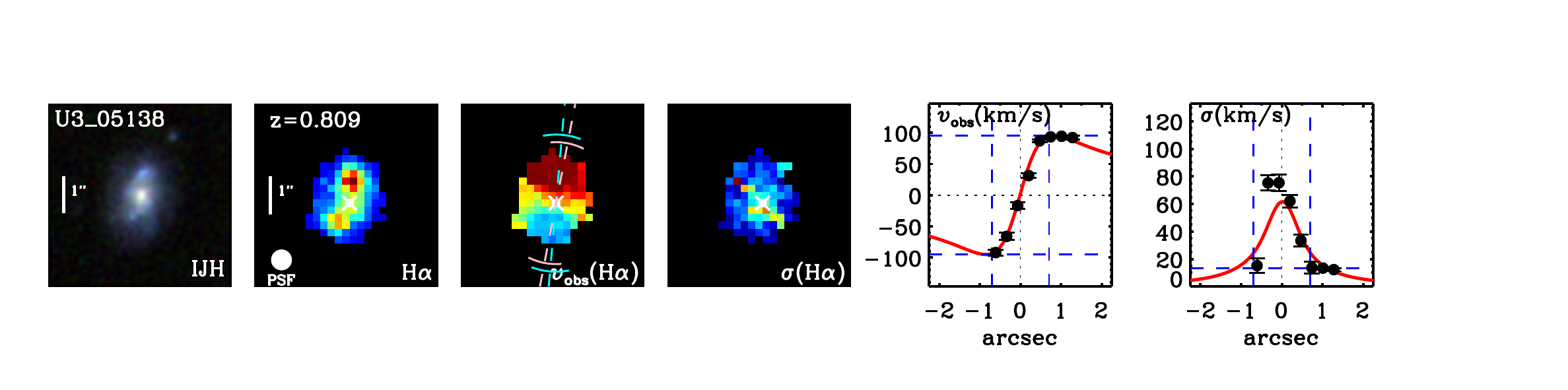}
\end{center}
\caption{\textit{cont.}; Kinematic maps and axis profiles for the high S/N disk galaxies in \kmostd first year data}
\end{figure*}
\addtocounter{figure}{-1}

\begin{figure*}
\begin{center}
\includegraphics[scale=0.8,  trim=1.5cm 0cm 1.0cm 0cm, clip, angle=90 ]{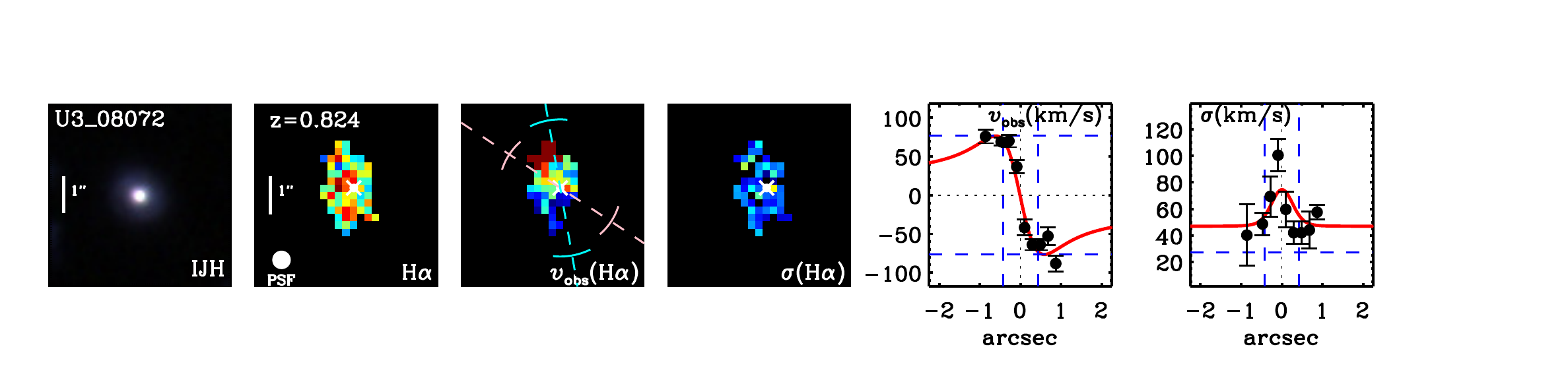}
\includegraphics[scale=0.8,  trim=1.5cm 0cm 1.0cm 0cm, clip, angle=90 ]{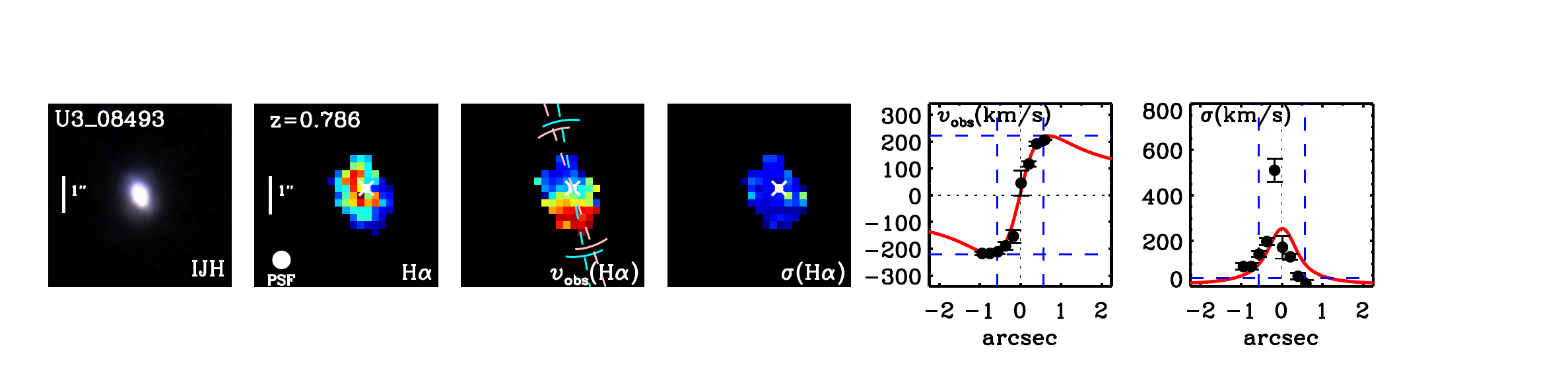}
\includegraphics[scale=0.8,  trim=1.5cm 0cm 1.0cm 0cm, clip, angle=90 ]{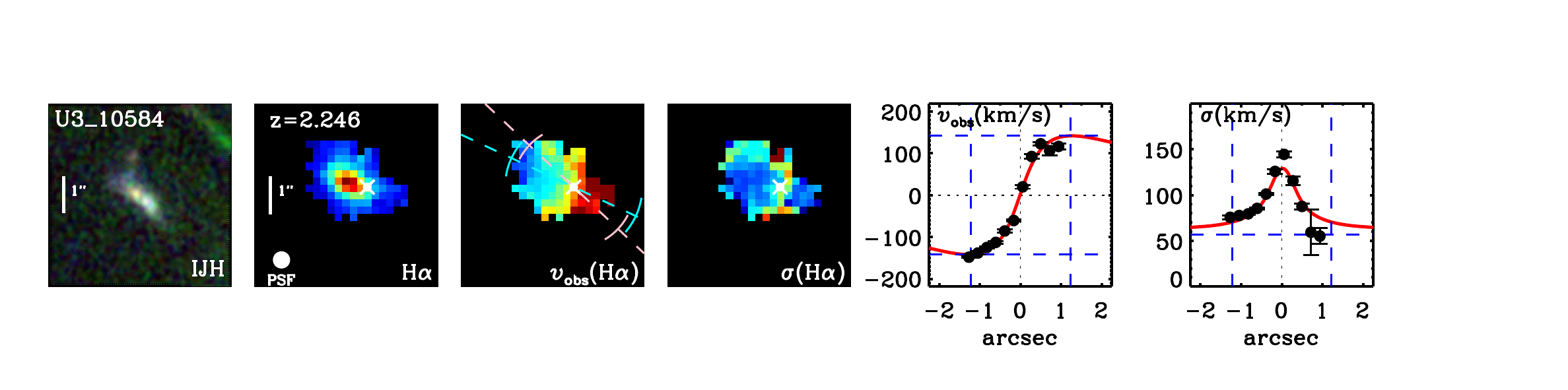}
\includegraphics[scale=0.8,  trim=1.5cm 0cm 1.0cm 0cm, clip, angle=90 ]{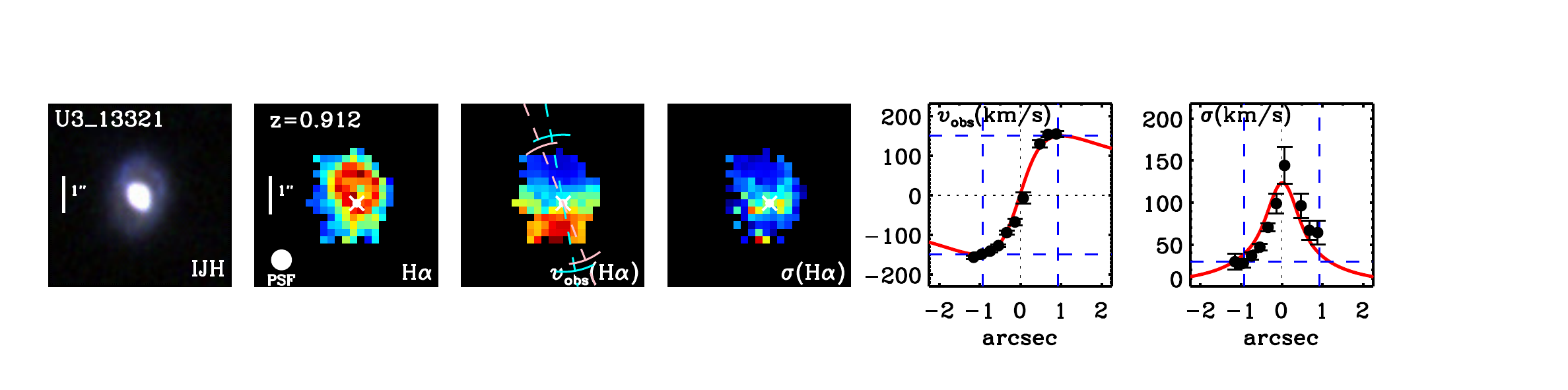}
\includegraphics[scale=0.8,  trim=1.5cm 0cm 1.0cm 0cm, clip, angle=90 ]{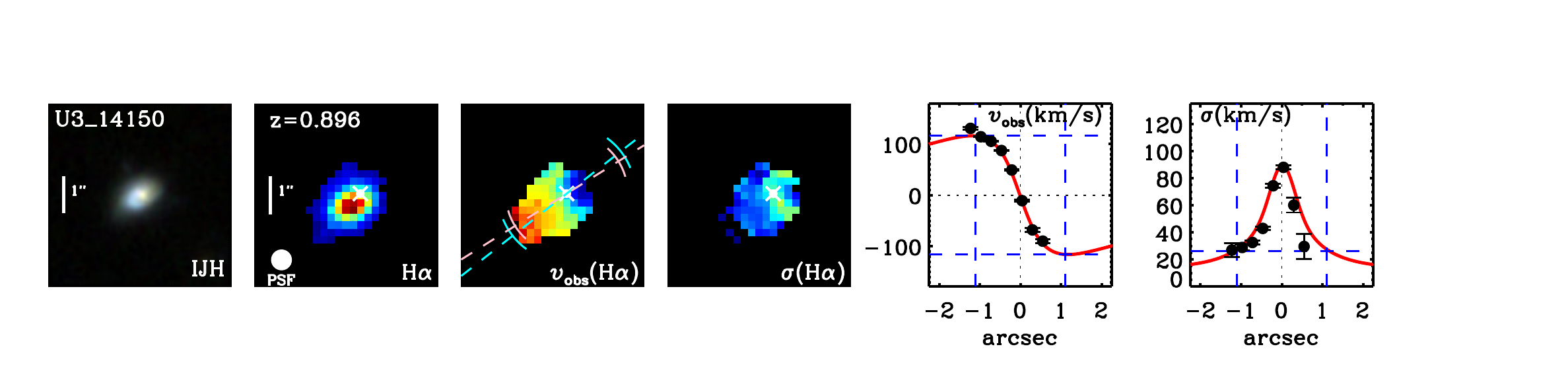}
\includegraphics[scale=0.8,  trim=1.5cm 0cm 1.0cm 0cm, clip, angle=90 ]{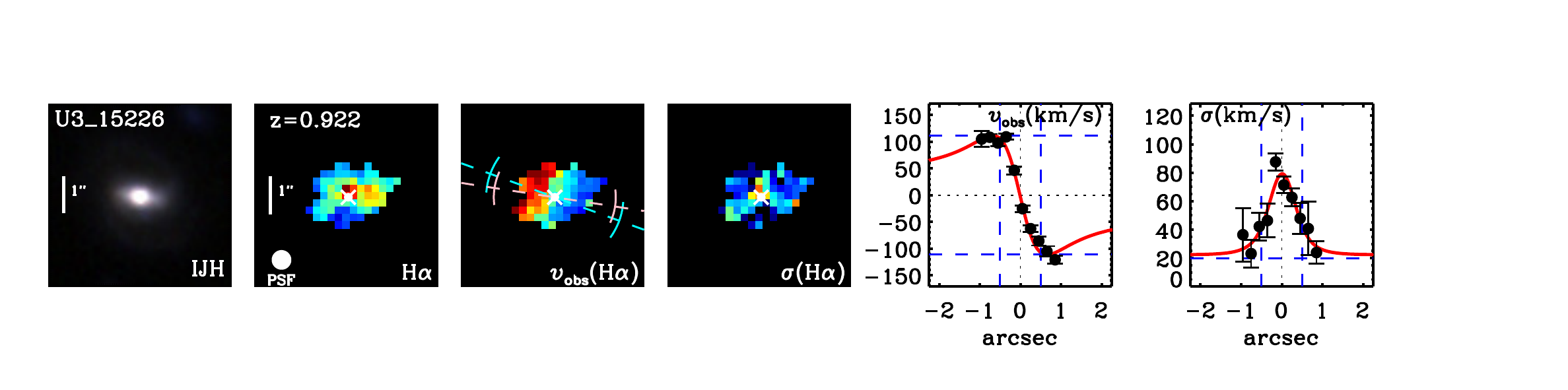}
\includegraphics[scale=0.8,  trim=1.5cm 0cm 1.0cm 0cm, clip, angle=90 ]{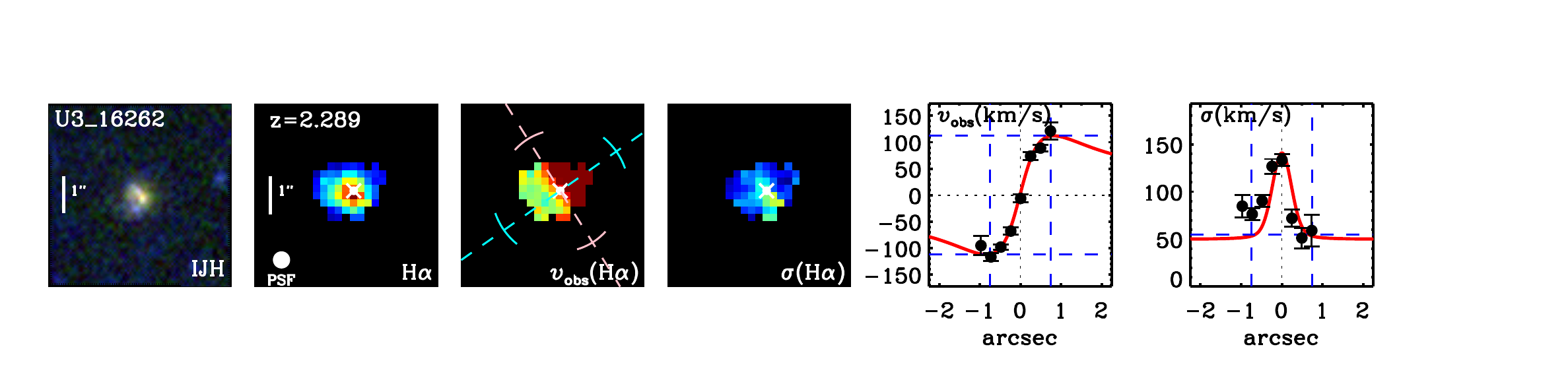}
\includegraphics[scale=0.8,  trim=0.5cm 0cm 1.0cm 0cm, clip, angle=90 ]{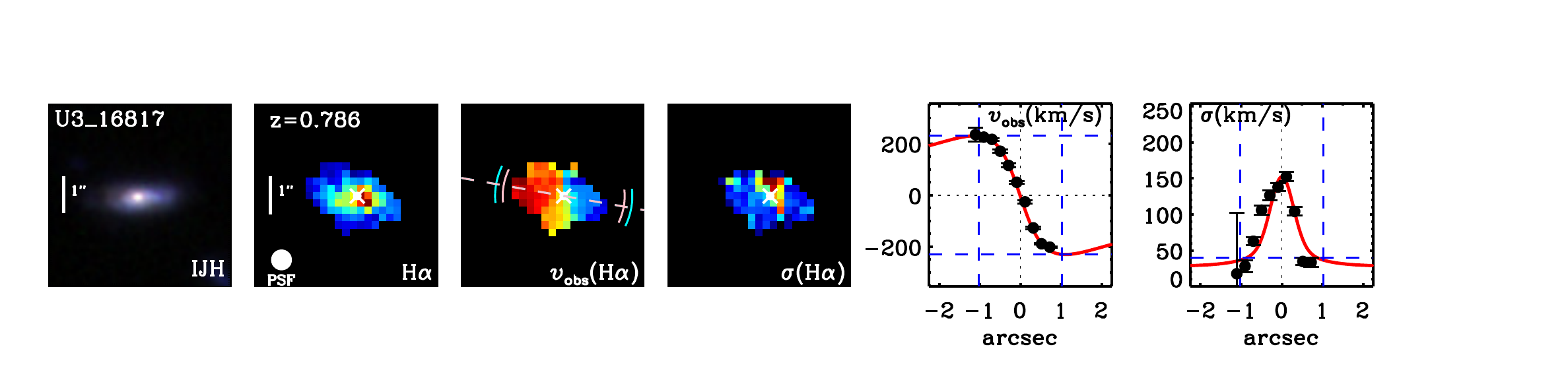}
\end{center}
\caption{\textit{cont.}; Kinematic maps and axis profiles for the high S/N disk galaxies in \kmostd first year data}
\end{figure*}
\addtocounter{figure}{-1}

\begin{figure*}
\begin{center}
\includegraphics[scale=0.8,  trim=1.5cm 0cm 1.0cm 0cm, clip, angle=90 ]{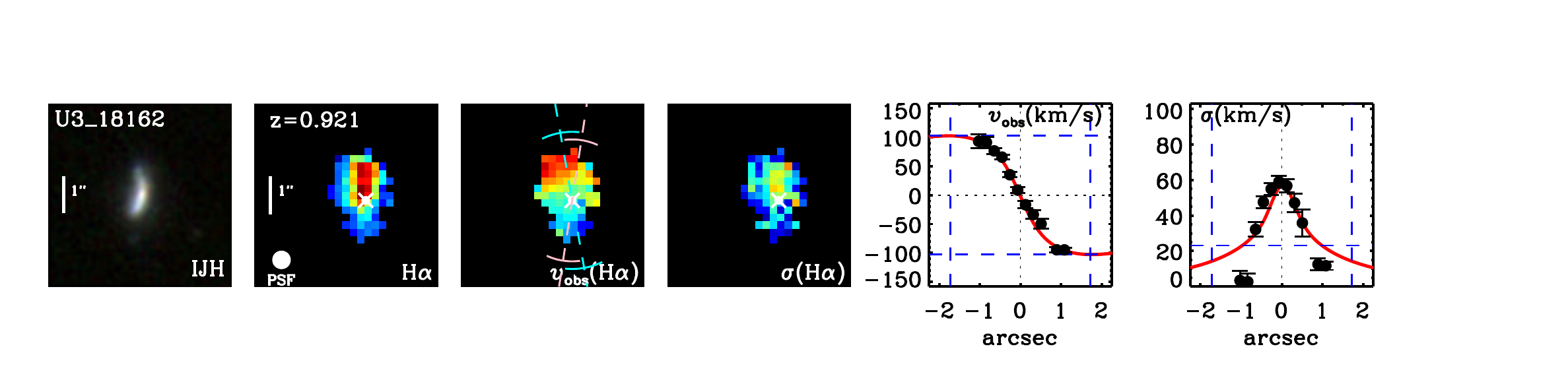}
\includegraphics[scale=0.8,  trim=1.5cm 0cm 1.0cm 0cm, clip, angle=90 ]{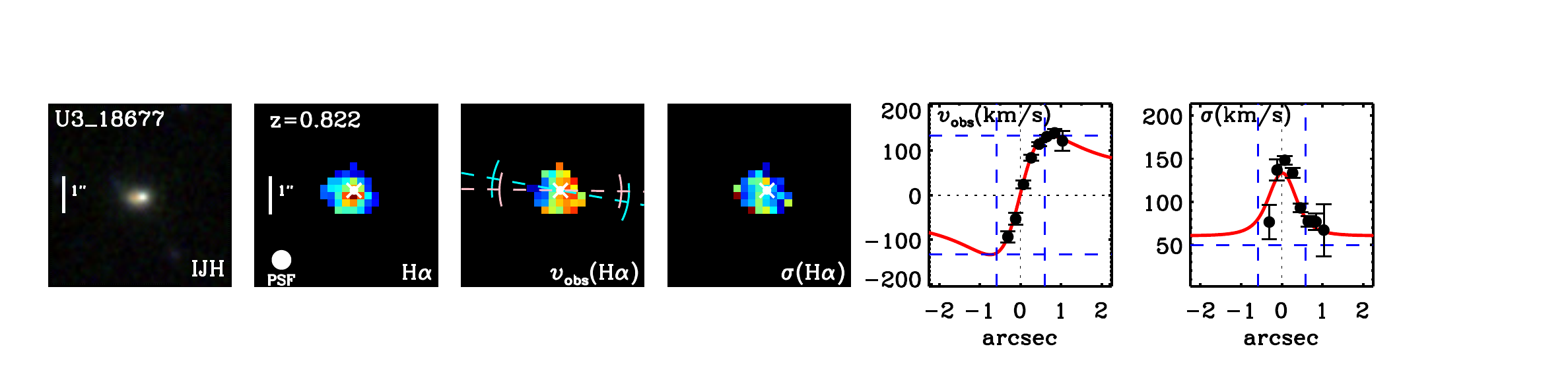}
\includegraphics[scale=0.8,  trim=1.5cm 0cm 1.0cm 0cm, clip, angle=90 ]{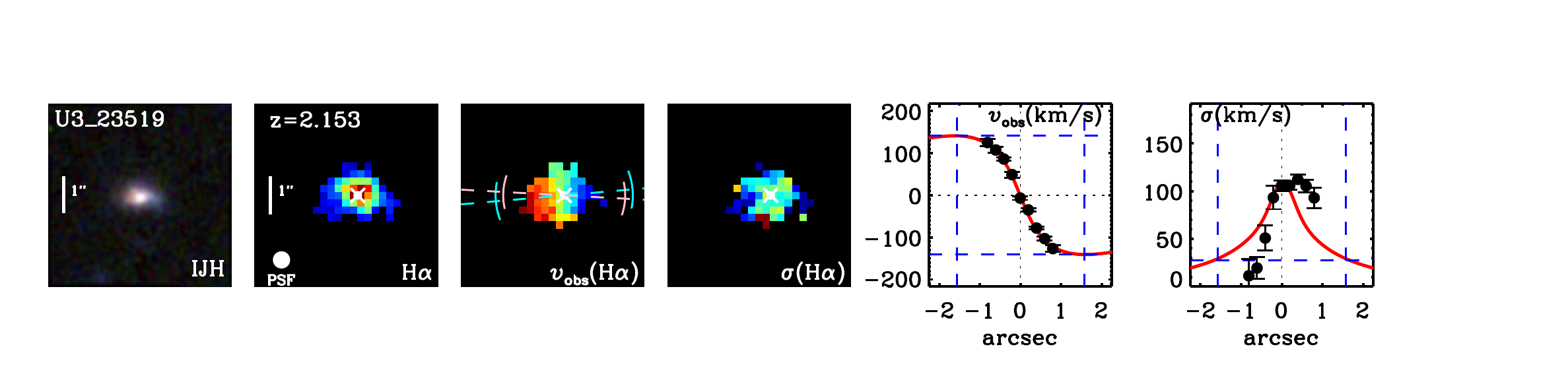}
\includegraphics[scale=0.8,  trim=1.5cm 0cm 1.0cm 0cm, clip, angle=90 ]{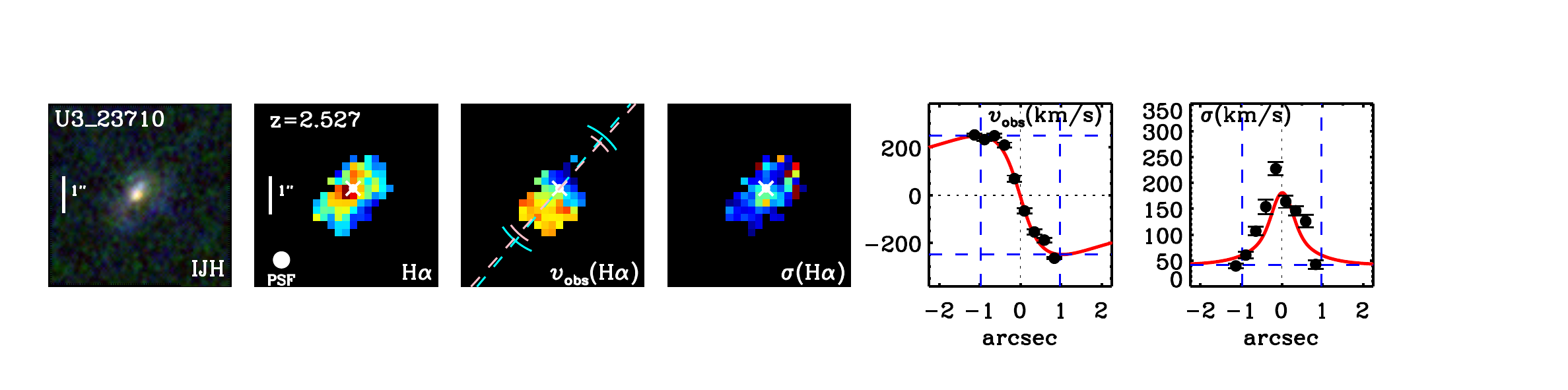}
\includegraphics[scale=0.8,  trim=1.5cm 0cm 1.0cm 0cm, clip, angle=90 ]{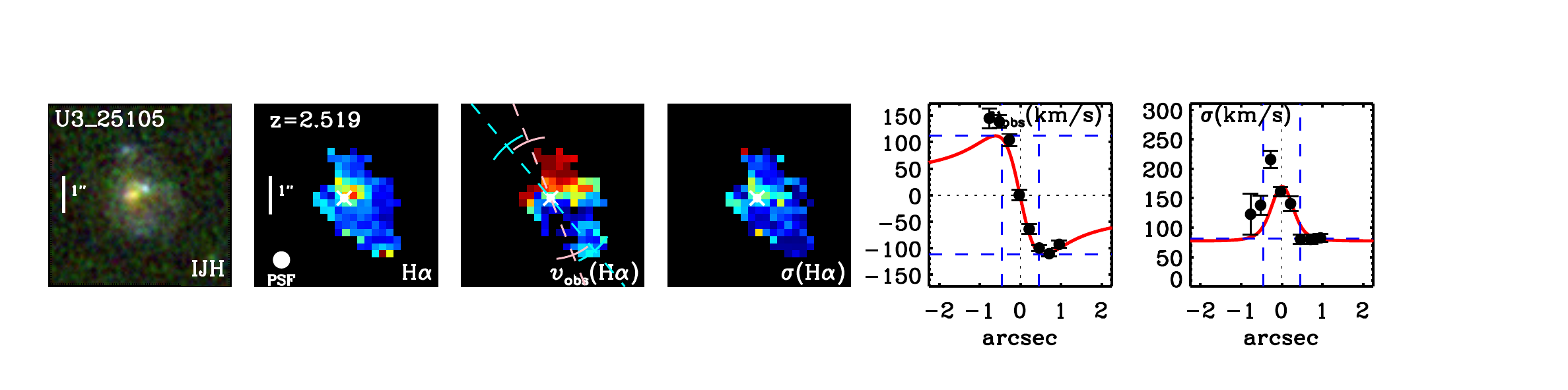}
\includegraphics[scale=0.8,  trim=0.0cm 0cm 1.0cm 0cm, clip, angle=90 ]{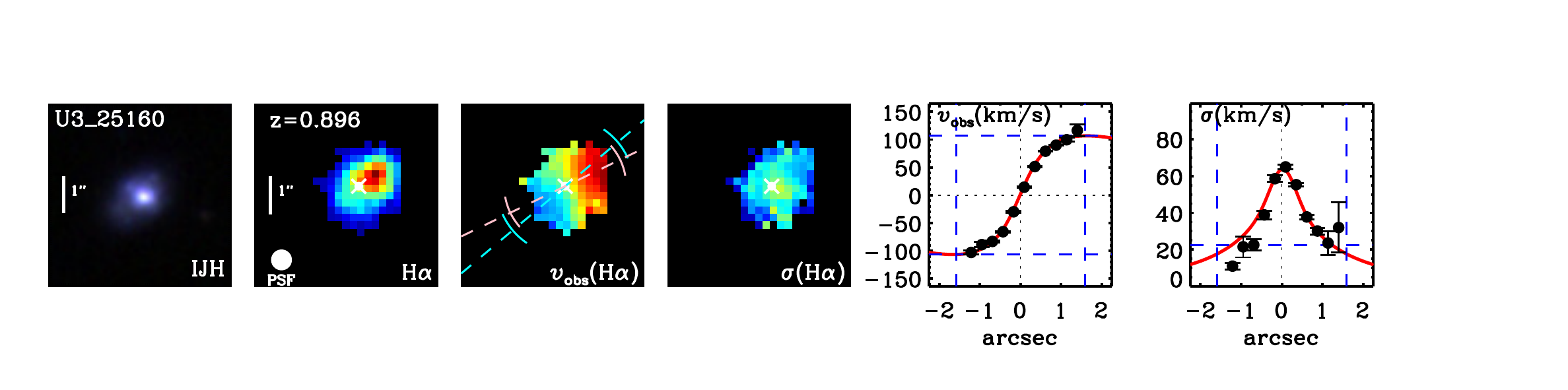}
\end{center}
\caption{\textit{cont.}; Kinematic maps and axis profiles for the high S/N disk galaxies in \kmostd first year data}
\end{figure*}
\addtocounter{figure}{-1}



\end{appendix}

\clearpage

\label{lastpage}

\end{document}